\documentclass[aps,pra,showpacs,twoside,10pt,floatfix,nofootinbib,longbibliography,twocolumn,10pt]{revtex4-2}
\usepackage[colorlinks=true, citecolor=blue, urlcolor=blue ]{hyperref}
\usepackage{epsfig,newlfont,amssymb,amsfonts,amsmath,bm,subfigure,palatino,mathtools,amsthm,braket,soul,enumitem,graphics,graphicx,times,physics, multirow, makecell}
\usepackage[normalem]{ulem}
\usepackage{tabularx}
\usepackage[table,xcdraw]{xcolor}

\begin{document}
\title{
Ergotropic Mpemba crossings in finite-dimensional quantum batteries
}

% \author{Triyas Sapui$^1$,  Tanoy Kanti Konar$^{1,2}$, Aditi Sen (De)$^1$}
% \affiliation{$^1$Harish-Chandra Research Institute, A CI of Homi Bhabha National Institute,  Chhatnag Road, Jhunsi, Prayagraj - 211019, India\\ 
% $^2$Institute of Theoretical Physics, Faculty of Physics, Astronomy, and Computer Science,\\ Jagiellonian University in Krakow, Stanisława \L{}ojasiewicza street 11, PL-30-348 Krak\'ow, Poland}

\author{Triyas Sapui$^{1,2}$}
\email{triyassapui@hri.res.in}
\author{Tanoy Kanti Konar$^{1,2,3}$}
\email{tanoy.kanti.konar@uj.edu.pl}
\author{Aditi Sen (De)$^{1,2}$}
\email{Corresponding author : aditi@hri.res.in}
\affiliation{$^1$Harish-Chandra Research Institute, Chhatnag Road, Jhunsi, Allahabad - 211019, India}
\affiliation{$^2$Homi Bhabha National Institute, Training School Complex, Anushakti Nagar, Mumbai 400 094, India}
\affiliation{$^3$Instytut Fizyki Teoretycznej, Wydzia\l{} Fizyki, Astronomii i Informatyki Stosowanej, Uniwersytet Jagiello\'nski, \L{}ojasiewicza 11, PL-30-348 Krak\'ow, Poland}

\begin{abstract}
The quantum Mpemba effect is a counterintuitive phenomenon in which a state initially farther from equilibrium relaxes more rapidly than one that starts nearer to equilibrium. In the context of finite-dimensional quantum batteries interacting with an environment, we introduce the notion of an ergotropic Mpemba crossing (EMC), defined by the intersection of ergotropy trajectories during the dynamics. For qubit batteries subjected to amplitude damping noise, we derive a condition for the occurrence of EMC in terms of the relative coherence of the initial states and fully characterize the region of state space that exhibits EMC with respect to a fixed reference state. Interestingly, our analysis reveals that under anisotropic Pauli noise, the emergence of EMC is jointly governed by the coherence and the energy of the initial states. To elucidate the physical origin of EMC, we decompose ergotropy into coherent and incoherent contributions and show that, in qubit systems, the coherent component plays a crucial role for EMC, an observation that strikingly does not extend to three-level batteries, \textcolor{black}{providing the dimensional role in observing EMC.} Further, by extending our analysis to non-Markovian environments, we demonstrate that, unlike the Markovian case, non-Markovian dynamics can give rise to multiple Mpemba crossings, with the total number of crossings always being odd. \textcolor{black}{Also, we propose a set-up showing that EMC can be utilized to suppress the ergotropic loss, occurred due to dissipation.} Moreover, analyzing the connection between the EMC and the conventional state Mpemba effect reveals that, for qubits, an EMC necessarily entails a state Mpemba crossing, while  this correspondence breaks down for qutrits, where EMCs may arise without any state Mpemba crossing.

%we examine the relationship between EMC and the conventional state Mpemba effect and demonstrate that for qubit systems, the EMC necessarily implies state Mpemba crossing, while it does not hold true in qutrit systems, where EMC can occur without state Mpemba crossings.

%analyze the \emph{ergotropic Mpemba effect} (EME) in finite-dimensional quantum systems, focusing on qubit and qutrit setups. We demonstrate that, in qubit systems, quantum coherence plays a crucial role in the emergence of ergotropic Mpemba crossings (EMC). Based on this insight, we identify the complete region of the state space that exhibits EMC with respect to a fixed reference state. From a physical perspective, we investigate the origin of EME by decomposing ergotropy into its coherent and incoherent contributions. We show that while incoherent ergotropy does not display EME in qubit systems, it can exhibit ergotropic Mpemba crossings in higher-dimensional systems, such as qutrits. We further extend our analysis to non-Markovian environments and show that, in contrast to the Markovian case, an odd number of Mpemba crossings can occur under non-Markovian dynamics. Finally, we examine the relationship between EME and the conventional state Mpemba effect, and demonstrate that for qubit systems the absence of the state Mpemba effect necessarily implies the absence of the ergotropic Mpemba effect.
 
\end{abstract}

\maketitle

\section{Introduction}
\label{sec:intro}

The Mpemba effect is a counterintuitive thermodynamic phenomenon in which a system prepared at a higher initial temperature relaxes to equilibrium faster than an initially colder one.
%a hotter system equilibrates faster than an initially colder one. 
It was first observed experimentally by Mpemba and Osborne~\cite{Mpemba1969} and a systematic theoretical framework for classical systems coupled to thermal environments was subsequently developed~\cite{Lu2017}. In the quantum domain, the Mpemba-like behavior has been explored primarily in nonequilibrium settings, such as random quantum circuits~\cite{liu_prl_2024,Turkeshi2025,Yu2025} and interacting many-body systems~\cite{Ares2023,Yamashika2024,Murciano2024,Ares2025}.  For open quantum systems, 
%the Mpemba effect often referred to as 
the phenomenon, often termed as \emph{thermal Mpemba effect} has been identified in both Markovian~\cite{Carollo2021,chatterjee_pra_2024,moroder_prl_2024,nava_prl_2024,Chatterjee2024,Qian2025,caldas2025,saliba2025,ulčakar2025} and non-Markovian dynamics~\cite{Strachan_prl_2025,Longhi_2025_systemenv_corr}. In the Markovian case, exponential acceleration of relaxation has been demonstrated via global rotations in a variety of models, including the Dicke model~\cite{Carollo2021}, spin chains~\cite{Kochsiek2022,Dong2025,wei2025,das2025}, quantum dots~\cite{Chatterjee2023,Wang2024,Nava2024}, fermionic systems~\cite{Wang2024}, and bosonic setups~\cite{Longhi2024,Longhi2025}, with relaxation typically quantified using distance measures such as the Hilbert–Schmidt and trace distances. The role of memory effects in the non-Markovian regime has also been extensively analyzed~\cite{Strachan2025,Li2025,zhang2026}
along with studies on the typicality of relaxation under environmental coupling~\cite{bao2026} (cf. ~\cite{li_arxiv_2025,bagui2025,solanki2025,lejeune2026}). 
Crucially, these theoretical insights have been supported by experimental realizations in platforms such as trapped-ion simulators ~\cite{joshi_prl_2024} and  nuclear magnetic resonance platforms~\cite{chatterjee2025,schnepper2025}.
%and nuclear magnetic resonance systems [38,39].
%Beyond theoretical developments, experimental realizations of the Mpemba effect have been achieved in trapped-ion simulators~\cite{joshi_prl_2024} and NMR platforms~\cite{chatterjee2025,schnepper2025}.
%Typacality of state relaxation under the environmntal effect has been explored in Ref.\cite{bao2026}.  
%More broadly, several variants of Mpemba-like phenomena have been reported in both nonequilibrium and open quantum systems~\cite{li_arxiv_2025,bagui2025,solanki2025,lejeune2026}.

In the quest to understand thermodynamic principles at the microscopic scale, quantum batteries, quantum-mechanical energy storage devices~\cite{Alicki2013,quantum_battery_review,Ferraro2026}, have emerged as a key platform for identifying genuine quantum advantages in energy storage and delivery. Over the past few years, extensive research has explored various aspects of quantum batteries, including collective and high-power charging protocols \cite{Binder2015,Campaioli2017,Ferraro2018,grazi2025}, the role of quantum correlations \cite{Andolina_prl_2019,Francica2020,Shi2022,vigneshwar2025}, fundamental bounds on capacity and power \cite{Farre2020,Yang2023}, and connections to quantum speed limits \cite{Mohan2021,Mohan_2022,Shrimali2024}. In realistic settings, however, quantum batteries inevitably interact with their environment, leading to dissipation and decoherence. This has motivated a growing body of work examining how environmental effects influence the charging and discharging dynamics of quantum batteries \cite{Ghosh2021,Zakavati2021,Arjmandi2022,sen2023,Liu2024,ahuja2025,Tirone2023,Tirone2024,Tirone2025,sarkar2025,Kamin2020,Santos2021,Xu2024,Morrone2023,Maryam2025,topological_quantumbattery,Vigneshwar2026}.

\textcolor{black}{The study of the Mpemba effect has attracted considerable attention due to its potential applications in emerging quantum technologies, where controlling the Liouvillian structure of open quantum systems can provide operational advantages. Closely related Mpemba-like phenomena have also been reported in several quantum thermodynamic settings, including self-contained quantum refrigerators~\cite{mondal2025} and quantum thermometry, where Mpemba crossings have been shown to offer an operational advantage over conventional equilibrium-based protocols~\cite{chattopadhyay2026} (see also Refs.~\cite{van_prl_2025,summer_prx_2026}). Beyond quantum thermodynamics, Mpemba-inspired protocols have recently found applications in fast quantum state preparation~\cite{feyisa2026} and thermodynamic computation~\cite{lejeune2026,moroder2026}, highlighting the broad significance of Mpemba physics in quantum technologies. \\
Within this context, a natural and intriguing question arises: \emph{Can quantum battery performance metrics, such as ergotropy, exhibit Mpemba-like behavior?} Addressing this question can identify pairs of initial states for which a battery with higher initial ergotropy discharges more slowly than another with lower initial ergotropy under environmental noise, thereby providing both fundamental insights into nonequilibrium quantum thermodynamics and practical guidance for designing robust quantum batteries. In this work, we answer this question affirmatively for quantum batteries composed of two- and three-dimensional systems subjected to both Pauli and non-Pauli noise.\\
While the ergotropic Mpemba effect has recently been reported for continuous-variable quantum batteries interacting with dissipative Markovian and non-Markovian environments~\cite{Medina2024, Li2025}, these findings do not naturally extend to discrete-variable systems. The charging mechanisms, state space, and dissipative dynamics of finite-dimensional quantum batteries differ fundamentally from those of continuous-variable platforms, preventing a direct generalization of the phenomenon. Since discrete-variable quantum batteries constitute one of the primary architectures that have been implemented in experimental platforms such as such as nuclear magnetic resonance (NMR) system~\cite{nmr_battery}, superconducting circuits~\cite{superconductQBexp}, establishing the existence of the ergotropic Mpemba effect and identifying its underlying mechanisms in finite-dimensional systems are essential.}

In this work,  focusing on a single-qubit battery subject to Markovian noise, modeled by generalized amplitude damping and Pauli channels, we derive conditions expressed in terms of properties of the initial states for the occurrence of EMC. For amplitude damping noise, we rigorously prove that two arbitrary initial states with ordered initial ergotropies cannot manifest EMC if their \(l_1\)-norm of coherence follows the same ordering. In contrast, for Pauli noise, the appearance of EMC is jointly governed by the relative coherence and energy of the initial states. To uncover the physical origin of this effect, we decompose ergotropy into coherent and incoherent contributions. For qubits, we find that the incoherent component decays exponentially, while the coherent one delays relaxation toward the steady state, thereby enabling EMC. Extending our analysis to higher-dimensional batteries, particularly qutrits, we demonstrate that EMC can arise even in the absence of coherent ergotropy, underscoring the essential role of additional energy levels. Furthermore, under non-Markovian amplitude damping noise, we identify the state-dependent conditions leading to EMC and prove that only an odd number of ergotropic crossings can occur. We examine the relationship between the ergotropic Mpemba effect and the conventional state Mpemba effect~\cite{carollo_prl_2021,moroder_prl_2024}. \textcolor{black}{Furthermore, we develop a protocol demonstrating how the EMC can be harnessed to mitigate the adverse effects of environmental noise on quantum battery performance. } We show that for qubit systems, the ergotropy Mpemba effect implies the state Mpemba effect, however, the converse does not generally hold for three-level systems.

This paper is structured as follows. Sec.~\ref{sec:single_qubit_mpemba} provides an overview of the ergotropic Mpemba effect and demonstrates the criteria to find ergotropy Mpemba crossing  for a two-qubit state with consideration of both amplitude damping and the anisotropic Pauli channel. In Sec.~\ref{sec:coherent_and _incoherent_ergotropy}, we illustrate that the occurrence of EMC due to the trade-off relationship between  the incoherent and coherent parts of the ergotropy itself. Sec. \ref{sec:qutrit_EME}, depicts the EMC in a qutrit system, specifically in a diagonal qutrit system. In Sec.~\ref{sec:nonmarkov_mpemba}, we demonstrate the emergence of EMC in a non-Markovian setup with multiple crossings. \textcolor{black}{We present a scheme of how ergotropic loss can be overcome using EMC in Sec.~\ref{sec:use_of_EMC}}.
Before concluding remarks in Sec.~\ref{sec:conclusion}, we connect the ergotropic Mpemba crossing with the state Mpemba effect in Sec. \ref{sec:connection_eme_state}.

\section{Criteria for detecting single-qubit ergotropic Mpemba effect}
\label{sec:single_qubit_mpemba}

The Mpemba effect naturally underscores the influence of the environment on physical systems. Such an effect can, in principle, be observed for any quantity associated with a quantum state, such as entanglement or coherence. While early studies focused on how the state itself approaches equilibrium, these derived quantities or resources need not follow the same relaxation behavior as the state (cf. \cite{carollo_prl_2021,moroder_prl_2024}).
%This makes it compelling to explore Mpemba-like phenomena in these quantum resources. 
%In this work, 
We here concentrate on  the ergotropic Mpemba effect in a finite-dimensional quantum battery. 
%We begin by presenting a detailed illustration of the ergotropic Mpemba effect.

\subsubsection{Concept of ergotropic Mpemba crossing}

Before presenting the results, we introduce the concept of the ergotropic Mpemba crossing, which serves as the central identifier for the Mpemba effect in an energy storage device.

\emph{Ergotropy.} The maximum extractable work from the battery through unitary operation, called \emph{ergotropy} ~\cite{Alicki2013} is defined as 
\begin{equation}
    \mathcal{E}(\rho)=\Tr[\rho H_B] - \min_{U} \Tr[U \rho U^{\dagger} H_B],
    \label{eq:ergotropy_def}
\end{equation}
where \(\rho\) is an arbitrary state of the battery Hamiltonian, $H_{B}=\sum_{i}\epsilon_i |\epsilon_i\rangle \langle \epsilon_i|$~\cite{Allahverdyan2004} where \(\epsilon_i\geq \epsilon_{i-1}\) for all \(i\). The second term in Eq. (\ref{eq:ergotropy_def}) represents the energy corresponding to the passive state, $\rho_{\pi}$, of $\rho$. If  $\rho=\sum_{i=1}^{n} p_i |p_i\rangle \langle p_i|$ where \(p_i\geq p_{i+1}\) for all \(i\), the passive state is given by $\rho_{\pi}=\sum_{i=1} p_{i} |\epsilon_i\rangle \langle \epsilon_i|$ where $\{|\epsilon_i \rangle\}$ is the set of energy eigenstates of the battery Hamiltonian. 
Using this, Eq. (\ref{eq:ergotropy_def})  reduces to \(\mathcal{E}(\rho) = \Tr[\rho H_B]-\sum p_i \epsilon_i\).

\textcolor{black}{\emph{Ergotropic Mpemba crossing.} The Mpemba-type behavior discussed here arises in the discharging dynamics of a quantum battery made up of qubit and qutrit systems. When two initial states with different ergotropies are coupled to a  bath, it may happen that the state with a higher ergotropy discharges faster than the one with a lower ergotropy. This phenomenon can be called the \emph{ergotropic Mpemba effect} and the corresponding intersection point of the two ergotropy curves at finite time is referred to as the \emph{ergotropic Mpemba crossing} (EMC).}
%It was first proposed by Medina \emph{et al.}, who showed that 
In a continuous-variable quantum battery, it was shown that a  highly charged state may discharge more rapidly, with the discharge rate being determined by the squeezing and displacement properties of the initial state~\cite{Medina2024}. 

\begin{figure*}
    \centering
    \includegraphics[width=\linewidth]{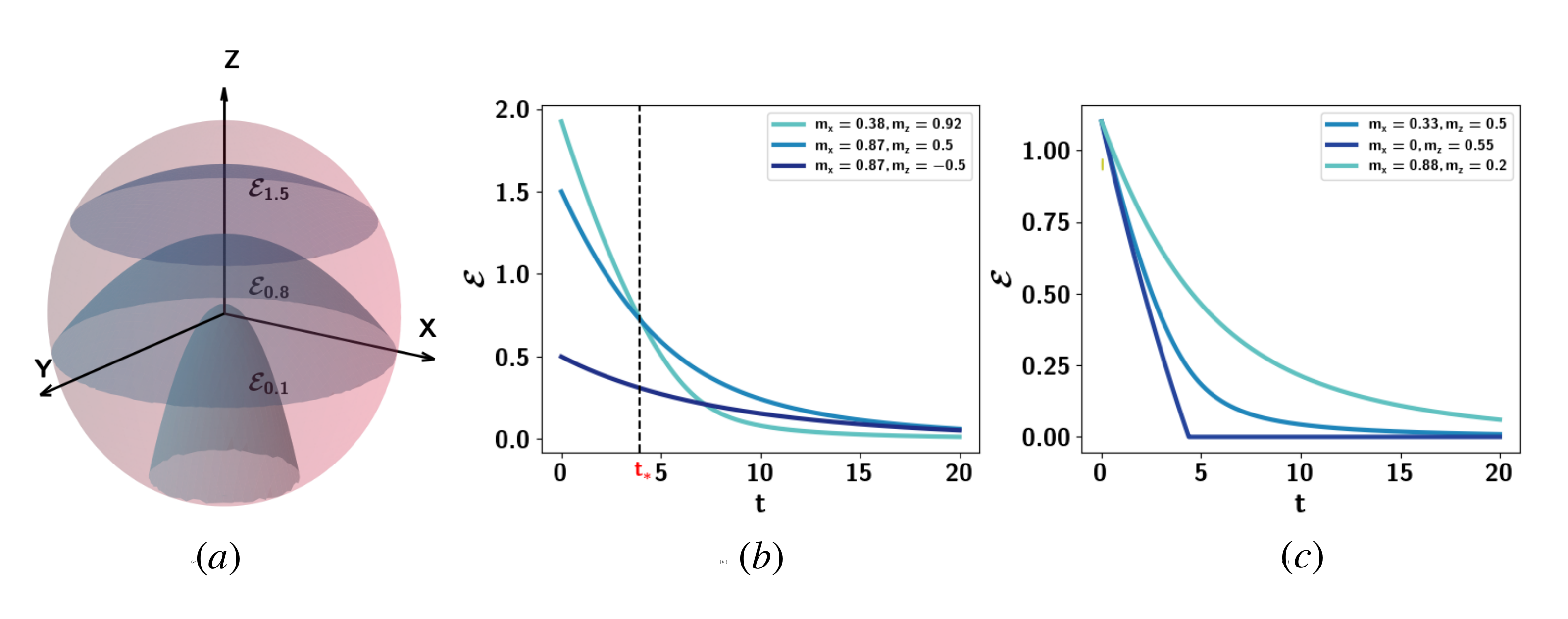}
    \caption{\textbf{ Ergotropic Mpemba crossing in a qubit system under ADC.} (a) Isoergotropic surfaces for different values of ergotropy are plotted on the Bloch sphere using  Eq.~(\ref{eq:iso_ergo}). This surface never intersects and forms a paraboloid surface in the Bloch sphere (see cf.~\cite{malavazi2025}). (b) Ergotropy (ordinate) against time (abscissa) for different initial ergotropic states. There are pairs of states that show ergotropic Mpemba crossings  at \(t=t_*\), which depend upon the state and the noise strengths. Also, there exist pairs of states that do not show crossings at any finite time. (c) Decay of ergotropy for isoergotropic states. There is no EMC if the initial state belongs to the isoergotropic surface and the state with highest \(m_z\) relaxes faster than the others. Other parameters of the systems are \(T=0\) and \(\gamma_-=0.1\).}
    \label{fig:iso_ergo_surface}
\end{figure*}

% \emph{A general framework for open quantum systems.} In realistic scenarios, any quantum system inevitably interacts with its surrounding environment. To describe the effective dynamics of the system alone, one traces out the environmental degrees of freedom, leading to the framework of open quantum systems. The evolution of such systems is commonly described by the Gorini–Kossakowski–Sudarshan–Lindblad (GKSL) ~\cite{breuer2002,nielsen2010,rivas2012} equation, also known as the quantum master equation (QME). While the QME can, in principle, accommodate both Markovian and non-Markovian dynamics, in this section we restrict our discussion to the Markovian case. The Markovian QME takes the form
% \begin{equation}
% \frac{d\rho}{dt}
% = -i[H_B,\rho]
% + \sum_k \gamma_k \left( L_k \rho L_k^{\dagger}
% - \frac{1}{2} \{ L_k^{\dagger} L_k, \rho \} \right),
% \label{eq:gksl}
% \end{equation}
% where $\rho$ is the reduced density matrix of the system obtained after tracing out the environment and \(H_B\) is the Hamiltonian of the system. The operators $L_k$ are the Lindblad jump operators that encode the action of the environment on the system, and $\gamma_k$ denote the corresponding decay rates determined by the system-environment coupling. In general, solving the QME explicitly is challenging, though exact solutions are available for certain specific models (cf.~\cite{}). A detailed analysis of the solution of the GKSL equation is provided in Appendix \ref{sec:vectorization}.

In our work, we consider that the battery is under the influence of two different Markovian noises -- (1) generalized amplitude damping channel (gADC) and (2) anisotropic Pauli channel. In the former case, the battery is influenced by a thermal bath with temperature, \(T\) and the governing Gorini–Kossakowski–Sudarshan–Lindblad (GKSL) master equation ~\cite{nielsen2010,breuer2002,rivas2012} reads 
\begin{equation}
    \frac{d\rho}{dt}=-i[H_B,\rho]+\gamma_{-}\mathcal{D}[\sigma_-]+\gamma_{+}\mathcal{D}[\sigma_+],
    \label{eq:adc_channel}
\end{equation}
where \(\mathcal{D}[A]=A\rho A^\dagger-\frac{1}{2}\{A^\dagger A,\rho\}\), \(\gamma_{\pm}\ge0\) are the decay  rates of the ADC, and \(\sigma_\pm = (\sigma_x \pm i \sigma_y)/2\) are the Lindblad jump operators. Here \(\gamma_+=\gamma_-e^{-2\beta h_z}\) and 
%such maps are called Davis map, i.e., 
the steady state of the Lindblad channel is a thermal state,  \(\rho_{ss}=\frac{e^{-\beta H_B}}{\Tr[e^{-\beta H_B}]},\) where \(\beta={1}/{k_B T}\) is the inverse temperature of the bath with \(k_B\) being the Boltzmann constant. If the bath temperature is taken to be vanishing, i.e.,  \(T=0\), implying \(\gamma_+=0\), the corresponding  steady state is the ground state of the Hamiltonian, \(H_B\).  

Unlike gADC, let us consider another channel, namely \emph{Pauli} channel. The evolution of the qubit is given as 
\begin{equation}
    \frac{d\rho}{dt}=-i[H_B,\rho]+\gamma_x\mathcal{D}[\sigma_x]+\gamma_y\mathcal{D}[\sigma_y]+\gamma_z\mathcal{D}[\sigma_z],
    \label{eq:pauli_channel}
\end{equation}
where the dissipation strength \(\gamma_{x,y,z}\ge 0\).
%and are dissipation strength of the environment. 
We consider  \(\gamma_x=\gamma_y=\gamma_{\perp}\) with arbitrary value of \(\gamma_z\). For this case, the dynamics is again phase covariant and is referred to as anisotropic Pauli noise. So, we can divide three regimes where \(\gamma_{\perp}>\gamma_z\), \(\gamma_{\perp}=\gamma_z\) and \(\gamma_{\perp}<\gamma_z\) which depicts the strength of the noise in each direction. Interestingly, for all the cases, the steady state is given by \(\mathbb{I}_2/2\).

In order to study EMC, let us consider ergotropies of two initial states, \(\rho_1\), and \(\rho_2\), as    $\mathcal{E}_1\equiv\mathcal{E}(\rho_1)$ and $\mathcal{E}_2\equiv\mathcal{E}(\rho_2)$ respectively, with $\mathcal{E}_1>\mathcal{E}_2$.
When the system is coupled to a dissipative bath, the ergotropy typically decreases with time and eventually saturates to a steady state value. We call the dynamics exhibiting EMC if there exists a finite time, $t=t_*$,  such that
\begin{equation}
\boxed{\mathcal{E}_1(t_*^-)>\mathcal{E}_2(t_*^-)
   \;\;\underset{t=t_*}{\longrightarrow}\;\;
   \mathcal{E}_1(t_*^+)<\mathcal{E}_2(t_*^+),}
\end{equation}
where $t_*^\pm$ denote times immediately before and after $t_*$, respectively.

\subsection{State-dependent condition for EMC under amplitude damping noise}
\label{sec:conadc}

%We investigate the ergotropic Mpemba Crossing in finite-dimensional quantum systems. 
Let us consider a single qubit Hamiltonian for a quantum battery,
%\textcolor{black}{} 
\[
H_B^{d=2} = h_z \sigma_z,
\]
where $h_z$ and $\sigma_z$ denote the strength of the external magnetic field and $\sigma_z$ is the Pauli operator respectively. To analyze the emergence of the Mpemba crossing, we prepare two different initial states, having two different sets of magnetization components along the three spatial directions, denoted as $m_i$  ($i\in {x,y,z}$), given by
\begin{equation}
    \rho^{d=2}(0)=\frac{1}{2}\left(\mathbb{I}_2+m_x\sigma_x +m_y\sigma_y+m_z\sigma_z\right),
\label{eq:qubit_initial_state}
\end{equation}
where $\mathbb{I}_2$ is the two-dimensional identity operator. 
%Using Eq.~(\ref{eq:ergotropy_def}), 
The ergotropy, in this case, leads to
\begin{equation}
    \mathcal{E}(\rho^{d=2}(0)) = h_z \left(m_z +\sqrt{m_x^2+m_y^2+m_z^2}\right).
    \label{eq:iso_ergo}
\end{equation}
The above quadratic dependence generates a paraboloid structure in the Bloch sphere when the left hand side is constant, defining \emph{isoergotropic} surfaces (all states on a given surface, which has a rotational symmetry about the $z-$axis, possess identical ergotropy), parametrized by the magnetization vector (see Fig.~\ref{fig:iso_ergo_surface}(a)).  Such surfaces help to understand the EMC phenomena more clearly. %\textcolor{black}{ 
%All states on a given surface possess identical ergotropy, and the rotational symmetry about the $z-$axis. 
From now on, we set $h_z=1$ which does not hamper results as \(h_z\) only scales the energy of the system. 
%\textcolor{black}{ We consider two sets of initial states,  \(\Lambda\in\{\rho_1^\Lambda,\rho_2^\Lambda\}\) and  \(\Gamma\in\{\rho_1^\Gamma,\rho_2^\Gamma\}\).  In the set \(\Lambda\), the two states possess different initial ergotropies, i.e., \(\mathcal{E}_1^\Lambda\ne \mathcal{E}_2^\Lambda\), specifically  \(\mathcal{E}_1^\Lambda> \mathcal{E}_2^\Lambda\). While, the set \(\Gamma\) consists of isoergotropic states such that  \(\mathcal{E}_1^\Gamma= \mathcal{E}_2^\Gamma\).}

%\emph{EMC under amplitude damping .} 
%After state preparation, 
\textcolor{black}{When the system is coupled to a thermal bath at temperature \(T=0\), the evolved state can be calculated easily by using the vectorization method described in Appendix~\ref{sec:vectorization} and \ref{app:single_qubit_evolution} and 
the ergotropy of the time-evolved state is given as 
\begin{eqnarray} 
\mathcal{E}(t,m_x,m_y,m_z)=
&& e^{-\gamma_- t} \Bigg [ \nonumber (1-e^{\gamma_- t} + m_z)\\+&&\sqrt{(1-e^{\gamma_- t} + m_z)^2+e^{\gamma_- t}(m_x^2+m_y^2)} \Bigg ],\nonumber\\
\label{eq:ergotropy_time}
\end{eqnarray}
which depends on the initial state parameters and the environmental dissipation rate. Note, further, that it is invariant under a rotation along \(z\)-axis, which we call it as ''phase covariant," and depends only on the magnitude of the transverse Bloch vector, \(m_x^2+m_y^2\), thereby remaining independent of its orientation in the \(xy\)-plane.}
%In Fig.~\ref{fig:iso_ergo_surface}(b), we plot the ergotropy dynamics of two arbitrary state with different ergotropy. 
We observe ergotropic Mpemba crossings at a finite time 
\(t_{*}\) at which the corresponding ergotropy curves intersect, as shown in Fig.~\ref{fig:iso_ergo_surface}(b). This demonstrates that the EMC can occur in finite-dimensional systems, in close analogy with earlier observations in continuous-variable quantum batteries~\cite{Medina2024}, although the phenomenon is not generic. \textcolor{black}{Indeed, for other choices of initial states, no such crossing is observed}, indicating that the emergence of EMC is highly sensitive to the initial state preparation (see Fig.~\ref{fig:iso_ergo_surface}(c)).

Motivated by these observations, a natural question arises: {\it ``What conditions on the initial states lead to the occurrence of an EMC?"} More specifically, {\it is there an underlying structural relationship between pairs of initial states that provides a criterion for the occurrence of the EMC?} To address these questions, we prove the following two lemmas.

% In contrast, when the initial states belong to the set \(\Gamma\), we find that the ergotropy curves do not intersect during the evolution (see Fig.~\ref{fig:iso_ergo_surface}(c)). Interestingly, we observe that among all isoergotropic states, the one with the largest magnetization along the \(z\) direction relaxes the fastest and saturates to its steady-state value first, consistent with the findings of Ref.~\cite{malavazi2025}. We refer to this phenomenon as \emph{ergotropic faster relaxation} (EFR).

\textbf{Lemma 1.} \emph{ For a pair of isoergotropic states, the state possessing a larger magnetization along the $z$ direction ($m_z$) exhibits a faster decay of ergotropy compared to the state with smaller  $m_z$.}
%For two isoergotropic states, the ergotropy of the state with a larger magnetization along the $z$ direction ($m_z$) decays faster than that of the state with a smaller $m_z$.}
\begin{figure*}
    \centering
    \includegraphics[width=\linewidth]{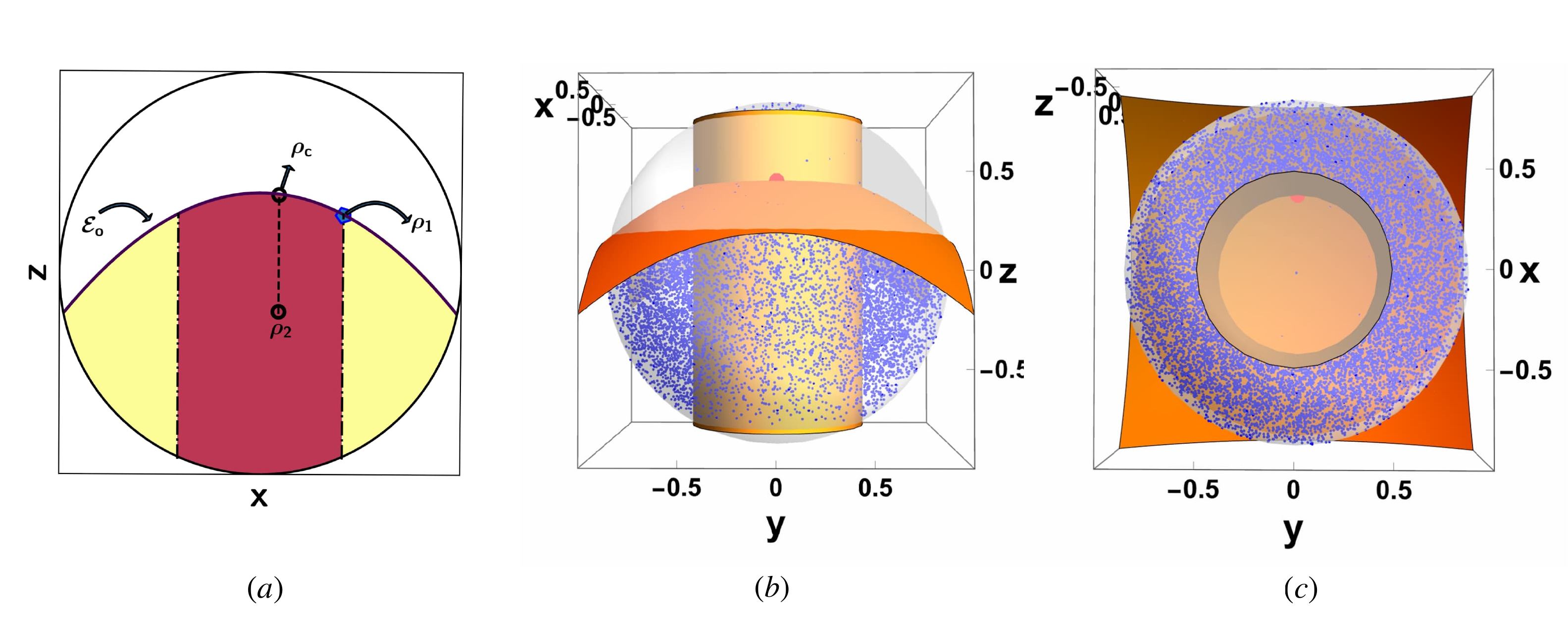}
    \caption{\textbf{No EMC region for AD channel.} (a) The non-EMC region is plotted in the \(xz\)-plane for a corresponding fixed state \(\rho_1\) under ADC. The no-EMC region is denoted by \textcolor{black}{the red region} with a dashed boundary, and \(\mathcal{E}_o\) denotes the isoergotropic line in the \(xz\)-plane.\textcolor{black}{We take \(\rho_1\) given by \(m_x=0.5,m_y=0.0,m_z=0.5\) (denoted by a small dot) and for \(\rho_2\), we generate random initial states from the Bloch sphere and see if there is any crossing.} (b) and (c)  Side and down views  of the EMC region  where the dots in the Bloch sphere are the states that show EMC with respect to a fixed state \(\rho_1\) having ergotropy \(\mathcal{E}_o\).\textcolor{black}{The orange surface denotes an isoergotropic surface corresponding to the state, \(\rho_1\) and the yellow cylinder consists of all the states that have the same coherence as the state \(\rho_1\).} Other parameters of the systems are \(h_z=1\), \(T=0\) and \(\gamma_-=0.01\).   }
    \label{fig:xz_plane}
\end{figure*}
\begin{proof}
Without loss of generality, we restrict our attention to states lying in the $xz-$plane, i.e., we set $m_y=0$, reducing Eq.  (\ref{eq:ergotropy_time}) as
\begin{align}
\mathcal{E}(t,m_x,m_z)
= e^{-\gamma_- t} \Big [ \nonumber & (1-e^{\gamma_- t} + m_z)+\\&\sqrt{(1-e^{\gamma_- t} + m_z)^2+e^{\gamma_- t}m_x^2} \Big ].
\label{eq:ergo_zxplane}
\end{align}
For isoergotropic states, the initial ergotropy satisfies
\begin{equation}
   \mathcal{E}_0\equiv\mathcal{E}(0,m_x,m_z) = m_z+\sqrt{m_x^2+m_z^2},
   \label{eq:iso_ergo_xz}
\end{equation}
where $\mathcal{E}_0$ is a constant. Rearranging this expression, $m_x^2=\mathcal{E}_0^2-2\mathcal{E}_0 m_z$, and substituting this into Eq.~(\ref{eq:ergo_zxplane}), we now prove that for a given \(t\) and \(\gamma_-\), the ergotropy monotonically decreases with the increase of \(m_z\) on this surface. This is true for all values of \(t\) and valid \(\gamma_-\). To do so, we compute
\begin{align}
    &\frac{\partial \mathcal{E}(t,m_z)}{\partial m_z}=e^{-\gamma_-t}\Bigg [1+\frac{2(1-e^{\gamma_- t} + m_z)-2e^{\gamma_-t}\mathcal{E}_0}{2\sqrt{(1-e^{\gamma_- t} + m_z)^2+e^{\gamma_- t}m_x^2}}\Bigg ]\nonumber\\&=e^{-\gamma_-t}\Bigg [ 1+\frac{e^{-\gamma_-t}(1+m_z)-1-\mathcal{E}_0}{\sqrt{(e^{-\gamma_-t}(1+m_z)-1)^2+e^{-\gamma_- t}m_x^2}}\Bigg]\nonumber\\&\qquad\qquad\qquad\qquad\equiv e^{-\gamma_-t}\Bigg[1+\frac{A}{B}\Bigg ].
    \label{eq:dervitaive_ergo}
\end{align}
% For $\mathcal{E}(t,m_z)$ to be a monotonically decreasing function of $m_z$ for all $m_z$, we require
% \[
% \frac{\partial \mathcal{E}(t,m_z)}{\partial m_z}<0
% \quad\Longleftrightarrow\quad
% A<0
% \ \text{and}\
% \left|\frac{A}{B}\right|>1.
% \]
First, note that \( A\equiv e^{-\gamma_-t}(1+m_z)-1-\mathcal{E}_0 <0, \)
which follows from the inequalities
\begin{eqnarray}
    &&e^{-\gamma_-t}(1+m_z)<1+m_z\nonumber\\
    &\Rightarrow& e^{-\gamma_-t}(1+m_z)-1<m_z<\mathcal{E}_0\nonumber\\
    &\Rightarrow& e^{-\gamma_-t}(1+m_z)-1-\mathcal{E}_0<0\nonumber\\
    &\Rightarrow& A<0,
\end{eqnarray}
where we have used $m_z<m_z+\sqrt{m_x^2+m_z^2}=\mathcal{E}_0$.
Secondly,
\begin{eqnarray}
    A^2-B^2&=\mathcal{E}_0^2-{2\mathcal{E}_0m_z}e^{-\gamma_-t}+2\mathcal{E}_0(1-e^{-\gamma_-t})\nonumber\\&
    -e^{-\gamma_-t}(\mathcal{E}_0^2-2\mathcal{E}_0m_z)\nonumber\\&
    =(1-e^{-\gamma_-t})(\mathcal{E}_0^2+2\mathcal{E}_0)>0,
\end{eqnarray}
implying that \(|\frac{A}{B}|>1\). These two results together imply
$\frac{\partial \mathcal{E}(t,m_z)}{\partial m_z}<0$, demonstrating that in the $xz-$plane, the ergotropy of an isoergotropic state with a larger $m_z$ relaxes faster than that of a state with a smaller $m_z$. Finally, because the amplitude damping channel is phase covariant, \textcolor{black}{the time-evolved ergotropy is symmetric under rotations about the $z-$axis}. Hence, \textcolor{black}{the isoergotropic curve generates a rotationally symmetric surface in the Bloch sphere, and the above conclusion holds for all states on that surface.}
\end{proof}

\textcolor{black}{\textbf{Lemma 1} establishes that no ergotropic Mpemba crossing (EMC) can occur between two states possessing the same initial ergotropy. In particular, among states lying on the same isoergotropic surface, those with a larger value of \(m_z\) relax more rapidly, as illustrated in Fig.~\ref{fig:iso_ergo_surface}(c) (see also Ref.~\cite{malavazi2025} and Appendix~A of Ref.~\cite{Malavazi_prxq_2025}). However, this result is restricted to states belonging to the same isoergotropic surface and, therefore, does not provide a general criterion for the occurrence of EMC between arbitrary pairs of initial states, as exemplified in Fig.~\ref{fig:iso_ergo_surface}(b). To address this more general scenario, we establish the following lemma.}

% Fig.~\ref{fig:iso_ergo_surface}(c) illustrates the relaxation dynamics of ergotropy for initial states lying on an isoergotropic surface 
% %Among these states, the one with the largest value of $m_z$ relaxes faster than all others with smaller $m_z$, in full agreement with Lemma~1 
% (see also Ref.~\cite{malavazi2025} and  Appendix~A of Ref.~\cite{Malavazi_prxq_2025}). We now prove how the properties of initial states with identical initial ergotropy govern the subsequent ergotropy dynamics.
%a rigorous proof of this behavior for
%the ergotropy dynamics when all initial states possess the same initial ergotropy.

\textbf{Lemma 2.} 
\emph{For two non-isoergotropic states with identical $m_x$, 
the state with the larger  magnetization in the \(z\)-direction,
\(m_z\), exhibits a slower decay of ergotropy compared to the state with a smaller \(m_z\).}

%the ergotropy of the state with a larger $m_z$ decays more slowly than that of the state with a smaller $m_z$.}

\begin{proof}
By fixing a value of $m_x=M_x$ in Eq.~(\ref{eq:ergo_zxplane}) and compute the derivative of the ergotropy with respect to $m_z$, we obtain
\begin{align}
    \frac{\partial \mathcal{E}(t,m_z)}{\partial m_z}
    &=e^{-\gamma_-t}\Bigg [1+\frac{1-e^{\gamma_- t} + m_z}{\sqrt{(1-e^{\gamma_- t} + m_z)^2+e^{\gamma_- t}M_x^2}}\Bigg ]\nonumber\\
    &\equiv e^{-\gamma_-t}\Bigg[1+\frac{C}{D}\Bigg ]
\end{align}
for which it can be shown that \(|\frac{C}{D}|<1\),
% \[
% \left|\frac{C}{D}\right|<1,
% \]
since 
\[
\left|1-e^{\gamma_- t} + m_z\right|
<
\sqrt{(1-e^{\gamma_- t} + m_z)^2+e^{\gamma_- t}M_x^2}.
\]
Hence,
\begin{equation}
    \frac{\partial \mathcal{E}(t,m_z)}{\partial m_z}>0,
\end{equation}
which shows that $\mathcal{E}(t,M_x,m_z)$ is a monotonically increasing function of $m_z$, when $m_x$ is taken to be fixed. Hence the proof.
\end{proof}

% \begin{figure}
%     \centering
    
%     \caption{Caption}
%     \label{fig:}
% \end{figure}

We are now ready to provide the conditions under which two initial states exhibit no Mpemba crossing in terms of their coherence.

\textbf{Theorem 1.} \emph{For an amplitude damping channel, two arbitrary initial states $\rho_1$ and $\rho_2$ with initial ergotropies $\mathcal{E}_1\ge\mathcal{E}_2$, do not exhibit the ergotropic Mpemba crossing if their coherence satisfy \[\mathcal{C}(\rho_1)\ge \mathcal{C}(\rho_2),\] 
where \(\mathcal{C}(.)\) is the \(l_1\)-norm of coherence in the energy eigenbasis of the battery Hamiltonian.}
%and an ergotropic Mpemba crossing is obtained if \(\mathcal{C}(\rho_1)< \mathcal{C}(\rho_2)\),

\begin{proof}
The proof follows from Lemmas 1 and 2, which together identify the surface of the spherocylinder and rely on the rotational symmetry of ergotropy about the $z$-axis under amplitude damping dynamics.

Let $\rho_1$ lie in the $xz$-plane with magnetization vector 
$\vec{m}_1=(m_{x_1},0,m_{z_1})$ and initial ergotropy $\mathcal{E}_1$, and similarly, the magnetization vector of \(\rho_2\) with initial ergotropy $\mathcal{E}_2$ be
%let the state
%$\rho_2$, have magnetization vector 
$\vec{m}_2=(m_{x_2},0,m_{z_2})$ with \(|m_{x_2}| < |m_{x_1}| \). %and ergotropy $\mathcal{E}_2$, where 
Here $\mathcal{E}_1>\mathcal{E}_2$ (see Fig.~\ref{fig:xz_plane}(a)). 
To show that $\mathcal{E}_1(t)>\mathcal{E}_2(t)$ for all $t>0$, consider a third state $\rho_c$ which have same \(m_x\) value as \(\rho_2\) lying on the isoergotropic curve corresponding to $\mathcal{E}_1$. Thus, $\rho_c$ has the same ergotropy as $\rho_1$, and its magnetization vector is 
$\vec{m}_c=(m_{x_2},0,m_{z_c})$ with $m_{z_c}>m_{z_1},m_{z_2}$. 
%\textcolor{black}{Note that $\rho_c$ shares the same $x$-component of magnetization as $\rho_2$.} 
From Lemma~1, we obtain
%\begin{equation}
    \(\mathcal{E}_1(t)>\mathcal{E}_c(t)\),
 %   \label{eq:res_first_lemma}
%\end{equation}
since both states lie on the same isoergotropic curve and $\rho_c$ has the larger $m_{z_c}$ value than \(m_{z_1}\). Similarly, Lemma~2 implies
%\begin{equation}
    \(\mathcal{E}_c(t)>\mathcal{E}_2(t)\).
   % \label{eq:res_second_lemma}
%\end{equation}
%because $\rho_c$ has larger $m_z$ than $\rho_2$ while sharing the same $m_x$. 
Combining them, we get
%Eqs.~(\ref{eq:res_first_lemma}) and (\ref{eq:res_second_lemma}), we find
\begin{equation}
    \mathcal{E}_1(t)>\mathcal{E}_c(t)>\mathcal{E}_2(t)
    \Rightarrow 
    \mathcal{E}_1(t)>\mathcal{E}_2(t).
\end{equation}
Since $\rho_2$ is arbitrary within the region satisfying \(\mathcal{E}_1 > \mathcal{E}_2\) and \(|m_{x_2}| < |m_{x_1}| \), the set of such states forms a region, denoted by $\mathcal{R}_{NE}$ in the $xz$-plane (see the red region of Fig.~\ref{fig:xz_plane}(a)). Therefore,
\begin{equation}
    \mathcal{E}_1(t)>\mathcal{E}_2(t)\quad \forall\rho_2\in \mathcal{R}_{NE}.
\end{equation}

Since the ADC is phase covariant, the ergotropic dynamics remain invariant under rotations about the $z$-axis. Rotating the region $\mathcal{R}_{NE}$ forms a three-dimensional volume $\mathcal{V}_{NE}$ in the Bloch sphere, which has the geometry of a spherocylinder (see Fig.~\ref{fig:xz_plane}(b) and (c)).
%, \textcolor{black}{with $\rho_1$ lying on the intersection of smooth and the curved surface.} 
Therefore, no Mpemba crossing occurs for any state $\rho_2$ within this region. Precisely, all states $\rho_2$ that lie inside the spherocylinder are characterized by a transverse distance from the $z$-axis given by $\sqrt{m_{x_2}^2+m_{y_2}^2}$, which is bounded above by the corresponding distance $\sqrt{m_{x_1}^2+m_{y_1}^2}$ associated with $\rho_1$. This relation can be expressed as
\begin{eqnarray}
        \sqrt{m_{x_2}^2+m_{y_2}^2}&\le& \sqrt{m_{x_1}^2+m_{y_1}^2}\nonumber\\
        \Rightarrow \mathcal{C}(\rho_2)&\le & \mathcal{C}(\rho_1),
\end{eqnarray}
where, in the last step, we have used the fact that the $l_1$-norm of coherence of a qubit in the $\sigma_z$ basis is given by $\sqrt{m_x^2+m_y^2}$.
%%, with $m_x$ and $m_y$ denoting the magnetization components along the $x$ and $y$ directions, respectively. %This completes the proof.

\end{proof}

\textit{EMC region in terms of initial state parameters.} The preceding analysis characterizes only the region in which the EMC is absent and does not explicitly identify the set of states that \emph{do} exhibit ergotropic Mpemba crossings with a fixed reference state, $\rho_1$, which we will now illustrate.
% \textcolor{red}{This discussion also answers that {\bf Theorem 1} provides a sufficient condition but not a necessary one.}
%. Let us now identify the properties of pairs of initial states 
%We now show that, for $\mathcal{E}_1\ge \mathcal{E}_2$, the $l_1$-norm of coherence provides a necessary and sufficient criterion for the occurrence of EME. Our argument relies on the fact that the time-dependent ergotropy is a monotonically decreasing function of time $t$.

%\textcolor{black}{Owing to the phase covariance of the amplitude damping channel, we again restrict our attention to states in the $xz$-plane} and 
Let us now analyze the asymptotic behavior of the ergotropy in the long-time limit, $t\to\infty$,  %This limit 
in Eq.~(\ref{eq:ergo_zxplane}) 
leading to 
\begin{eqnarray}
   && \mathcal{E}(t,m_x,m_z)\nonumber\\
%= e^{-\gamma_- t} \Big [ \nonumber  (1-e^{\gamma_- t} + m_z)\nonumber\\
%&& \qquad\qquad\qquad +\sqrt{(1-e^{\gamma_- t} + m_z)^2+e^{\gamma_- t}m_x^2} \Big ]\nonumber\\
&& \simeq  e^{-\gamma_- t} (1+m_z-e^{\gamma_- t})
+\sqrt{1+e^{-\gamma_- t}(m_x^2-2(1+m_z))}\nonumber\\
&&\simeq e^{-\gamma_- t}(1+m_z)
+\frac{1}{2}e^{-\gamma_- t}(m_x^2-2(1+m_z))\nonumber\\
&&=\frac{1}{2}m_x^2e^{-\gamma_- t}
\end{eqnarray}
This shows that, at long times, the ergotropy decays exponentially and depends solely on the transverse magnetization $m_x$.
Consider a state $\rho_2$ lying outside the spherocylinder but under the isoergotropic surface, corresponding to the reference state 
%bounded by the ergotropy,} \(\mathcal{E}_1\) of the reference state 
$\rho_1$. If \(\mathcal{C}(\rho_2) > \mathcal{C}(\rho_1)\), 
%. In this case, \(\rho_2\) has 
$|m_{x_2}| > |m_{x_1}|$, it implies that at sufficiently large times, the ergotropy of $\rho_2(t)$ exceeds that of $\rho_1(t)$. However, by assumption, the initial ordering satisfies $\mathcal{E}_1>\mathcal{E}_2$ and hence, we have
%. This situation can be summarized as
\(
\mathcal{E}_1>\mathcal{E}_2
\;\;{\longrightarrow}\;\;
\mathcal{E}_1(t\to\infty)<\mathcal{E}_2(t\to\infty),
\)
which necessarily implies the existence of a finite crossing time $t=t_*$. 
%Since $\rho_2$ is an arbitrary state outside the spherocylinder, this argument generalizes to all the states outside the spherocylinder, obeying the relation $\mathcal{E}_2\le \mathcal{E}_1$. Hence, all such states exhibit ergotropic Mpemba crossings with $\rho_1$. Moreover, all these points outside the spherocylinder satisfy
%\begin{equation}
% \(   \mathcal{C}(\rho_1)<\mathcal{C}(\rho_2)\).
%\end{equation}
\textcolor{black}{ Therefore, for \(d=2\), the \(l_1\)-norm of coherence provides a criterion for the occurrence of the ergotropic Mpemba crossing, in agreement with the states shown in Fig.~\ref{fig:iso_ergo_surface}(b). In particular, the lighter-colored curve corresponds to the state with the highest initial ergotropy and exhibits an EMC with two other states because its initial coherence is larger than theirs. In contrast, the remaining two states possess the same initial coherence and, consequently, their ergotropy curves never intersect. This demonstrates that, under the condition \(\mathcal{E}_1>\mathcal{E}_2\), the initial coherence serves as a necessary and sufficient criterion for the occurrence of an EMC. Conversely, when \(\mathcal{E}_1<\mathcal{E}_2\), an ergotropic Mpemba crossing occurs provided that \( \mathcal{C}(\rho_2)<\mathcal{C}(\rho_1). \) This behavior is illustrated in Fig.~\ref{fig:xz_plane}(b), where every state lying inside the cylindrical region above the isoergotropic surface exhibits an EMC with the chosen reference state.}

\emph{Note.}  Although the entire above analysis is discussed 
%For brevity, we have the theorem in 
at the zero-temperature limit \(T=0\), the corresponding Mpemba crossing region remains unchanged, and the result can be straightforwardly generalized to finite temperatures \(T\neq 0\), i.e., for gADC.

% The abouve \textbf{Theorem} characterizes the region in state space where the EME does not occur, relative to a reference state $\rho_1$ with higher ergotropy $\mathcal{E}_1$ compared to another state $\rho_2$ with ergotropy $\mathcal{E}_2$, where $\mathcal{E}_1>\mathcal{E}_2$. It is evident from the theorem that the region of no EME depends explicitly on the properties of the initial states which provides the following corollary.

% \textbf{Corollary 1.} 
% \emph{For an amplitude damping channel, two initial states $\rho_1$ and $\rho_2$ (with $\mathcal{E}_1>\mathcal{E}_2$) do not exhibit the ergotropic Mpemba crossing effect if the $l_1$-norm of coherence of $\rho_1$ is greater than or equal to that of $\rho_2$.}

% \begin{proof}

% \end{proof}

% \emph{Note.} The coherence-based criterion stated above provides a necessary condition for the absence of EME, but it is not sufficient. That is, while the absence of EME implies $\mathcal{C}(\rho_1)\ge \mathcal{C}(\rho_2)$ for $\mathcal{E}_1>\mathcal{E}_2$, the converse does not necessarily hold.

It is straightforward to observe that \textbf{Theorem 1} applies to all states irrespective of their purity. 
%\textcolor{black}{In particular, for isoergotropic pure states, all such states lie on a circle characterized by a fixed value of the magnetization component $m_z$.} 
For arbitrary initial pure states given by the form \(\ket{\psi(\theta,\phi)}=\cos\frac{\theta}{2}\ket{0}+e^{i\phi}\sin\frac{\theta}{2}\ket{1}\), one can derive an explicit condition with respect to the state parameters when the ergotropic Mpemba crossings are absent.

\textbf{Corollary 1.}  \emph{Two arbitrary pure states  \(\ket{\psi_1(\theta_1,\phi_1)}\) and \(\ket{\psi_2(\theta_2,\phi_2)}\) when disturbed by gADC cannot exhibit the EMC, when  \(\theta_1+\theta_2\ge \pi\).}

\begin{proof}
%From \textbf{Theorem 1}, the region $\mathcal{R}_{NE}$ intersects the $xz-$plane along an arc, as illustrated schematically in Fig.~\ref{}. 
If \(\rho_1\) lies on the surface of the Bloch sphere, it represents a pure state and can be written as \(|\psi(\theta_1,\phi_1)\rangle\) (see Fig. \ref{fig:xz_plane})(a). Let the pure state which also lies on the Bloch sphere surface, and is connected to \(|\psi(\theta_1,\phi_1)\rangle\) along a vertical line be denoted as  \(|\psi(\theta_2,\phi_2)\rangle\). These two states form the corners of the region $\mathcal{R}_{NE}$.  Further, \(|\psi(\theta_1,\phi_1)\rangle\) and \(|\psi(\theta_2,\phi_2)\rangle\) are mirror images of each other with respect to the plane \(z=0\). This geometric symmetry directly implies \(\theta_1+\theta_2\ge\pi\).

% \textcolor{black}{The boundary states of $\mathcal{R}_{NE}$,   are the pure states which  can be parameterized by   \(\ket{\psi_1(\theta_1,\phi_1)}\) and \(\ket{\psi_2(\theta_2,\phi_2)}\),  lying on the upper and lower semicircles of the Bloch sphere cross section, respectively (as can also seen from Theorem 1).
% %From Fig.~\ref{}, it follows that 
% Since these two arcs are mirror-symmetric with respect to the plane, $z=0$, 
% %
% it implies
% \(\theta_1+\theta_2\ge\pi.\)} 
\end{proof}

\textcolor{black}{{\it Note.} The competition between the energy of the state in Eq. (\ref{eq:timeevolve2vec}) and passive state energy in Eq. (\ref{eq:passive_state_energy}) does not play any role for observing the EMC.}

\emph{Mpemba crossing time.}  Let us now determine the precise time at which the ergotropy curves cross, i.e., the Mpemba crossing time, \(t_*\). In general, it is hard to calculate for any two arbitrary qubit. Here, we find out \(t_*\) for any two arbitrary pure states, which is given as 
% \begin{equation}
% e^{\alpha\gamma t_\ast} = 
% 4\left[\frac{
%   \alpha \left(1 + \cos\theta_{1}\cos\theta_{2}\right) + \left(\cos\theta_{1} + \cos\theta_{2}\right) 
% }{
% \left(\cos\theta_{1} + \cos\theta_{2}\right)
% \left( 4 + \alpha\left(\cos\theta_{1} + \cos\theta_{2}\right) \right)
% }\right]
% \label{eq:}
% \end{equation}
\textcolor{black}{
\begin{equation}
 t_{\ast} =\frac{1}{a\gamma} \ln \left( 
4\left[\frac{
  a \left(1 + \cos\theta_{1}\cos\theta_{2}\right) + \left(\cos\theta_{1} + \cos\theta_{2}\right) 
}{
\left(\cos\theta_{1} + \cos\theta_{2}\right)
\left( 4 + a\left(\cos\theta_{1} + \cos\theta_{2}\right) \right)
}\right]\right),
\label{eq:qubit_tstar}
\end{equation}}
where \(a=1+2n\) and \(\gamma_+=\gamma n\), \(\gamma_-=\gamma (1+n)\). \(n=\frac{1}{e^{2h_z \beta}-1}\) is the mean number of bosons in the thermal state of the bath at temperature \(T\). \textcolor{black}{One can immediately check that when  \(\theta_1+\theta_2\ge\pi\), the denominator of the expression inside the logarithm is either \(0\) or negative; as a result, no solution for \(t_{*}\) exists. For the detailed derivation of \(t_{*}\), see Appendix~\ref{sec:order_parameter}}. Therefore, this condition provides a sharp criterion for identifying the class of pure states that never exhibit an EMC. Conversely, whenever the crossing occurs, Eq.~(\ref{eq:qubit_tstar}) yields the exact value of the crossing time.  Note that the 
ergrotropic profiles of \(\rho_1\) and \(\rho_2\) with time  can indicate the strength of the ergotropic Mpemba. 
% crossing points can also indicate how \textcolor{black}{fast} Mpemba crossing occurs for pure states. 
To quantify this,  we consider the definition of  ``Mpemba parameter"~\cite{Furtado2025} in the context of ergotropy 
%to capture EMC analogous to the standard Mpemba effect 
(see Appendix~\ref{sec:order_parameter} and Fig. \ref{fig:qubit_colourplot}).
% \begin{figure}
%     \centering
%     \includegraphics[width=0.9\columnwidth]{qubit_mpemba.pdf}
%     \caption{This plot is for a single qubit ergotropy while amplitude damping noise is acting on it. Every curve show ergotropy at time $t$ for diffrent initial value of $\theta$. T=0, $\gamma=0.03$ and $\gamma_z=0$.}
%     \label{fig:qubit_mpemba}
% \end{figure}

\subsection{Constraints on initial states for EMC against an anisotropic Pauli channel}

%Till now we have considered only a specific type of quantum channel, namely amplitude damping channel which is specific type of Davis map, in which the steady state of the dynamics is a thermal state. Now, 

% and all the values of dissipation strength are equal which is called as depolarising channel and the fixed state of the dynamics is \(\mathbb{I}_2/2\). For our work, we only consider first type of dissipation channel.

Let us now move to the scenario when the battery is exposed to the anisotropic Pauli channel. 
%we obtain an ergotropic Mpemba crossings, suggesting apart from Davis map, other quantum channel can showcase the EMC. 
Unlike ADC, we obtain a trade-off relation between the dissipation strength,  initial energy and coherence of the initial states in order to obtain EMC in the following. Let $E_{1(2)}$ be the energy of the state $\rho_{1(2)}$.
%Unlike ADC, the initial energy of the state plays role in the occurrence of Mpemba crossings. Such relationship between the parameters can be coined into the following theorem:

\textbf{Theorem 2.} 
\emph{For an anisotropic Pauli channel, the occurrence of ergotropic Mpemba crossings between two states $\rho_1$ and $\rho_2$ with $\mathcal{E}_1>\mathcal{E}_2$ and \(m_z>0\), depends on the relative strengths of the transverse and longitudinal noise rates -- 
%$\gamma_\perp$ and $\gamma_z$. 
if $\gamma_\perp<\gamma_z$, an EMC occurs when $E_1<E_2$ while in the region with  $\gamma_\perp>\gamma_z$,
%Conversely, if 
 EMC can be observed, provided $\mathcal{C}(\rho_1)<\mathcal{C}(\rho_2)$.}

\begin{proof}
    For the detailed proof, see Appendix~\ref{app:pauli}.
\end{proof}

\textcolor{black}{Remark. The above Theorem does not hold when the initial states have \(m_{z} <0\). In this situation,  the boundary of the EMC region is jointly determined by the isoergotropic curve and  a distinct condition, namely
\(
\frac{m_{x_1}^2}{m_{z_1}}
>
\frac{m_{x_2}^2}{m_{z_2}}\),
where \(m_{x_i}\) and \(m_{z_i}\) (\(i=1,2\)) are the magnetization of \(\rho_i\)s. It explicitly relates the Bloch-vector components of the two initial states. Therefore, in this regime, the occurrence of the EMC is determined by the specific parameters of the initial states rather than solely by their initial energy and coherence. For more details, see  Appendix~\ref{app:pauli}.}

% \textbf{Theorem.}  \emph{A Mpemba crossing in anisotropic Pauli channel can be obtained }

% \textbf{Theorem.} \emph{For a anisotropic Pauli channel with \(\gamma_x=\gamma_y=\gamma\) when \(\gamma<\gamma_z\) then whether we will get a  Mpemba crossing in ergotropy for two states $\rho_1$ and $\rho_2$ will depend on the energy (i.e if $\mathcal{E}_1(0)>\mathcal{E}_2(0)$  then $E1<E2$ and the opposite for $\mathcal{E}_1(0)<\mathcal{E}_2(0)$ )  of the two states but for the case of \(\gamma>\gamma_z\) coherence (i.e. if $\mathcal{E}_1(0)>\mathcal{E}_2(0)$ then $\mathcal{C}(\rho_1)<\mathcal{C}(\rho_2)$ and the opposite for $\mathcal{E}_1(0)<\mathcal{E}_2(0)$)  of the two states play a decisive role in mpemba crossing.}

% So, When $\mathcal{E}_1(0)>\mathcal{E}_2(0)$ then $m_{x1}<m_{x2}$ must be satisfied for a crossing to exist. So $\mathcal{C}(\rho_1)<\mathcal{C}(\rho_2)$ must be satisfied. \end{proof}

Physically, the dependency of the EMC under the change of noise strength can be understood from the action of the noise on the states. 
%Intuitively, it can be argued as follows: 
The anisotropic noise squeezes the Bloch sphere along the \(xy\)-plane while the transverse noise is a dephasing channel, squeezes the Bloch sphere along the \(z\)-axis. While \(\gamma_\perp>\gamma_z\), the anisotropic part dominates, it tries to destroy the coherence of the state. Hence, the coherence plays a pivotal role for the occurrence of ergotropic  Mpemba crossing while in \(\gamma_\perp<\gamma_z\), the coherence decays faster and the system squeezes along the \(z\)-axis, which depends upon \(m_z\), only, in turn, depends on the initial energy of the systems.

%\begin{proof}
\begin{figure}
    \centering
    \includegraphics[width=\columnwidth]{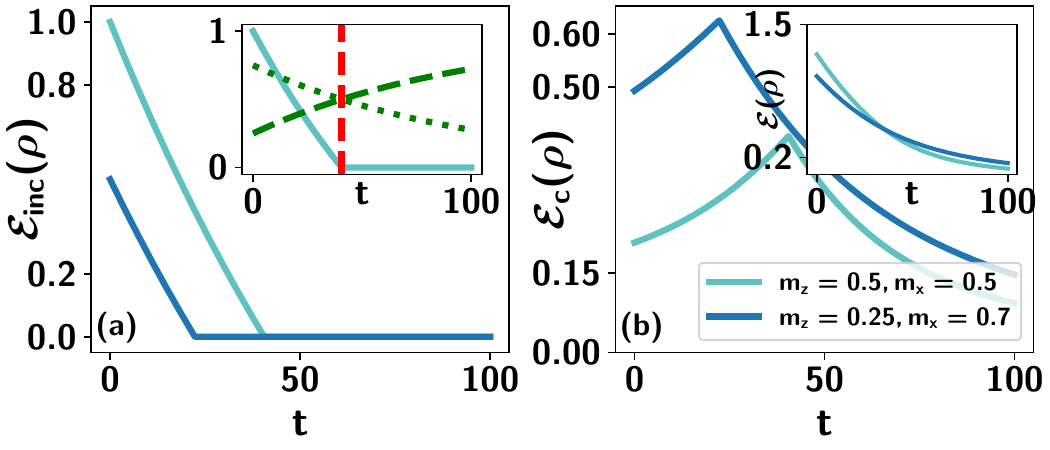}
    \caption{\textcolor{black}{\textbf{Incoherent and coherent ergotropies against time, \(t\) for ADC.} (a) Incoherent ergotropy, \(\mathcal{E}_{inc} (\rho)\) (ordinate) vs time, \(t\). The incoherent ergotropy decreases exponentially, which does not show any crossings. In the inset, we plot the populations of the ground (dashed line) and excited (dotted line) states for the first state with the incoherent part (solid line), showing the incoherent part vanishes exactly when the two populations cross, i.e., the populations of two levels become equal.  (b) Coherent ergotropy, \(\mathcal{E}_{c} (\rho)\) against \(t\). It increases upto a certain time, \(t_{c}\), and then  decreases sharply. This increasing behavior  makes the delay in total ergotropy, leading to  the ergotropic Mpemba crossing. The inset shows the total ergotropies corresponding to the two initial states \(\rho_1\) and \(\rho_2\).} Other parameters are same as in Fig. \ref{fig:xz_plane}.
    %Other parameters of the systems are \(h_z=1\), \(T=0\) and \(\gamma_{-} =0.01\). 
    }
    \label{fig:qubit_inco}
\end{figure}

\begin{figure*}
    \centering
    \includegraphics[width=\textwidth]{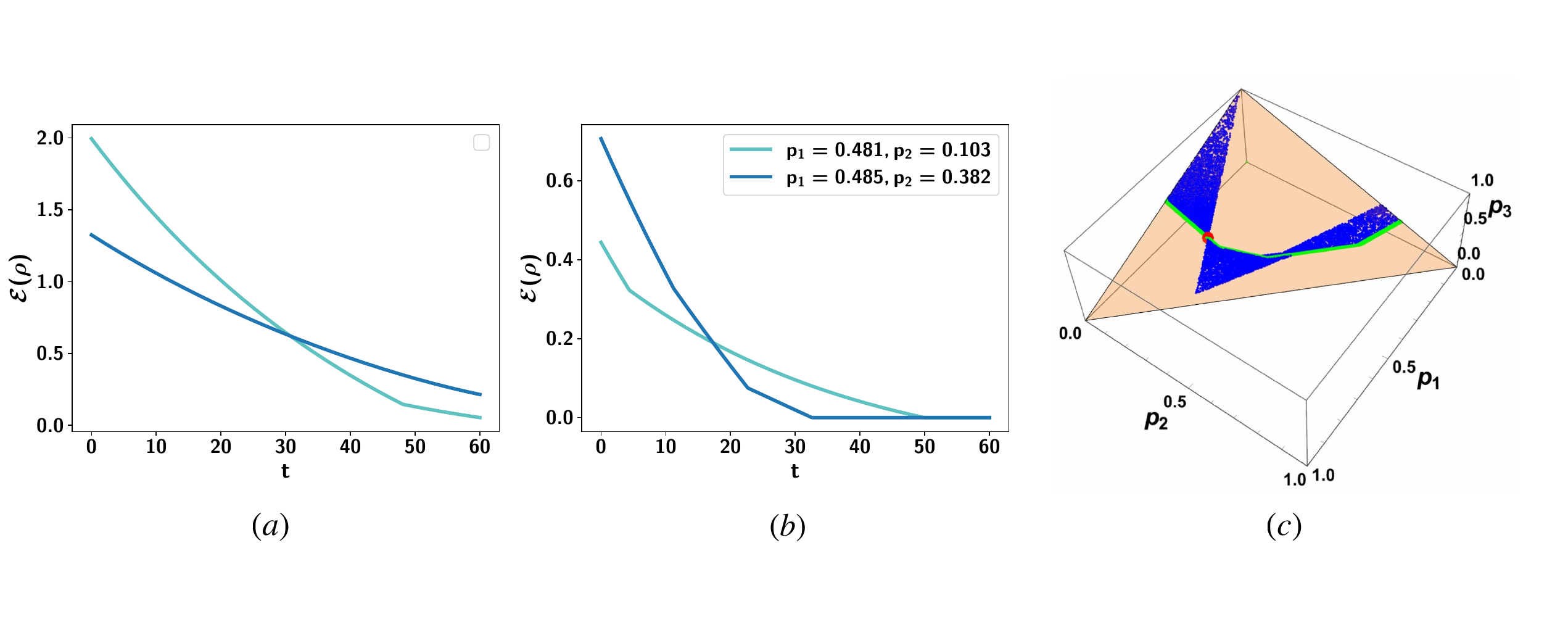}
    \caption{\textbf{EMC for a qutrit battery under ADC channel.} (a) Ergotropy dynamics against time \(t\) for two arbitrary pure states to highlight that EMC can be observed in higher-dimensional systems. \textcolor{black}{In particular, we choose two pure states,  \(|\psi_1\rangle=0.995|0\rangle-0.098|1\rangle+0.017|2\rangle\) and \(|\psi_2\rangle=-0.588|0\rangle-0.079|1\rangle+0.013|2\rangle\).} (b) \(\mathcal{E}(\rho)\) vs \(t\) for two pairs of diagonal initial states,  \(  \rho^{d=3}(0)=p_1\ket{2}\bra{2}+p_2\ket{1}\bra{1}+(1-p_1-p_2)\ket{0}\bra{0}\). Note that, unlike two-level quantum battery,  the incoherent ergotropy alone can show the ergotropic Mpemba crossing. The  set of state parameters chosen are $\{p_1=0.481, p_2=0.103\}$ (light solid line) and $\{p_1=0.485,p_2=0.382\}$ (dark solid line). (c) The EMC region of all the diagonal states is plotted for a fixed diagonal reference state \textcolor{black}{\(\rho=0.5|2\rangle\langle 2|+0.2|1\rangle\langle1|+0.3|0\rangle\langle 0|\)} (marked as solid (red) circles). The (blue) dots are the states which show EMC with respect to the fixed state, and the (green) solid line represents the isoergotropic surface. \textcolor{black}{Note that the qualitative feature of the EMC region in (c) remains similar for other choices of the set of \(\{p_1, p_2\}\).} }
    \label{fig:qutrit_random_state}
\end{figure*}

\section{Incoherent and coherent ergotropy in ergotropic Mpemba effect}
\label{sec:coherent_and _incoherent_ergotropy}

To gain insight into the physical origin of the EMC, we analyze the finer structure of ergotropy by decomposing it into its coherent and incoherent contributions~\cite{Francica2020}. The incoherent ergotropy quantifies the maximum extractable work achievable without altering the coherence of the density matrix, whereas the coherent ergotropy accounts for the additional work that can be extracted by modifying the coherence of the state. Accordingly, the total ergotropy of a quantum state $\rho$ can be written as
\begin{equation}
    \mathcal{E}(\rho)=\mathcal{E}_{\text{inc}}(\rho)+\mathcal{E}_{\text{c}}(\rho).
\end{equation}
For $T\neq0$, gADC modifies the population of the energy eigenstates, i.e., the population of the  excited state decreases while the same for the ground state increases. At long times, the system approaches a thermal state in which the ground state is more populated than the excited state. Consequently, there exists a finite time $t=t_s$ at which the ground state population becomes higher than the excited state, causing the incoherent ergotropy to vanish. This behavior can be understood from the analytical form of the incoherent ergotropy dynamics  
\textcolor{black}{\begin{equation}
\mathcal{E}_{\text{inc}}=2\theta(m_z)\theta(t_s-t)\frac{1}{1+2n}\big [e^{-(1+2n)\gamma t}(1+m_z(1+2n))-1\big],
\end{equation}}
where $\theta(\cdot)$ is the Heaviside step function, defined as $\theta(t)=1$ for $t>0$ and $\theta(t)=0$ for $t<0$, and  correspondingly, the time,
\begin{eqnarray}
    t_s=\frac{1}{(1+2n)\,\gamma}\,\ln\!\left(1+m_z (1+2n) \right),
\end{eqnarray}
%which denotes the the time 
after which the incoherent ergotropy vanishes, \textcolor{black}{ expressed by the \(\theta(t_s-t)\) term}~\footnote{Note that at \(t=0\), \(\mathcal{E}_{inc} = 2 \theta(m_z) m_z\). The factor "2" is present (which is typically absent) as \(H_B = h_z \sigma_z\) instead of \(H_B = \frac{1}{2}h_z \sigma_z\).}. It follows immediately that $\mathcal{E}_{\mathrm{inc}}$ decays exponentially in time and depends only on the
initial population imbalance \(m_z\). \textcolor{black}{Once we have the incoherent ergotropy, we can get the coherent part by subtracting it from the total ergotropy.} At early times, the coherent contribution to ergotropy may increase, indicating a transient generation of coherent ergotropy. After a characteristic time \textcolor{black}{which is the same as \(t_s\) only for  initial states belonging to the upper hemisphere of the Bloch sphere}, the coherent ergotropy begins to decay and eventually vanishes (see Fig.~\ref{fig:qubit_inco}(b)). This temporary buildup of coherent ergotropy effectively delays the relaxation process, causing the total ergotropy to approach its steady-state value more slowly.

More specifically, we find that for two-dimensional systems, the incoherent ergotropy alone cannot give rise to ergotropic Mpemba crossings (see Fig.~\ref{fig:qubit_inco}(a)). Further, \textcolor{black}{it vanishes exactly at the time \(t_s\) where the populations of the ground and the first excited state coincide (see the inset of Fig.~\ref{fig:qubit_inco}(a)).} In contrast, the coherent contribution exhibits a qualitatively distinct and nontrivial dynamical behavior; an exchange in relaxation behavior occurs in which the coherent ergotropy relaxes more slowly, even though the incoherent part decays faster (see Fig.~\ref{fig:qubit_inco}(a)). Consequently, for an EMC to occur, the coherent and incoherent contributions to ergotropy must be ordered differently for the two initial states $\rho_1$ and $\rho_2$. 

Beyond the generalized amplitude damping channel, a similar qualitative behavior is also observed for the anisotropic Pauli channel (see Appendix~\ref{app:pauli}).
%In this case as well, a behavioral exchange between the coherent and incoherent contributions to ergotropy occurs for pairs of states that exhibit EMC. Unlike the gADC, however, the coherent ergotropy does not display nonanalytic behavior in time, instead, it decreases sharply. Meanwhile, the coherent part of the ergotropy relaxes more slowly, while the incoherent part relaxes faster. The combined effect of these contrasting relaxation behaviors gives rise to ergotropic Mpemba crossings, as illustrated in Fig.~\ref{fig:qubit_inco_pauli}(a) and (b).
%\textcolor{cyan}{ As a result, the  incoherent ergotropy alone cannot give rise to ergotropic Mpemba crossing  (see Fig.~\ref{fig:qubit_inco}(a)). In contrast, the coherent ergotropy exhibits a qualitatively different and nontrivial dynamical behavior. As it is clearly visible from Fig.~\ref{fig:qubit_inco} and Fig.~\ref{fig:qubit_inco_pauli} for a EMC to exist the order of the incoherent and coherent part of  ergotropy must be differently ordered for the two states \(\rho_1\) and \(\rho_2\).} 
These analyses indicate that the emergence of the EMC is driven by the interplay between coherent and incoherent ergotropies. However, we will also show that such an argument does not hold when the dimension of the battery increases.

\section{qutrit system ergotropic Mpemba effect}
\label{sec:qutrit_EME}
So far, we have restricted our analysis to qubit systems in which the geometric simplicity  of the Bloch sphere plays a crucial role in enabling an analytical characterization of the EMC. Extending such an analysis to arbitrary higher dimensions \(d\) is, however, significantly more challenging. As a first step towards understanding higher-dimensional quantum batteries, we now consider a three-level (\(d=3\)) system described by the Hamiltonian\(H_B^{d=3}=h_zS_z\),
where \(S_z\) is the generalized Pauli operator along the \(z\)-direction in three dimensions, and \(h_z\) denotes the strength of the external magnetic field. 

% An arbitrary initial state in this three-dimensional Hilbert space can be written in the form
% \begin{equation}
%     \rho^{d=3}(0)=\frac{1}{3}\left(\mathbb{I}_3+\sum_{i=1}^{8}r_i\hat{\Gamma}_i\right),
% \end{equation}
% where \(r_i=\Tr[\rho^{d=3}(0)\hat{\Gamma}_i]\) with \(\hat{\Gamma}_i\) being the Gell--Mann matrices, forming a basis for the Lie algebra \(\mathrm{SU}(3)\) and satisfying \(\Tr[\hat{\Gamma}_i\hat{\Gamma}_j]=2\delta_{ij}\).

After preparing the initial state, the system is subjected to an amplitude damping channel, and the corresponding Lindblad operators governing the dynamics are \(L_j=\ket{0}\bra{1},\ket{0}\bra{2},\ket{1}\bra{2}\), where \(\{\ket{0},\ket{1},\ket{2}\}\) represents the set of energy eigenstates of the battery Hamiltonian, ordered according to increasing energy, and \(\gamma\) is the associated (time-independent) decay rate for all the transitions. \textcolor{black}{ As in the qubit case, the environment induces spontaneous decay from the excited states toward the ground state. Consequently, the ground state of \(H_B^{d=3}\) is the unique steady state of the dynamics, and the ergotropy decays to zero in the long-time limit. Such Lindbladian superoperator can be cast into a block-diagonal form, \(\mathcal{L}_{p}\oplus \mathcal{L}_{c}\), where \(\mathcal{L}_p\) denotes the population block and \(\mathcal{L}_c\) represents the coherence block (cf. Ref.~\cite{moroder_prl_2024}). It indicates that the population and coherence dynamics are independent of each other.}

%Note that this structure implies that if the initial state lies entirely within one of these blocks, the time-evolved state remains confined to the same block throughout the evolution.

% \begin{figure}
%     \centering
%     \includegraphics[width=0.9\columnwidth]{qutrit_diagonal_dyn.pdf}
%     \caption{ergotropy of two diagonal states ${0.48125839,0.10284593,0.415896}$ and ${0.48538582,0.38238276,0.132231}$ with time  }
%     \label{fig:qutrit_diagonal_dyn}
% \end{figure}

We find that, like in the qubit case, an ergotropic Mpemba crossing in a qutrit system can also be observed (see Fig.~\ref{fig:qutrit_random_state}(a)), confirming the phenomenon in higher-dimensional systems. \textcolor{black}{In particular, we choose two initial pure states, given as \(\ket{\psi_1}=0.995|0\rangle-0.098|1\rangle+0.017|2\rangle\), and  \(\ket{\psi_2}=-0.588|0\rangle-0.079|1\rangle+0.013|2\rangle\). As shown in Fig.~\ref{fig:qutrit_random_state}(a), we show that EMC is observed for two arbitrary pure states at a critical time \(t_*\). However, the analysis is not analytically tractable even for the entire three-dimensional pure state space.\\
To carry out the investigation analytically, let us restrict all qutrit initial states with diagonal density matrices of the  form, }
%we resort to all the qutrit states with diagonal entries.  In this situation, a useful criterion can be derived when the initial states are restricted to be of the form
%\begin{equation}
  \(  \rho^{d=3}(0)=p_1\ket{2}\bra{2}+p_2\ket{1}\bra{1}+(1-p_1-p_2)\ket{0}\bra{0}\),
 %   \label{eq:qutrit_initial}
%\end{equation}
(where \(0\le p_1,p_2\le 1\)). \textcolor{black}{Interestingly, we find the region that shows EMC even for the diagonal state, thereby revealing the contrasting behavior of qubit and qutrit systems.  In particular, the dynamics of the ergotropy can be visualized within a probability simplex, as illustrated in Fig.~\ref{fig:qutrit_random_state}(c). Numerically,  we obtain this region by observing the sign change of the ergotropy difference with a fixed diagonal reference state, \(\rho=0.5|2\rangle\langle 2|+0.2|1\rangle\langle1|+0.3|0\rangle\langle 0|\), marked as a (red) solid circle.} Note that the state remains diagonal throughout the evolution (see Appendix \ref{App:qutrit} ), which leads to the ergotropy being purely incoherent, i.e., $\mathcal{E}_{\mathrm{c}}=0$ and $\mathcal{E}_{\mathrm{inc}}\neq 0$. In this context, it is noteworthy that, in contrast to the qubit case, an \emph{incoherent} ergotropic Mpemba crossing can occur in a qutrit system (see Fig.~\ref{fig:qutrit_random_state}(b)). The presence of additional energy levels leads to distinct relaxation rates between different pairs of levels, which, in turn, delays the saturation of ergotropy during the dynamics. As a consequence, ergotropic Mpemba crossings arise in qutrits, with their occurrence determined by the initial population distribution among the energy levels.

% Note, however, that unlike the qubit case, the EMC can be observed even when the initial state has no coherence (see Fig.~\ref{fig:qutrit_diagonal_dyn}).

\begin{table}[h]
\centering
\begin{tabular}{c c}
\hline
\textbf{Ordering of probabilities} & \textbf{Ergotropy} \\
\hline
$p_1 < p_2 < p_3$ & $0$ \\[4pt]
$p_1 < p_3 < p_2$ & $2p_2 + p_1 - 1$ \\[4pt]
$p_2 < p_1 < p_3$ & $p_1 - p_2$ \\[4pt]
$p_2 < p_3 < p_1$ & $3p_1 - 1$ \\[4pt]
$p_3 < p_1 < p_2$ & $3(p_1 + p_2) - 2$ \\[4pt]
$p_3 < p_2 < p_1$ & $4p_1 + 2p_2 - 2$ \\
\hline
\end{tabular}
\caption{\textcolor{black}{Ergotropy for different arrangements of populations in diagonal qutrit states is presented where \(p_3 = 1- p_1 - p_2\) and \(0\le p_i\le 1\) where \(i\in\{1,2,3\}\)}.}
\label{tab:qutrit_isoergo}
\end{table}

We find that, analogous to the qubit case, isoergotropic surfaces play a crucial role in identifying the no-EMC (\(\mathcal{R}_{NE}\)) region within the probability simplex. 
% In particular, the isoergotropic surfaces form the boundary of the region where EMCs occur.
For diagonal qutrit states, there are six distinct ways to order the level populations in increasing order and each ordering corresponds to a different functional form of the ergotropy in terms of the independent probabilities $p_1$ and $p_2$, as summarized in Table~\ref{tab:qutrit_isoergo}. A point lies on a given isoergotropic surface only if it satisfies both the left- and right-hand conditions of the corresponding expressions listed in Table~\ref{tab:qutrit_isoergo}. 
%\adi{During the dynamics induced by the amplitude damping channel, the state populations explore the entire probability simplex. 
We numerically identify a subset of states represented by the dark-shaded region in Fig.~\ref{fig:qutrit_random_state}(c) that exhibit EMC with a fixed reference state (marked as a solid circle). Notably, as in the qubit case, the boundary of this region is formed by isoergotropic states, indicated by the green curve. This observation highlights that a detailed understanding of the isoergotropic surfaces is essential for characterizing the EMC region associated with a given reference state.

\section{Non-Markovian ergotropic Mpemba effect}
\label{sec:nonmarkov_mpemba}

\begin{figure}
    \centering
    \includegraphics[width=\columnwidth]{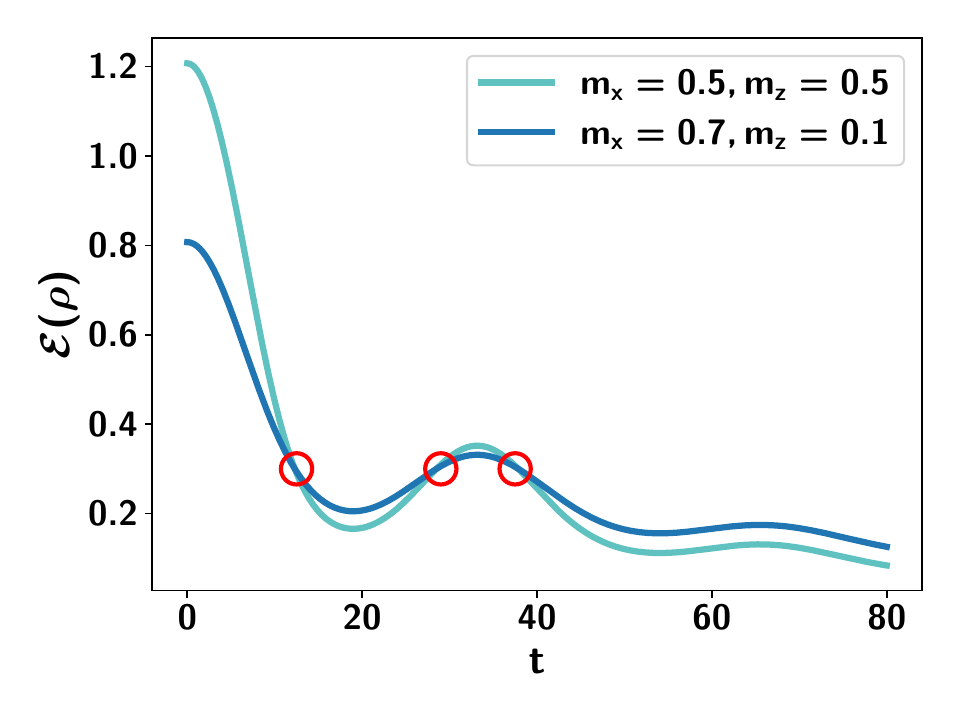}
    \caption{\textcolor{black}{\textbf{Dynamics of Ergotropy under non-Markovian noise.} For a qubit battery, the EMC can be observed under non-Markovian noisy channel. Interestingly,  an odd number of crossings {(depicted by the hollow (red) circles)} is observed in the transient regime,  and these curves never intersect after a certain time, referred to as \emph{quasiergotropic} Mpemba effect \cite{Li2025}. All other system parameters are $\lambda=0.03$, $\Delta=0.13$ and $\gamma=0.3$.}}
    \label{fig:qubit_nonmarkovian}
\end{figure}
%Until now, our analysis of EMC has been restricted to Markovian dynamics, in which the ergotropy decays monotonically in time. It is therefore natural to ask whether EMC can persist in the non-Markovian regime, where information can flow back from the environment to the system and when the system exhibits collapse and revival of ergotropy due to such memory effects. Let us now exhibit that even in the non-Markovian regime, ergotropic Mpemba crossing occurs. 
Relaxing the assumption of Markovian dynamics, under which ergotropy decays monotonically in time, we now address the following question: {\it "Can the ergotropic Mpemba crossing persist in the non-Markovian regime, where information backflow from the environment to the system leads to memory effects and induces collapse and revival of ergotropy?"}
%We have observed a EMC in Markovian dynamics where the ergotropy drops exponentially as in the Markovian evolution. in the non-Markovian case, due to the back-flow of information from the bath to the system, collapse and revival of ergotropy can be observed as reported in \cite{Kamin2020,Choquehuanca2024,Li2025}.
%the dynamics at time \(t=t_n\) only depends on the previous instant of time \(t=t_{n-1}\) and the information of the initial system is lost with time. On the other hand, 
%In this paper, we presented a similar scenario in qubit system. 
%We consider a qubit system coupled to a bosonic bath. 
\begin{figure*}
    \centering
    \includegraphics[width=\linewidth]{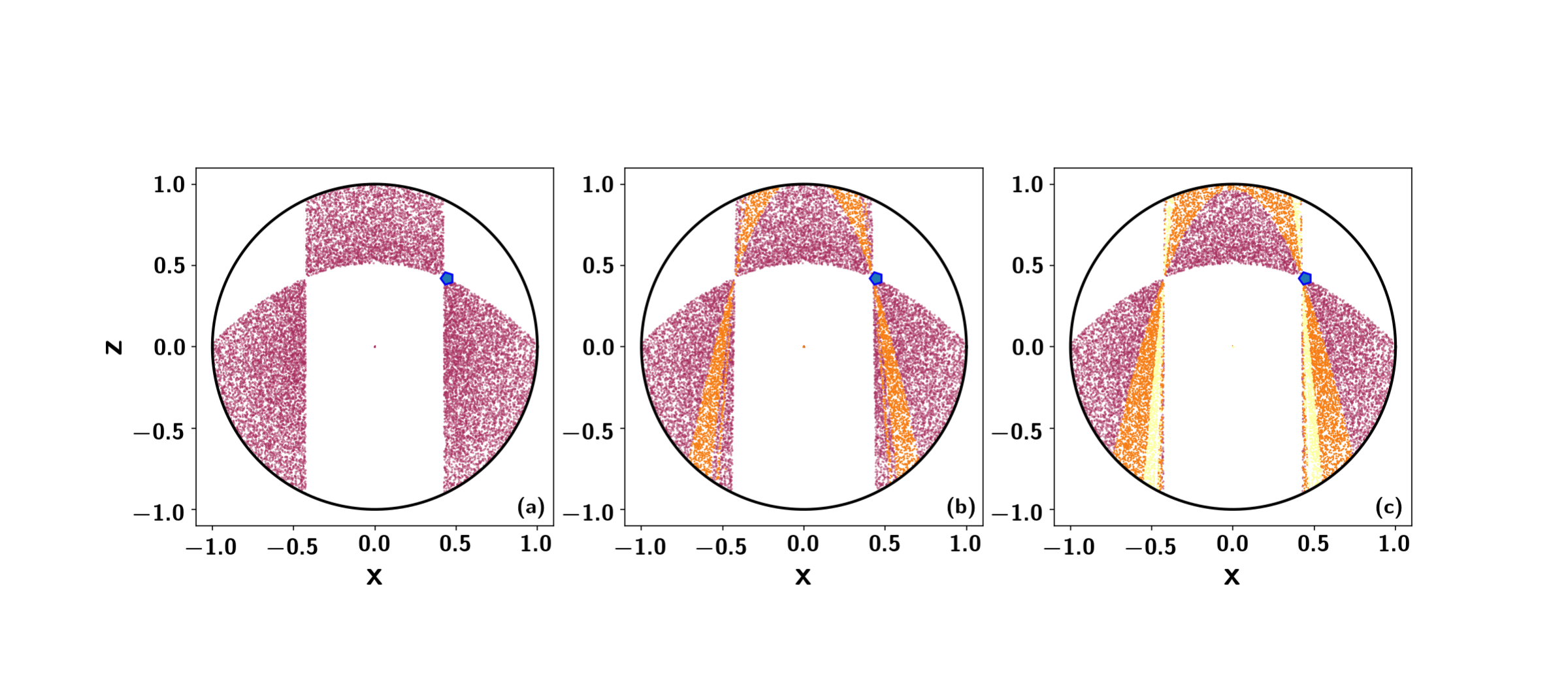}
    \caption{ \textcolor{black}{\textbf{Ergotropic Mpemba crossing regions  in \(xz\)-plane with respect to a fixed state \(\rho_1\)\textcolor{black}{\((m_x=0.43,m_y=0.0,m_z=0.41)\)} (given by blue solid circle).} Different shades of color represent the number of crossings in the transient regime between the fixed state and the states in the EMC region. The number of crossings is found by counting the number of sign changes ocurred in \(\mathcal{E}(\rho_1,t)-\mathcal{E}(\rho_2,t)\).} The red dots represent the states with which the state corresponding to the reference state show a single EMC, the orange (lighter shade) dots correspond to the states for which  three EMCs are observed while the yellowish (more lighter dots) dots are states with five EMCs. The non-Markovianity of the evolution is increased from left to right, specifically the parameters are chosen as (a) \((\gamma,\lambda)=(0.05,0.2)\) (b) \((\gamma,\lambda)=(0.3,0.03)\) and (c) \((\gamma,\lambda)=(1,0.03)\). Here $\Delta=0.1$. }
    \label{fig:odd_crossings}
\end{figure*}
To address this, let us consider the battery Hamiltonian of the system (\(H_B\)), environment (\(H_E\)) and system-environment interaction (\(H_{BE}^{\text{int}}\)), given by 
\begin{equation}
    H= \underbrace{h_z \sigma_z}_{H_B} + \underbrace{\sum_k \omega_k b_k^\dagger b_k}_{H_E} +\underbrace{\sum_k \left( g_k \sigma_+ b_k + g_k^* \sigma_- b_k^\dagger \right)}_{H_{BE}^{\text{int}}},
    \label{eq:non_markov_hint}
\end{equation}
where the bath consists of harmonic oscillators having frequency  \(\omega_k\) of the $k$-th mode,   \(b_k\)(\(b_k^\dagger\)) is the annihilation (creation) operator for the $k$-th mode of the bath,  and the coupling strength between the system and the mode \(k\) of the bath is denoted as \(g_k\). In the interaction picture, the interaction Hamiltonian reads as
%\begin{equation}
  \(  H^{\text{int}} = \sum_k  g_k \sigma_+ b_k e^{i(2h_z - \omega_k)t} + \text{h.c.}\).
%    \label{eq:non_markov_hint}
%\end{equation}
We consider the bath to have a Lorentzian spectral density, given by 
%\begin{equation}
\(J(\omega)=\frac{1}{2\pi}\,\frac{\gamma \lambda^{2}}{(2h_z-\Delta-\omega)^2+\lambda^{2}}\),
%\end{equation}
\textcolor{black}{where \(\lambda\) is the width of the spectrum, \(\gamma\) is the effective coupling and \(\Delta\) is the detuning.} It is known that for \(\gamma/\lambda\ll1\),  the dynamics is Markovian, and for \(\gamma/\lambda\gg1\), it is non-Markovian \cite{Lorenzo2013,Kamin2020}.

Initializing the system in an arbitrary pure state and the environment in the vacuum state, i.e., 
%\begin{equation}
 \( \ket{\psi(0)} = (\alpha \ket{0} + \beta \ket{1}) \otimes \ket{0}_E\), where \(\alpha\), and \(\beta\) are the \textcolor{black}{complex amplitudes of the excited and the ground states, respectively,} with \(\abs{\alpha}^2+\abs{\beta}^2=1\).
%\end{equation}
\textcolor{black}{The evolution of the total system-environment can be obtained analytically as}
%is given by
\begin{equation}
    \ket{\psi(t)} = \alpha \nu \ket{0} \otimes \ket{0}_E + \sum_k \eta_k(t) \ket{1} \otimes \ket{1_k}_E + \beta \ket{1} \otimes \ket{0}_E,
\label{eq:nm_ansatz}
\end{equation}
%where \(\nu(t)\) is given by
where \(\ket{1_k}_E\) denotes the state of the environment with a single excitation in the \(k\)-th mode, and \(\nu\) is  a time dependent function, given as
%\begin{equation}
 \(\nu = e^{-\frac{\lambda t}{2}} \left( \cosh\left( \frac{\zeta t}{2} \right) + \frac{\lambda - i\Delta}{\zeta} \sinh\left( \frac{\zeta t}{2} \right) \right) \) with \(\zeta^2 = (\lambda - i\Delta)^2 - 2\gamma \lambda\).
 After tracing out the bath, the state of the battery at arbitrary  time becomes
\begin{equation}
    \rho(t) = \begin{pmatrix}
\abs{\alpha \nu}^2 & (\alpha\nu) \beta \\
(\alpha\nu)^* \beta^* & 1 - \abs{\alpha\nu}^2
\end{pmatrix}.
\label{eq:evolved_non_markov}
\end{equation}
%Now, one can observe that the 
%Now such results can be easily extended for the mixed state, i.e., if the initial state become mixed apart from pure state. 
Instead of pure state of the system, if one considers a mixed initial state, by using linearity of the channel, the evolved  state of the system can be written as
\begin{equation}
    \rho(t) = \lambda_+ \rho_+(t) + \lambda_- \rho_-(t),
    \label{eq:mixed_evolved_non_markov}
\end{equation}
where \(\lambda_{\pm} = \frac{1}{2}\left(1 \pm |\vec{m}|\right)\) are the eigenvalues of the initial state,
%which is given as
%\begin{equation}
 \(\rho_0=\lambda_+ \ket{\psi_+}\bra{\psi_+}
+ \lambda_- \ket{\psi_-}\bra{\psi_-}\),
%\end{equation}
with \textcolor{black}{\(\ket{\psi_+} = \cos \frac{\theta}{2} \ket{0}
+ e^{i\phi}\sin \frac{\theta}{2}\ket{1}\) and  \(\ket{\psi_-} = \sin \frac{\theta}{2} \ket{0}
- e^{i\phi}\cos \frac{\theta}{2}\ket{1}\) } being the eigenstates of the initial state, \textcolor{black}{having \(0\leq \theta \leq \pi\) and \(0 \leq \phi \leq 2 \pi\)}. 
%which are given by \(\ket{\psi_\pm} = \cos \frac{\theta}{2} \ket{0}
%\pm e^{i\phi}\sin \frac{\theta}{2}\ket{1}\), 
Considering any arbitrary mixed state, \(|\vec{m}|=\sqrt{m_x^2+m_y^2+m_z^2}\) and \(\rho_\pm(t)\) is the evolved state when the initial state is \(\ket{\psi_\pm}\) respectively. 
%For a mixed state of a 
% In the case of a two-dimensional system, 
% %\[
% starting from \(\rho_0 = \frac{1}{2}\left( \mathbb{I}_2 + \vec{m}\cdot\vec{\sigma} \right)
% = \lambda_+ \ket{\psi_+}\bra{\psi_+}
% + \lambda_- \ket{\psi_-}\bra{\psi_-},
% \), where
% \(
% \lambda_{\pm} = \frac{1}{2}\left(1 \pm |\vec{r}|\right),
% \)
% and
% \(
% \ket{\psi_+} = \cos\!\left(\frac{\theta}{2}\right)\ket{0}
% + e^{i\phi}\sin\!\left(\frac{\theta}{2}\right)\ket{1},
% \) and 
% \(
% \ket{\psi_-} = \sin\!\left(\frac{\theta}{2}\right)\ket{0}
% - e^{i\phi}\cos\!\left(\frac{\theta}{2}\right)\ket{1}.
% \), 
% % %Because of linearity of the quantum channel we can add up the solutions for
% % $\ket{\psi_+}$ and $\ket{\psi_-}$.
% % \[
% % \rho(t) = \lambda_+ \rho_+(t) + \lambda_- \rho_-(t),
% % \]
% % where
% % \[
% % \rho_+(0) = \ket{\psi_+}\bra{\psi_+}, \qquad
% % \rho_-(0) = \ket{\psi_-}\bra{\psi_-}.
% % \]
The ergotropy of the battery starting from any arbitrary mixed state can be computed as
\begin{eqnarray}
&&\mathcal{E}
=
\left(
|\nu|^2 \left( 1 + |\vec m| \cos\theta \right) - 1
\right)    \nonumber\\
&& + \sqrt{
\left(
|\nu|^2 \left( 1 + |\vec m| \cos\theta \right) - 1
\right)^2
+ |\vec m|^2 \sin^2\theta \, |\nu|^2
},
\label{eq:nmarkovianmixed}
\end{eqnarray}
% \begin{eqnarray}
% &&\mathcal{E}
% =
% h_z \left(
% |\nu|^2 \left( 1 + |\vec m| \cos\theta \right) - 1
% \right)    \nonumber\\
% && + h_z \sqrt{
% \left(
% |\nu|^2 \left( 1 + |\vec m| \cos\theta \right) - 1
% \right)^2
% + |\vec m|^2 \sin^2\theta \, |\nu|^2
% },
% \label{eq:nmarkovianmixed}
% \end{eqnarray}
which turns out to be independent of the phase, \(\phi\), due to the phase covariant nature of the ergotropy dynamics. Analyzing the behavior of \(\mathcal{E}\),  two regimes in the non-Markovian evolution emerge,  namely (i) the \emph{transient} and (ii) the \emph{steady state}. In the \emph{transient} regime (see Fig.~\ref{fig:qubit_nonmarkovian}),  an oscillatory behavior of the ergotropy, a signature of non-Markovianity, can be observed, while at the long-time limit, the oscillations become less pronounced. Interestingly, we find that  unlike Markovian case, there are multiple crossing points in the transient time  (cf. \cite{Li2025}) although  a specific time \(t=t_m\) exists at which the final crossing occurs, after which the two curves never intersect again, named the \emph{quasiergotropic Mpemba effect}.
% We plot the ergotropy with time for several initial pure states, which depend only on the angle \(\theta\) with \(\mathcal{E}_1>\mathcal{E}_2\) .
%Because of the phase covariant nature of the dynamics there is no \(\phi\) dependence in the formula of ergotropy.
% \textbf{Proposition 2} \emph{In non-Markovian dynamics, ergotropy for two different values of $\theta$ can cross each other sometimes before showing mpemba effect. Which has been described as quasi-ergotropy mpemba effect before.}

{\it Even vs odd crossing.} Like the Markovian case, we can also identify the region in the Bloch sphere of the initial states that exhibit Mpemba crossings in the non-Markovian evolution as well. In order to identify this, we again use the phase covariance of the ergotropy and hence, it is sufficient to consider the ergotropy in the \(xz\)-plane. We probe into the large-time behavior of ergotropy in the non-Markovian evolution, which is given as
\begin{eqnarray}    
 \mathcal{E} &\sim &  \left(
|\nu|^2 \left( 1 + |\vec m| \cos\theta \right) - 1
\right)+
1
-
|\nu|^2 \left( 1 + |\vec m| \cos\theta \right) \nonumber\\&&
\qquad\qquad+
\frac{|\nu|^2}{2}
|\vec m|^2 \sin^2\theta\nonumber\\
&\Rightarrow&
\mathcal{E} \sim
\frac{|\nu|^2}{2} m_x^2,
\label{eq:nm_inf_approx}
\end{eqnarray}
where we have considered \(t\) to be large. Note that the ergotropy depends upon \(\nu\) and the initial magnetization, \(m_x\). In turn, the ergotropy behavior depends upon the \(l_1\)-coherence of the state as \(m_x^2\) is the coherence in the \(xz\)-plane. From Eq. (\ref{eq:nm_inf_approx}), it is clear that at large time,  \(m_{x_1}^2 > m_{x_2}^2\) implies that \(\mathcal{E}(\rho_1(t\to\infty))\ge \mathcal{E}(\rho_2(t\to\infty))\) which implies that they cross each other an even number of times. On the other hand, \(m_{x_1}^2 \leq m_{x_2}^2\), there is at least an odd number of crossings \textcolor{black}{calculated by counting the number of sign changes in \(\mathcal{E}(\rho_1, t) - \mathcal{E}(\rho_2, t)\)}, emerged as EMC in non-Markovian evolution (see Fig.~\ref{fig:qubit_nonmarkovian}). Also, we determine the region of states that shows different odd numbers of crossings in the transient regime (see Fig.~\ref{fig:odd_crossings}). In Fig.~\ref{fig:odd_crossings}, we illustrate the number of EMC for different initial states while keeping one reference state fixed \textcolor{black}{(as mentioned in the caption).} 
%The results show that states exhibiting a larger number of crossings occupy a significantly smaller region of the state space.
Interestingly, the number of states displaying a single crossing increases with the degree of non-Markovianity, as depicted in Fig.~\ref{fig:odd_crossings}(a). In contrast, the population of states exhibiting multiple crossings grows with increasing non-Markovianity, as indicated by the variation in color intensity in Fig.~\ref{fig:odd_crossings}(b) and (c).

We find out that there is no even number of crossings. In order to prove that there is no even number of crossings possible, we use the following two lemmas which are given as follows:

\textbf{Lemma 3.} \emph{For two isoergotropic states, the ergotropy of the state with a larger magnetization along the $z$-direction decays faster than that of the state with a smaller $m_z$ for  non-Markovian amplitude damping channel.}

\begin{figure}
    \centering
    \includegraphics[width=\columnwidth]{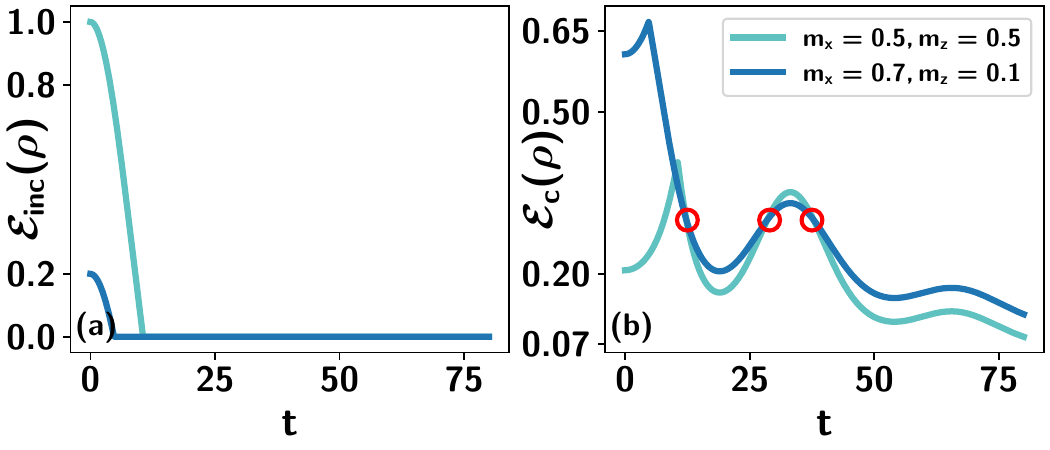}
    \caption{\textcolor{black}{\textbf{Dynamical profile of incoherent and coherent ergotropies under a non-Markovian channel.} (a) Incoherent and (b) coherent ergotropies (ordinate) against time, \(t\) (abscissa). Note that the hollow (red) circles in (b) depict the  Mpemba crossings in the coherent ergotropy as also observed in Fig.~\ref{fig:qubit_nonmarkovian}, while such crossings are absent in incoherent ergotropy in (a).}  
    %The incoherent ergotropy decreases exponentially, which does not show any crossings. (b) Coherent ergotropy is plotted against \(t\). The coherent ergotropy shows an oscillation in the transient time and it decreases slowly. The contrast of decrement of coherent and incoherent ergotropy is observed as the ergotropic Mempba effect. 
    Other system parameters are the same as in Fig. \ref{fig:qubit_nonmarkovian}. 
    %$\lambda=0.03$, $\Delta=0.13$ and $\gamma=0.3$.
    }
    \label{fig:qubit_nonmarkovian_inco}
\end{figure}

\textbf{Lemma 4.} \emph{For two non-isoergotropic states with identical $m_x$, the ergotropy of the state with a larger $m_z$ decays more slowly than that of the state with a smaller $m_z$.}

For the proof of the Lemmas, see Appendix~\ref{app:proof_lemma34}.

Analogous to \textbf{Theorem~1}, Lemmas~\textbf{3} and \textbf{4} together establish the existence of a region $\mathcal{R}_{NE}$ in the $xz$ plane of the Bloch sphere, consisting of states that do not exhibit any ergotropic Mpemba crossings during the evolution. The presence of this region excludes the possibility of an even number of crossings in the non-Markovian regime.

\emph{Coherent and incoherent ergotropy in non-Markovian dynamics.} As observed in the Markovian case, the coherent and incoherent contributions to the ergotropy, $\mathcal{E}_c$ and $\mathcal{E}_{\mathrm{inc}}$, display qualitatively similar roles in non-Markovian dynamics for qubits. \textcolor{black}{Like the Markovian case, the expression of incoherent ergotropy can be calculated exactly from the expression of the state (in Eq.~(\ref{eq:mixed_evolved_non_markov})), given by
\begin{equation}
\mathcal{E}_{inc}=2\left(|\nu|^2\bigl(1+|\vec{m}|\cos\theta\bigr)-1\right),
\end{equation}
which is a function of \(\nu\). }  The incoherent contribution decays rapidly, whereas the decay of the coherent ergotropy occurs over a longer timescale. During the transient regime, oscillatory behavior emerges as a signature of memory effects (see Fig.~\ref{fig:qubit_nonmarkovian_inco}(b)).  Remarkably, in the non-Markovian dynamics, we observe an exchange of dominance between the coherent and incoherent contributions to ergotropy, which leads to the emergence of the \emph{quasi-ergotropic Mpemba effect}. In this regime, a trade-off between $\mathcal{E}_c$ and $\mathcal{E}_{\mathrm{inc}}$ is required for the appearance of Mpemba crossings. These findings indicate that ergotropic Mpemba crossings\footnote{\textcolor{black}{Since \(\nu\) is a nontrivial function of time containing hyperbolic functions and the system parameters,  a closed-form analytical expression for this crossing time is, in general, not readily obtainable.}} arise from a combined and subtle interplay between the incoherent and coherent components of ergotropy.
\begin{figure}
    \centering
    \includegraphics[width=0.8\linewidth]{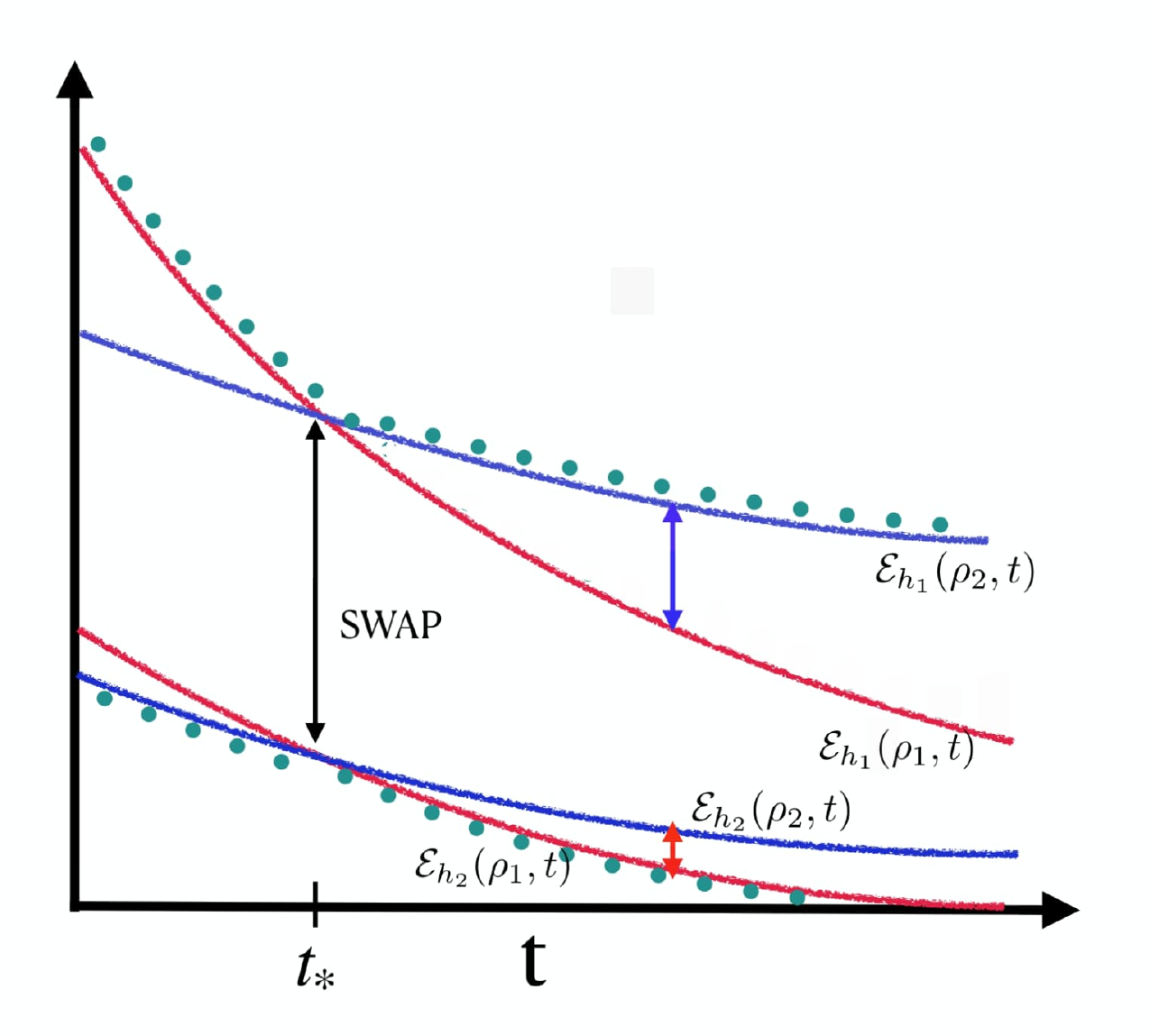}
    \caption{ \textcolor{black}{ Initially, the two qubits are prepared in the states \(\rho_{1}^{0}\) and \(\rho_{2}^{0}\), evolving under the local magnetic fields \(h_1\) and \(h_2\), respectively. At the crossing time \(t=t_*\), a SWAP unitary operation is applied, exchanging the quantum states of the two qubits. Consequently, the first qubit evolves from the state \(\rho_{2}^{t_*}\) under the magnetic field \(h_1\), while the second qubit evolves from the state \(\rho_{1}^{t_*}\) under the magnetic field \(h_2\). The green dotted lines represent the ergotropy of the qubits as a result of the application of the SWAP unitary.} For \(t>t_*\), the ergotropy of the first qubit increases, as indicated by the blue vertical arrow, whereas that of the second qubit decreases, as indicated by the red vertical arrow. Owing to the asymmetry in the local magnetic fields, the increase in the ergotropy of the first qubit exceeds the corresponding decrease in the second qubit, resulting in a net enhancement of the total ergotropy.}
    \label{fig:swap_two_qubits}
\end{figure}

\textcolor{black}{\section{Mitigating ergotropic loss using EMC}
\label{sec:use_of_EMC}
In this section, we present a protocol to mitigate the ergotropic loss by exploiting the ergotropic Mpemba crossing. To illustrate the idea, consider two qubits initially prepared in the states \(\rho_1^0\) and \(\rho_2^0\), respectively, both evolving under an amplitude-damping channel. The ergotropy of each state is given in Eq.~(\ref{eq:general_ergotropy_dyn}). Assume that these two states exhibit an ergotropic Mpemba crossing at \(t=t_*\), with \(\mathcal{E}(\rho_1^0)>\mathcal{E}(\rho_2^0).\) Now suppose that  two qubits evolve under different local magnetic fields \(h_1\) and \(h_2\), where \(h_1>h_2\). It is important to note that the magnitude of the magnetic field only rescales the ergotropy and does not alter the corresponding isoergotropic surfaces on which it is located at any time \(t\), which are determined by \(m_z+\sqrt{m_x^2+m_y^2+m_z^2}=\mathrm{constant}.\) Consequently, at the crossing time \(t=t_*\), the evolved states \(\rho_1^{t_*}\) and \(\rho_2^{t_*}\) remain on the same isoergotropic surface despite evolving under different magnetic field strengths. At this instant, we apply a SWAP unitary~\cite{Linden2010,Campisi2015},
\begin{equation}
U=
\begin{bmatrix}
1&0&0&0\\
0&0&1&0\\
0&1&0&0\\
0&0&0&1
\end{bmatrix},
\end{equation}
which exchanges the quantum states of the two qubits. After the SWAP operation, \(\rho_2^{t_*}\) evolves under the stronger magnetic field \(h_1\), whereas \(\rho_1^{t_*}\) evolves under the weaker field \(h_2\), as illustrated in Fig.~\ref{fig:swap_two_qubits}. The energy exchanged during the SWAP operation is given by~\cite{Allahverdyan2010}
\begin{eqnarray}
\Delta E
&=&
\mathrm{Tr}
\!\left[ H_B\, \rho_2(t_*)\otimes\rho_1(t_*) \right] - \mathrm{Tr} \!\left[ H_B\, \rho_1(t_*)\otimes\rho_2(t_*) \right] \nonumber\\ &=& (h_1-h_2) \, \mathrm{Tr} \!\left[ \sigma_z \bigl( \rho_2(t_*)-\rho_1(t_*) \bigr) \right],
\end{eqnarray}
where \(H_B= h_1\sigma_z\otimes\mathbb{I}_2 + \mathbb{I}_2\otimes h_2\sigma_z.\) Since \(h_1>h_2\) and \(\mathrm{Tr} \!\left[ \sigma_z \bigl( \rho_2(t_*)-\rho_1(t_*) \bigr) \right] <0\) (by Lemma 1), it follows that \(\Delta E<0\), implying that energy is released during the protocol. 
%Moreover, as established by Lemma~1, states located farther from the \(z\)-axis exhibit a slower decay of ergotropy. 
Consequently, after the SWAP operation, the qubit evolving under the stronger magnetic field gains more ergotropy for \(t>t_*\) than the amount lost by the qubit evolving under the weaker magnetic field. This protocol, therefore, illustrates how ergotropic Mpemba crossings may be exploited to mitigate the dissipation-induced loss of extractable work and enhance the operational performance of finite-dimensional quantum batteries.}

\begin{figure}
    \centering
    \includegraphics[width=\columnwidth]{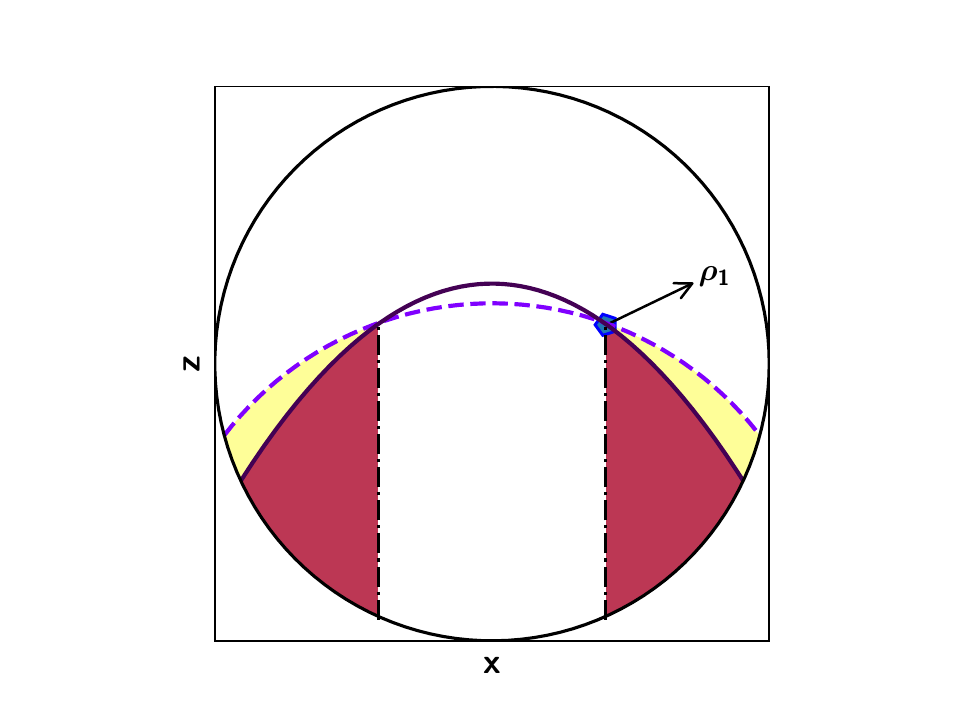}
    \caption{\textbf{Comparison between the ergotropic Mpemba crossing and the state Mpemba effect for a qubit evolving in the $xz$-plane under  ADC.} The solid curve denotes the isoergotropic line, while the dashed curve represents states that are equidistant from the steady state. The darker shaded region corresponds to states, exhibiting both EMC and the state Mpemba effect, whereas the lighter shaded region indicates states that display the state Mpemba effect without showing EMC. The remaining parameters are fixed as $T=0$, $\gamma_-=0.01$, and the reference state $\rho_1$ is chosen with $m_x=0.4$ and $m_z=0.15$.
 }
    \label{fig:trace_ergo_adc}
\end{figure}

\section{Connection between EMC and state Mpemba effect}
\label{sec:connection_eme_state}

It is well known that the dynamical behavior of a quantum state and that of its associated observables can differ significantly; for instance, cloning a quantum state is fundamentally distinct from cloning the expectation value of an operator. Motivated by the conjecture that Mpemba crossings may arise from properties of specific observables rather than from intrinsic features of the quantum state itself~\cite{summer_prx_2026}, we analyze the relationship between trace-distance dynamics\footnote{The trace distance between two quantum states $\rho_1$ and $\rho_2$ is defined as half of the trace norm of their difference, and is given by
\( \mathcal{D}(\rho_1,\rho_2) = \frac{1}{2}\, \Tr\left[ \sqrt{(\rho_1-\rho_2)^\dagger(\rho_1-\rho_2)} \right]. \)
} and ergotropy. In order to perform the analysis, we keep the reference state, \(\rho_1\) fixed for ergotropy and state evolution, i.e., we want to find out the region of state that show EMC and state Mepmba effect with a fixed reference state \(\rho_1\). 
% So far, we have focused exclusively on ergotropic Mpemba crossings. It is, however, instructive to explore the relationship between EMC and the conventional Mpemba effect, which is typically defined in terms of how the \emph{state itself} approaches the steady state. In general, an observable and the quantum state may display qualitatively different dynamical behaviors, and the Mpemba effect is often better understood as a property of a specific observable rather than of the state as a whole.
% In this section, we examine the interplay between the absence of EMC and the absence of the usual state Mpemba crossings (MC). In particular, we ask the following question: \emph{Is it possible to observe Mpemba crossings in the trace distance but not in ergotropy?} We answer this question affirmatively, while showing that the converse does not hold.

%\emph{Amplitude damping channel.}
We begin by considering the \textcolor{black}{ADC} with initial states restricted to the $xz$-plane, \textcolor{black}{given by  \(\rho=
    \frac{1}{2}
    \begin{bmatrix}
    1+m_z & m_x\\
    m_x & 1-m_z
    \end{bmatrix}\),
where \(m_z\) and \(m_x\) denote the magnetizations along the \(z\)- and \(x\)-directions, respectively. } In this setting, the set of states that are equidistant from the steady state is described by
\textcolor{black}{
\begin{equation}
\sqrt{\frac{
m_x^2
+ \left(1 + m_z\right)^2
}{4}}
= \kappa,
\end{equation}}
where $\kappa$ is a constant. This relation defines a curve in the $xz$-plane (see the dashed curve in Fig.~\ref{fig:trace_ergo_adc}). The time-dependent trace distance from the steady state is given by
\begin{equation}
   \mathcal{D}(t) = \frac{
e^{-a t \gamma}
\sqrt{
\frac{1}{4}
+ m_x^2 a^2\frac{ e^{a \gamma t}}{4}
+ \frac{am_z}{2}\!\left(1 + \frac{am_z}{2}\right)
}
}{a},
\end{equation}
where $a=1+2n$. To identify the region of states exhibiting the state Mpemba effect with a reference state \(\rho_1\), we analyze the long-time limit $t\to\infty$. In this regime, the trace distance with the steady state simplifies to
\(
\mathcal{D}(t)\sim\frac{|m_x|}{2}\, e^{- \frac{t \gamma}{2}},
\)
revealing an exponential decay governed by the transverse magnetization $|m_x|$, which quantifies coherence in the $xz$-plane. Analogous to the behavior of ergotropy, two states satisfying $\mathcal{C}(\rho_1)<\mathcal{C}(\rho_2)$ exhibit the state Mpemba effect. Consequently, the region of states displaying the state Mpemba effect is bounded by the equidistance curve together with the coherence of the chosen reference state (as shown in Fig.~\ref{fig:trace_ergo_adc}).

% The isodistance curve in the $xz$ plane intersects the corresponding isoergotropic curve, and states satisfying the condition $\mathcal{C}(\rho_1)<\mathcal{C}(\rho_2)$ for a fixed value of $\kappa$ exhibit a state Mpemba effect in terms of the trace distance.

%\emph{Anisotropic Pauli channel.} 
A similar analysis can be carried out for the anisotropic Pauli channel. In this case, the set of states that share the same trace distance from the steady state is described by
%\begin{equation}
 \(   \sqrt{\frac{ m_x^2}{4}
+ \frac{m_z^2}{4}
}=\kappa,\)
%\end{equation}
where $\kappa$ is a constant. The time evolution of the trace distance is given by
\textcolor{black}{
\begin{equation}
   \mathcal{D}(t)= e^{-2t(3\gamma_\perp+\gamma_z)}
\sqrt{
e^{8t\gamma_\perp}\frac{ m_x^2}{4}
+ \frac{e^{4t(\gamma_\perp+\gamma_z)} m_z^2}{4}
}.
\end{equation}}
In the long-time limit, when $\gamma_z > \gamma_\perp$, the decay of the trace distance is approximately \( e^{-4 \gamma_\perp t}\,\frac{|m_z|}{2}, \) whereas for $\gamma_\perp > \gamma_z$, it behaves as \( e^{-2t(\gamma_z+\gamma_\perp)}\,\frac{|m_x|}{2}. \) Thus, both the initial energy (through $m_z$) and the coherence (through $m_x$) play a role in determining the occurrence of Mpemba crossings for states.

\textcolor{black}{\emph{Qutrit systems: State vs ergotropic Mpemba effect.} For both noise models, we observe that in a qubit system, all states exhibiting ergotropic Mpemba crossings (EMC) also display the state Mpemba effect, although the converse is not generally true (as shown in Fig.~\ref{fig:trace_ergo_adc}). To examine whether this behavior persists in higher dimensions, we extend our analysis to qutrit systems. Let us first exhibit that the state Mpemba effect cannot occur for diagonal qutrit states. For diagonal qutrits evolving under the amplitude damping channel governed by Eq.~(\ref{eq:gksl}), its trace distance from the stationary state is given by
\begin{equation}
    \mathcal{D}(t)=e^{-\gamma t}(p_1+p_2),
\end{equation}
where \(p_1\) and \(p_2\) denote the initial populations of the excited levels and $\gamma$ is the dissipation strength. For any two arbitrary initial states with populations (\(p_1^A,p_2^A\)) and (\(p_1^B,p_2^B\)), \(\mathcal{D}(t)\) decays exponentially with time at the same rate, and hence, the ordering of trace distances is preserved for all time, implying no intersection for any value of the population. Consequently, diagonal states in the energy basis do not exhibit the state Mpemba effect for any pair of initial states. Nevertheless, as shown in Fig.~\ref{fig:qutrit_random_state}(b), ergotropic Mpemba crossings can still arise for such diagonal qutrit states.}

These findings demonstrate that Mpemba crossings may occur in the trace distance without corresponding ergotropic Mpemba crossings, and conversely, ergotropic Mpemba crossings may arise without state Mpemba crossings in systems with dimension greater than two. Hence, there is no universal one-to-one correspondence between the two phenomena; rather, any apparent correlation is strongly system-dependent.

%\textbf{Lemma 3.} 

% \textbf{Lemma 3.} \emph{For two isoergotropic states, the ergotropy of the state with a larger magnetization along the $z$ direction decays faster than that of the state with smaller $m_z$ for non-Markovian evolution.} 

% \textcolor{red}{For the proof of the lemma, see Appendix~\ref{app:proof_lemma34}.\\
% On the other hand, the opposite is true for non-isoergotropic states, as shown in Lemma 2 for Markovian dynamics. In other words, one can prove that  {\it for two non-isoergotropic states with identical $m_x$, the ergotropy of the state with a larger $m_z$ decays more slowly than that of the state with a smaller $m_z$} (see Appendix E)}

%Because of that 
%It also implies that there can be no even number of ergotropic Mpemba crossing in the non-Markovian case.

% There are some states which show no-EMC but can show MC. Like isoergotropic surface we can define isodistance states from the steady state which intersects the isoergotropic surface as shown in Fig.~\ref{fig:trace_ergo_adc} and the equation of such states is given as 
% \begin{equation}
% \sqrt{
% \frac{1}{4}
% + \frac{m_x^2}{4}
% + m_z\!\left(\frac{2}{4} + \frac{m_z}{4}\right)
% }
% = \kappa,
% \end{equation}

% The states corresponding to the same trace distance for adc from stationary state
% is given by
% \[
% \frac{
% \sqrt{
% \frac{1}{4}
% + \frac{m_x^2 a^2}{4}
% + m_z\!\left(\frac{a}{2} + m_z\!\left(\frac{a}{2}\right)^2\right)
% }
% }{a}
% = c.
% \]

\section{Conclusion}
\label{sec:conclusion}

The quantum Mpemba effect is the quantum analogue of the classical Mpemba effect, wherein a system initially farther from equilibrium can relax to its equilibrium state faster than a system that is initially closer to equilibrium. Such counterintuitive behavior is of practical interest in quantum technologies, as it can potentially be exploited to enhance the performance of quantum devices.

Motivated by the broader relevance of nonequilibrium relaxation phenomena, we investigated the ergotropic Mpemba effect in finite-dimensional quantum systems, with a particular focus on two- and three-level systems. Concentrating on ergotropic Mpemba crossings (EMC), defined as instances where two ergotropy curves intersect at least once during the dynamics, we derived a criterion for the occurrence of such crossings. Specifically, we showed that when the system is governed by a generalized amplitude damping channel (gADC), the occurrence of EMC is entirely dictated by the relative coherence of the initial states. This allows us to characterize the complete set of states that do not exhibit crossings with a fixed reference state and to show that this set forms a spherocylinder in state space, with the reference state lying on its surface.
\textcolor{black}{Precisely, we found an explicit condition \(\theta_1 + \theta_2 \geq \pi\) expressed in terms of state parameters \((\theta_i, \phi_i)\) (\(i=1,2\)) for a pair of pure initial states under which the EMC cannot occur. This criterion is further corroborated through the behavior of the Mpemba parameter.}
%and to characterize its geometry.
%the EMC criterion depends directly on the initial coherence of the states. Based on this criterion, we identified the complete set of states that do not exhibit EMC with a fixed reference state $\rho$, and 
%demonstrated that this set has the geometry of a spherocylinder with this reference state lying on its surface. 
Beyond the gADC, we also established the conditions for EMC in the presence of an anisotropic Pauli channel and found that both coherence and energy jointly control the emergence of EMC.
%which does not belong to the Davies class.
To elucidate the physical origin of ergotropic Mpemba crossings, we analyzed the coherent and incoherent contributions to ergotropy. In two-dimensional systems, we revealed that the incoherent ergotropy decays exponentially in time and does not exhibit Mpemba behavior, whereas the coherent ergotropy displays a transient increase followed by a rapid decay, indicating that the delayed relaxation of ergotropy is governed by the coherent contribution. In contrast, for three-level systems, we demonstrated that incoherent ergotropy alone can exhibit EMC due to the presence of multiple relaxation pathways between energy levels with different rates.

We further extended our analysis to non-Markovian dynamics, where we showed that, unlike the Markovian case, multiple ergotropic Mpemba crossings may occur. In particular, we proved that the total number of crossings is always odd. 
%As in the Markovian scenario, we identified the corresponding set of initial states that exhibit EMC with a fixed reference state. Finally, 
Further, we explored the relationship between ergotropic Mpemba crossings and the conventional state Mpemba effect. For two-dimensional systems, we observed that the presence of ergotropic Mpemba crossings implies the presence of state Mpemba crossings, although the converse need not hold. Our results provide criteria for identifying initial states that enable faster energy extraction in finite-dimensional systems. \textcolor{black}{However,  such phenomena are expected to be observed in other systems also, e.g., battery operation with non-Gaussian states~\cite{adhikary2026}, offering practical guidance for the realization and optimization of quantum batteries in laboratory settings. An intriguing direction for future research is to explore the connection between the ergotropic Mpemba effect and the quantum speed limit. Recent studies have shown that far-from-equilibrium quantum states with higher speed efficiency can relax faster than states initially closer to equilibrium, thereby establishing a link between the quantum state Mpemba effect and fundamental bounds on quantum evolution~\cite{Srivastav_pra_2025}. Extending such a connection to the EMC could provide new insights into the finite-time optimization of quantum batteries and nonequilibrium quantum thermodynamics. We therefore believe that the  results may stimulate further investigations  with implications for both the foundations of quantum thermodynamics and the development of practical quantum energy-storage devices.}
 
\acknowledgements
We acknowledge support from the project entitled ``Technology Vertical - Quantum Communication'' under the National Quantum Mission of the Department of Science and Technology (DST)  (Sanction Order No. DST/QTC/NQM/QComm/$2024/2$ (G)). This research was carried out and financed within the framework of the second Swiss Contribution MAPS (Grant No. 230870).

\appendix
\section{Vectorization of GKSL master equation}
\label{sec:vectorization}
Unlike unitary evolution, open-system dynamics are not invertible and instead form a semigroup. The evolution of the system is governed by the GKSL  master equation,   given by
\begin{equation}
\frac{d\rho}{dt}
= -i[H_B,\rho]
+ \sum_k \gamma_k \left( L_k \rho L_k^{\dagger}
- \frac{1}{2} \{ L_k^{\dagger} L_k, \rho \} \right),
\label{eq:gksl}
\end{equation}
where $\rho$ is the reduced density matrix of the system obtained after tracing out the environment and \(H_B\) is the Hamiltonian of the system. The operators $L_k$ are the Lindblad jump operators that encode the action of the environment on the system, and $\gamma_k$ denotes the corresponding decay rates determined by the system-environment coupling. Now, if the dynamics are governed by a GKSL master equation, i.e., the dynamics is Markovian, the corresponding quantum map is completely positive and divisible, and the time evolution can be written as
\begin{equation}
    ||\rho_t\rangle\rangle = e^{\mathcal{L} t} ||\rho_0\rangle\rangle,
    \label{eq:evol_rho}
\end{equation}
where $||\rho_0\rangle\rangle$ is the vectorised initial state and $\mathcal{L}$ denotes the Liouvillian superoperator. If the system consists of $N$ number of qubits, then the Liouvillian superoperator becomes a $4^N \times 4^N$ matrix, consisting of complex eigenvalues, and generally, the left and right eigenvectors of the Liouvillian are not connected by complex conjugation. The Liouvillian operator can be obtained from Eq. (\ref{eq:gksl}) which is given by 
\begin{eqnarray}
\mathcal{L}
= &&-i [ H_B \otimes \mathbb{I} - \mathbb{I} \otimes H_B^{\mathrm{T}} ]\nonumber \\
&& + \sum_j \frac{\gamma_j}{2} 
\left( 
2 L_j \otimes L_j^{*}
- L_j^{\dagger} L_j \otimes \mathbb{I}
- \mathbb{I} \otimes L_j^{\mathrm{T}} L_j^{*}
\right).
\end{eqnarray}
Now, the solution of the evolved state is given as
\begin{equation}
||\rho_t\rangle\rangle = ||\rho_{ss}\rangle\rangle + \sum_{i \neq 1} c_i(0)\, e^{\lambda_i t} ||r_i\rangle\rangle,
\end{equation}
where $||\rho_{ss}\rangle\rangle$ is the steady state and $r_i$ are the right eigenmatrices of the matrix \(\mathcal{L}\) and $c_i(0)=\Tr[l_i\rho_0]$ with $l_i$ being the left eigenmatrices of \(\mathcal{L}\). These eigenmatrices can be obtained by vectorization of the density matrix $\rho$~\cite{Gyamfi2020}. \textcolor{black}{We use this vectorization to obtain the analytical expressions of the density matrix at any time \(t\) in a compact form (eg. see Eqs. (\ref{eq:ergotropy_time}), (\ref{eq:timeevolve2vec}) and (\ref{eq:timeevolve3vec})).}

\begin{widetext}

\section{The evolution of the system under gADC}
\subsection{Single qubit battery evolution}
\label{app:single_qubit_evolution}
The evolution of the single qubit battery can be obtained via  vectorization of the Lindbladian. In this case, the Lindbladian, \(\mathcal{L}\) becomes a matrix of dimension $4\times4$, given by
\begin{equation}
\mathcal{L}=\begin{pmatrix}
 -\gamma_- & 0 & 0 & \gamma_+ \\
 0 &-\dfrac{\gamma_++\gamma_-}{2} - 2ih_z & 0 & 0 \\
 0 & 0 & -\dfrac{\gamma_++\gamma_-}{2} + 2ih_z & 0 \\
 \gamma_- & 0 & 0 & -\gamma_+
\end{pmatrix}.
\end{equation}

To find the evolved state, we need to find the right eigenvectors and the corresponding eigenvalues of $\mathcal{L}$  are given as
\begin{equation}
\lambda_1 = 0,\quad
\lambda_2 = -\gamma_+ - \gamma_-,\quad
\lambda_{3,4}=\tfrac{1}{2}\bigl(-\gamma_+-\gamma_-\pm4ih_z\bigr).
\end{equation}
and the corresponding right eigenmatrices read
\begin{align}
r_1 &= \frac{1}{\frac{\gamma_+}{\gamma_-}+1}\begin{pmatrix}\dfrac{\gamma_+}{\gamma_-} & 0\\[6pt] 0 & 1\end{pmatrix},\qquad
r_2 = \begin{pmatrix}-1 & 0\\[6pt] 0 & 1\end{pmatrix},\qquad
r_3 = \begin{pmatrix}0 & 0\\[6pt] 1 & 0\end{pmatrix},\qquad
r_4 = \begin{pmatrix}0 & 1\\[6pt] 0 & 0\end{pmatrix}, 
\end{align}
while the corresponding left eigenmatrices are
\begin{align}
l_1 &= \begin{pmatrix}1 & 0\\[6pt] 0 & 1\end{pmatrix},\qquad
l_2 = \frac{1}{\frac{\gamma_{-}}{\gamma_{+}}+1}\begin{pmatrix}-\dfrac{\gamma_{-}}{\gamma_{+}} & 0\\[6pt] 0 & 1\end{pmatrix},\qquad
l_3 = \begin{pmatrix}0 & 1\\[6pt] 0 & 0\end{pmatrix},\qquad
l_4 = \begin{pmatrix}0 & 0\\[6pt] 1 & 0\end{pmatrix}.
\end{align}
Let the initial mixed state be of the form Eq.~(\ref{eq:qubit_initial_state}).
% \(\ket{\Psi(0)}=\cos\frac{\theta}{2}\ket{0}+e^{i\phi}\sin\frac{\theta}{2}\ket{1}\) where, \(\ket{0}(\ket{1})\) is the ground (excited) state of the battery, \(H_B\).
The energy of the state at time \(t\) is given by 
\begin{eqnarray}
&&E(\rho(t))
= \frac{h_z\,(1-s+m_za)}{a s},
\label{eq:timeevolve2vec}
\end{eqnarray}
The eigenvalues of the evolved state are given as 
\begin{align}
\lambda_- &=\frac{1}{2}-\frac{1}{2as}
\sqrt{\bigl(1-s+m_za\bigr)^2+a^2s(m_x^2+m_y^2)},\\[6pt]
\lambda_+ &=\frac{1}{2}+\frac{1}{2as}
\sqrt{\bigl(1-s+m_za\bigr)^2+a^2s(m_x^2+m_y^2)},
\end{align}
where \(\gamma_+=\gamma\,n\), \(\gamma_{-}=\gamma\,(n+1)\), $a=1+2n$ and $s=e^{at\gamma}$ with \(n\) being the bosonic occupation number. 
% The energy of $\rho(t)$ is % \begin{equation} % E(t)=\frac{h_z\,(1-s+\cos\theta\,a)}{as}, % \end{equation} % and the energy of the passive state is % \begin{equation} % E_{\text{passive}}=-\frac{h_z}{as} % \sqrt{\bigl(s-1-a\cos\theta\bigr)^2+a^2se^{-4t\gamma_3}\sin^2\theta}. % \end{equation}
So the energy of the passive state at time \(t\) is given by
\begin{eqnarray}
&&E(\rho_{\pi}(t))
= -\frac{h_z}{a s}
\sqrt{\bigl(1-s+m_za\bigr)^2+a^2s(m_x^2+m_y^2)}.
\label{eq:passive_state_energy}
\end{eqnarray}
Now using the eigenvalues of the evolved state, the time-dependent ergotropy is given as
\begin{eqnarray}
&&\mathcal{E}(t)
= \frac{h_z\,(1-s+m_za)}{a s}+ \frac{h_z}{a s}
\sqrt{\bigl(1-s+m_za\bigr)^2+a^2s(m_x^2+m_y^2)}.
\label{eq:general_ergotropy_dyn}
\end{eqnarray}

\subsection{Single qutrit battery evolution} 
\label{App:qutrit}

To study the dynamics of the ergotropy of the three-level battery, we mainly consider the states with diagonal entry. The Lindbladian superoperator of qutrit evolution is given by the following matrix:
\begin{equation}
\begin{pmatrix}
-2\gamma & 0 & 0 & 0 & 0 & 0 & 0 & 0 & 0 \\
0 & - i h_z - \frac{3\gamma}{2} & 0 & 0 & 0 & 0 & 0 & 0 & 0 \\
0 & 0 & -2 i h_z - \gamma & 0 & 0 & 0 & 0 & 0 & 0 \\
0 & 0 & 0 & i h_z - \frac{3\gamma}{2} & 0 & 0 & 0 & 0 & 0 \\
\gamma & 0 & 0 & 0 & -\gamma & 0 & 0 & 0 & 0 \\
0 & 0 & 0 & 0 & 0 & - i h_z - \frac{\gamma}{2} & 0 & 0 & 0 \\
0 & 0 & 0 & 0 & 0 & 0 & 2 i h_z - \gamma & 0 & 0 \\
0 & 0 & 0 & 0 & 0 & 0 & 0 & i h_z - \frac{\gamma}{2} & 0 \\
\gamma & 0 & 0 & 0 & \gamma & 0 & 0 & 0 & 0
\end{pmatrix},
\label{Eq:qutrit_supop}
\end{equation}
where we have considered all the transition rates due to the environments being fixed to \(\gamma\) and the temperature of the bath to be \(T=0\).  Owing to the block structure of the Lindblad generator, the state remains diagonal throughout the evolution, and the evolved state is given as
\begin{equation}
\rho(t) = R_1
- p_1 R_8 \, e^{-(2\gamma t)}
-(p_1+p_2) R_9 \, e^{-\gamma t},
\label{eq:timeevolve3vec}
\end{equation}
with
\begin{align}
    R_1 =
\begin{pmatrix}
0 & 0 & 0 \\
0 & 0 & 0 \\
0 & 0 & 1 
\end{pmatrix}, \qquad R_8 =
\begin{pmatrix}
-1 & 0 & 0 \\
0 & 1 & 0 \\
0 & 0 & 0 
\end{pmatrix}, \qquad \text{and } \qquad R_9 =
\begin{pmatrix}
0 & 0 & 0 \\
0 & -1 & 0 \\
0 & 0 & 1 
\end{pmatrix}.
\end{align}
\textcolor{black}{Note that the Lindblad superoperator in Eq. (\ref{Eq:qutrit_supop}) has some off-diagonal terms, and it can be decomposed into block-diagonal form \(\mathcal{L}=\mathcal{L}_{p}\oplus \mathcal{L}_{c}\) as described in Sec.~\ref{sec:qutrit_EME} where \(\mathcal{L}_{p}\) governs the population dynamics in the system while \(\mathcal{L}_{c}\) dictates the coherence evolution in the system. In our case, as the initial state is always a diagonal state, the evolved state resides in the \(\mathcal{L}_{p}\) block, with no transfer to the coherence subspace, \(\mathcal{L}_c\), and consequently, the dynamical state remains diagonal throughout the evolution.} 
% The ergotropy of the given state at any time, \(t\) is given as

% The eigenvalues of this matrix are 
% \[
% \left\{
% 0,\;
% -\frac{1}{2} i (2 h_z - i \gamma),\;
% - i (2 h_z - i \gamma),\;
% \frac{1}{2} i (2 h_z + i \gamma),\;
% i (2 h_z + i \gamma),\;
% -\frac{1}{2} i (2 h_z - 3 i \gamma),\;
% \frac{1}{2} i (2 h_z + 3 i \gamma),\;
% -2 \gamma,\;
% -\gamma
% \right\}
% \]

% The right eigenmatrices are 

% \[
% R_1 =
% \begin{pmatrix}
% 0 & 0 & 0 \\
% 0 & 0 & 0 \\
% 0 & 0 & 1 
% \end{pmatrix}.
% \]

% \[
% R_2 =
% \begin{pmatrix}
% 0 & 1 & 0 \\
% 0 & 0 & 0 \\
% 0 & 0 & 0 
% \end{pmatrix}.
% \]

% \[
% R_3 =
% \begin{pmatrix}
% 0 & 0 & 0 \\
% 1 & 0 & 0 \\
% 0 & 0 & 0 
% \end{pmatrix}.
% \]

% \[
% R_4 =
% \begin{pmatrix}
% 0 & 0 & 1 \\
% 0 & 0 & 0 \\
% 0 & 0 & 0 
% \end{pmatrix}.
% \]

% \[
% R_5 =
% \begin{pmatrix}
% 0 & 0 & 0 \\
% 0 & 0 & 0 \\
% 1 & 0 & 0 
% \end{pmatrix}.
% \]

% \[
% R_6 =
% \begin{pmatrix}
% 0 & 0 & 0 \\
% 0 & 0 & 1 \\
% 0 & 0 & 0 
% \end{pmatrix}.
% \]

% \[
% R_7 =
% \begin{pmatrix}
% 0 & 0 & 0 \\
% 0 & 0 & 0 \\
% 0 & 1 & 0 
% \end{pmatrix}.
% \]

% \[
% R_8 =
% \begin{pmatrix}
% -1 & 0 & 0 \\
% 0 & 1 & 0 \\
% 0 & 0 & 0
% \end{pmatrix}.
% \]

% \[
% R_9 =
% \begin{pmatrix}
% 0 & 0 & 0 \\
% 0 & -1 & 0 \\
% 0 & 0 & 1 
% \end{pmatrix}.
% \]
\end{widetext}

% \begin{figure}
%     \centering
%     \includegraphics[width=0.9\columnwidth]{qubit_pauli.pdf}
%     \caption{.}
%     \label{fig:qubit_pauli}
% \end{figure}

\section{Mpemba parameter for EMC}
\label{sec:order_parameter}

%\textcolor{black}{One of the well-studied aspects of the Mpemba effect is how fast the Mpemba crossing occurs as there are works which derive the conditions of fast Mepmba  crossings. Based on the preceding discussion,} 

%\textcolor{black}{\emph{Mpemba crossing time.} Here, 
\textcolor{black}{Let us now  derive the Mpemba crossing time in more detail and discuss its applicability under different parameter regimes. For the amplitude-damping channel, the ergotropy dynamics of an arbitrary pure state is given by
\begin{align}
\mathcal{E}_\theta(t)
&\nonumber = \frac{(1-s+a\cos\theta)}{as}\\& + \frac{1}{as} \sqrt{(s-1-a\cos\theta)^2+a^2s\sin^2\theta},
\end{align}
where \(s=e^{a\gamma t}\), with \(\gamma\) denoting the dissipation strength, \(a=1+2n\),  \(n\) is the bosonic occupation number, and \(\theta\) specifies the position of the pure state on the Bloch sphere. If an ergotropic Mpemba crossing occurs at a finite time \(t=t_\ast\), then the ergotropies of two initial states, characterized by the Bloch sphere coordinates \(\theta_1\) and \(\theta_2\), must coincide. This condition yields}
\textcolor{black}{
\begin{eqnarray}
&&\mathcal{E}_{\theta_1}(t_\ast)-\mathcal{E}_{\theta_2}(t_\ast)=0 \nonumber \\
&& \Rightarrow
(\cos\theta_1-\cos\theta_2)
+\frac{1}{a}
\left(\sqrt{y_1}-\sqrt{y_2}\right)=0,
\end{eqnarray}
where \(y_{1(2)}^2=(s-1-a\cos\theta_{1(2)})^2+a^2s\sin^2\theta_{1(2)}\).Using this relation, we obtain
\begin{eqnarray}
&&\sqrt{y_1}-\sqrt{y_2}=a(\cos\theta_2-\cos\theta_1)\nonumber\\
&&\nonumber \Rightarrow a(\cos\theta_1+\cos\theta_2)
-as(\cos\theta_1+\cos\theta_2)
-2(s-1)\\&&\qquad\qquad\qquad\qquad- a(\cos\theta_1-\cos\theta_2) +2\sqrt{y_2}=0\nonumber\\
&&\nonumber \Rightarrow a^2s^2(\cos\theta_1+\cos\theta_2)^2
-4a^2s\cos\theta_2\cos\theta_1
+
\\&&\qquad\qquad4as(s-1)(\cos\theta_1+\cos\theta_2)
-4a^2s =0\nonumber\\
&&\Rightarrow
s
=
\frac{
4\Bigl(a(1+\cos\theta_1\cos\theta_2)
+(\cos\theta_1+\cos\theta_2)\Bigr)
}{
(\cos\theta_1+\cos\theta_2)
\Bigl(4+a(\cos\theta_1+\cos\theta_2)\Bigr)
}.
\end{eqnarray}}
\textcolor{black}{Substituting \(s=e^{a\gamma t_\ast}\), we find
\begin{align}
t_\ast
=
\frac{1}{a\gamma}
\ln\left(
4\left[
\frac{
a\left(1+\cos\theta_{1}\cos\theta_{2}\right)
+\left(\cos\theta_{1}+\cos\theta_{2}\right)
}{
\left(\cos\theta_{1}+\cos\theta_{2}\right)
\left(4+a\left(\cos\theta_{1}+\cos\theta_{2}\right)\right)
}
\right]
\right),
\label{eq:crossing_time}
\end{align}
which is precisely Eq.~(\ref{eq:qubit_tstar}).}

Going beyond a qualitative study, we quantify the EMC by addressing the natural question of how strong the effect is when an EMC occurs. To quantify this strength,  a quantity called ``Mpemba parameter'', analogous to that employed in the standard Mpemba effect~\cite{Furtado2025} was introduced. It is defined as
\begin{equation}
    \mathcal{O}
    =
    \frac{
    \int_{\mathcal{E}_2>\mathcal{E}_1}
    \left|\mathcal{E}_2(t)-\mathcal{E}_1(t)\right| \, dt
    }{
    \int_{0}^{t_f}
    \left|\mathcal{E}_2(t)-\mathcal{E}_1(t)\right| \, dt
    },
    \label{eq:Mpemba_param}
\end{equation}
where $\mathcal{E}_1(t)$ and $\mathcal{E}_2(t)$ denote the ergotropies of the initially higher- and lower-ergotropy states, respectively. 
%The denominator in  Eq.~(\ref{eq:Mpemba_param}) denotes the integral of the absolute value of the difference between the ergotropy of  two states with time while the numerator represents the integral of the abosolute value of the difference of ergotropy after the crossing at $t=t_*$. \textcolor{black}{ 
The quantity $\mathcal{O}$ measures the fraction of the total enclosed area between the curves $\mathcal{E}_1(t)$ and $\mathcal{E}_2(t)$ that lies in the region where the initially lower-ergotropy state temporarily overtakes the higher-ergotropy one, i.e., prior to and after the ergotropic Mpemba crossing at $t=t_*$. Note that $0 \le \mathcal{O} \le 1$ and vanishes in the absence of any crossing. Values of $\mathcal{O}$ close to zero correspond a very weak Mpemba effect, whereas values close to unity indicate Strong Mpemba effect. Thus, $\mathcal{O}$ provides a useful quantitative indicator of the strength of the ergotropic Mpemba effect.

\begin{figure}
    \centering
    \includegraphics[width=\columnwidth]{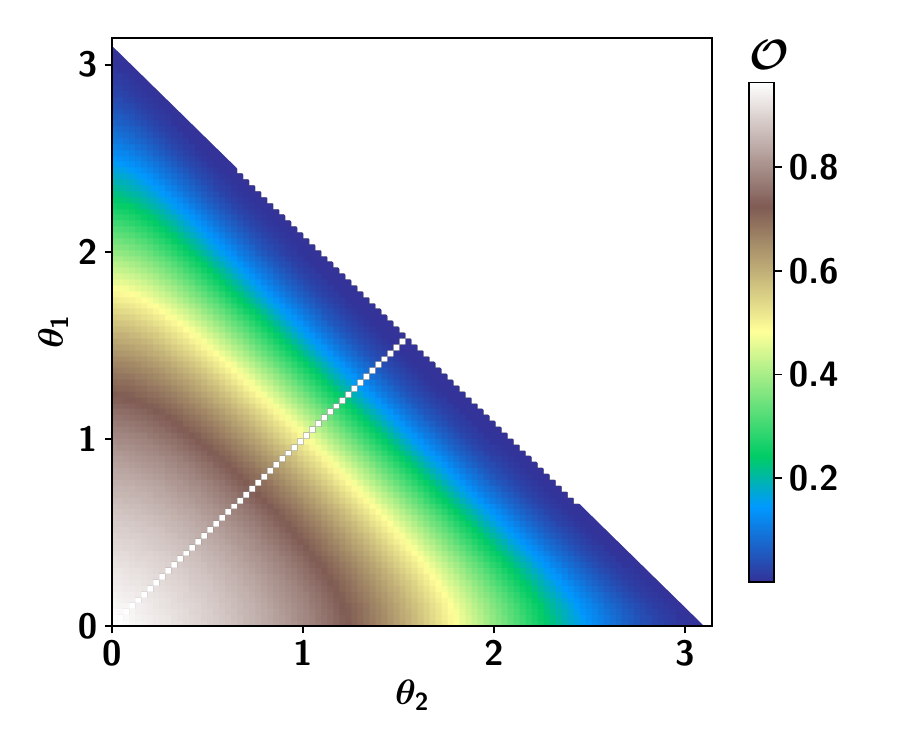}
    \caption{\textcolor{black}{Strength of ergotropic Mpemba crossing, quantified by $\mathcal{O}$ is plotted in the {$(\theta_1,\theta_2)$- plane} for amplitude damping noise on a single qubit. No ergotropic Mpemba crossing region for the pure states can be found in the white region, which is given by the relation \(\theta_1+\theta_2\ge \pi\). The white diamond line passing through the origin depicts \(\theta_1=\theta_2\), i.e., ergotropy dynamics of the two states being exactly the same. Other parameters are \(T=0\) and $\gamma_-=0.03$.}}
    \label{fig:qubit_colourplot}
\end{figure}

In Fig.~\ref{fig:qubit_colourplot}, we present the Mpemba parameter $\mathcal{O}$ as a function of the state parameter $\theta$, restricting our analysis to pure initial states. It is evident that when $\theta_1+\theta_2\ge \pi$, no Mpemba crossing occurs, as indicated by the white region in Fig.~\ref{fig:qubit_colourplot}. This behavior is fully consistent with \textbf{Corollary~1}.  Furthermore, the color gradient in the contour plot captures the variation of the Mpemba parameter. %\textcolor{black}{We observe that for small values of both $\theta_1$ and $\theta_2$, the parameter $\mathcal{O}$ approaches unity, signaling a fast Mpemba crossing. In contrast, when $\theta_1\approx \theta_2$, the value of $\mathcal{O}$ is significantly reduced, corresponding to slower crossings.}

For the single-qubit case, the slowest decaying modes (when $\gamma=0.0$) overlap with the initial pure state is given by $\sin\theta/2$. For the initial state $\theta=0$, corresponding to the ground state $|0\rangle$, this overlap vanishes. Consequently, compared to a generic state on the Bloch sphere, the $|0\rangle$ state exhibits an exponential speedup in relaxation. As a result, a crossing in the relaxation dynamics is observed between the $|0\rangle$ state and any other state of the form given in Eq.~(\ref{eq:qubit_initial_state}) with arbitrary $\theta$.
%\textcolor{black}{ This phenomenon is referred to as the \emph{strong Mpemba effect}. For other values of $\theta$, the overlap remains finite, leading to a non-exponential speedup and giving rise to the \emph{weak Mpemba effect}.}

\begin{figure*}
    \centering
    \includegraphics[width=\linewidth]{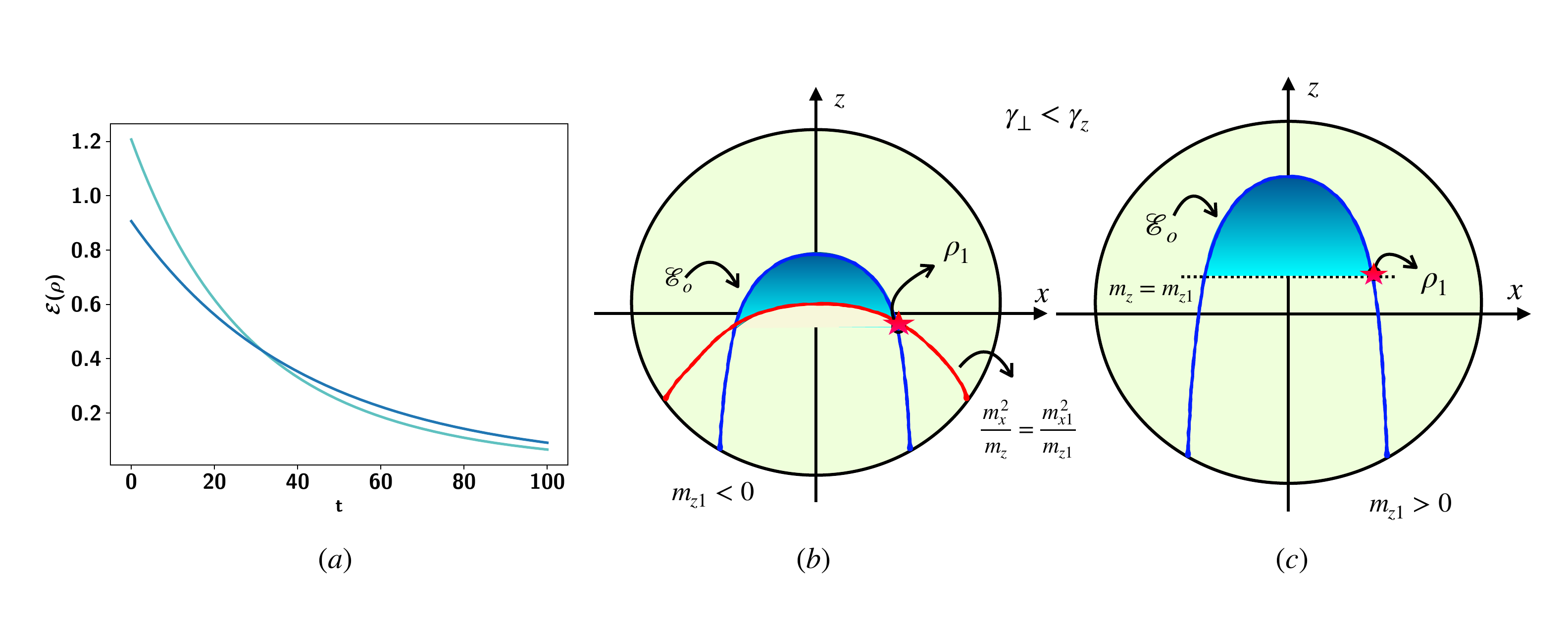}
    \caption{\textbf{EMC region for anisotropic Pauli channel in the \(xz\)-plane.} \textcolor{black}{(a) \(\mathcal{E}(\rho)\) vs \(t\) for two mixed states with \(\{m_x=0.5,m_z=0.5\}\), \(\{m_x=0.8,m_z=0.1\}\) and other system parameters \(\gamma_\perp=0.01\) and \(\gamma_z=0.001\). (b) A schematic presentation of the EMC (marked in blue) region for a fixed state \(\rho_1\) that resides on the lower hemisphere of the Bloch sphere with \(m_{z1}<0\), marked by a star. (c) A similar EME region is presented by blue region for the state, \(\rho_1\) belongs to the upper hemisphere, i.e, \(m_z>0\).}}
    \label{fig:pauli_eme_region}
\end{figure*}

\section{Anisotropic Pauli noise}
\label{app:pauli}
In the case, when the battery is exposed to the anisotropic Pauli channel, we obtain an ergotropic Mpemba crossings (see Fig.~\ref{fig:pauli_eme_region} (a)), suggesting apart from the Davis map, other quantum channels can showcase the EMC. We obtain a trade-off relationship between the dissipation strength and the initial energy and coherence of the initial states. Unlike ADC, the initial energy of the state plays a role in the occurrence of Mpemba crossings, which give rise to the following theorem: 

\textbf{Theorem 2.}  \emph{For an anisotropic Pauli channel, the occurrence of ergotropic Mpemba crossings between two states $\rho_1$ and $\rho_2$ with $\mathcal{E}_1>\mathcal{E}_2$ and \(m_z>0\), depends on the relative strengths of the transverse and longitudinal noise rates, $\gamma_\perp$ and $\gamma_z$. 
If $\gamma_\perp<\gamma_z$, an ergotropic Mpemba crossing occurs when $E_1<E_2$, where $E_{1(2)}$ denotes the energy of the state $\rho_{1(2)}$. 
Conversely, if $\gamma_\perp>\gamma_z$, EMC occurs when $\mathcal{C}(\rho_1)<\mathcal{C}(\rho_2)$.}

\begin{proof}
To prove the theorem, we again exploit the phase-covariant nature of the anisotropic Pauli channel, which allows us to restrict the analysis to the evolution in the $xz$ plane. The time-dependent ergotropy for an arbitrary initial state is given by
\begin{equation}
   \mathcal{E}(t,m_z,m_x)
= e^{-4 \gamma_\perp t}
\bigg [ m_z + \sqrt{ m_z^2 + e^{4t(\gamma_\perp-\gamma_z)} m_x^2 } \bigg ].
\label{eq:pauli_ergotropy}
\end{equation}

To identify the conditions for EMC, we analyze the long-time behavior of the ergotropy. In the case of $\gamma_\perp < \gamma_z$, the ergotropy at large \(t\) can be expressed as
\begin{eqnarray}
    \mathcal{E}(t)
&\approx& e^{-4 \gamma_\perp t}
\left [
m_z + |m_z|
\left(
1 + \frac{e^{4t(\gamma_\perp-\gamma_z)}}{2}\frac{m_x^2}{m_z^2}
\right)
\right ]\nonumber\\&\approx& (m_z+|m_z|)\, e^{-4 \gamma_\perp t}
+ \frac{e^{-4 \gamma_z t}}{2}\frac{m_x^2}{|m_z|}.
\label{eq:pauli_mzg}
\end{eqnarray}
Now in the case of \(m_z>0\), \(\mathcal{E}(t\to\infty) \sim 2m_z e^{-4 \gamma_\perp t}\) which tells that the decay of ergotropy depends solely on the initial magnetization along the $z$ direction. For two initial states satisfying $\mathcal{E}_1(0)>\mathcal{E}_2(0)$, the structure of the isoergotropic curves in the $xz$ plane implies $m_{z1}<m_{z2}$. Consequently, although $\rho_1$ initially possesses higher ergotropy, its ergotropy decays faster at long times, leading to
\[
\mathcal{E}_1(t)<\mathcal{E}_2(t)\quad \text{for sufficiently large } t,
\]
which guarantees the existence of an ergotropic Mpemba crossing at an earlier time. Since the energy of the initial state depends only on $m_z$, namely $E=\Tr[\rho(0)H_B]=m_z/2$, the crossing condition can be expressed as
\[
E_1<E_2.
\]

% and for \(m_z<0\) \(\mathcal{E}(p)\sim\frac{e^{-4 \gamma_z t}}{2}\frac{m_x^2}{|m_z|}\)

 On the other hand, when \textcolor{black}{\(\gamma_\perp>\gamma_z\)}, in the long- time limit $t\to\infty$ we obtain 
\begin{eqnarray}
    \mathcal{E}(t)
&\approx& m_z\, e^{-4 \gamma_\perp t}
+ e^{-2t(\gamma_\perp+\gamma_z)} |m_x|
\\&& \qquad\qquad+\frac{1}{2}
e^{-4 \gamma_\perp t}
e^{2t(\gamma_z-\gamma_\perp)}
\frac{m_z^2}{|m_x|}\nonumber\\&\approx&e^{-2t(\gamma_\perp+\gamma_z)} |m_x|.
\end{eqnarray}
In this case, the asymptotic decay of ergotropy is governed by the transverse magnetization $m_x$, and therefore by the coherence of the initial state. Using arguments analogous to those above, one finds that an ergotropic Mpemba crossing occurs whenever the initially higher-ergotropy state has smaller coherence, i.e.,
\[
\mathcal{C}(\rho_1)<\mathcal{C}(\rho_2).
\]

This completes the proof.
\end{proof}

In Fig.~\ref{fig:pauli_eme_region}(b) and (c), we depict the regions exhibiting ergotropic Mpemba crossings for a fixed reference state $\rho_1$, considering initial states with $m_z<0$ and $m_z>0$ in the $xz$-plane, respectively, \textcolor{black}{when \(\gamma_\perp < \gamma_z\)}. For $m_z>0$, we find that the line of constant magnetization $m_z=\mathrm{constant}$ forms the boundary of the EMC region, with the reference state $\rho_1$ located at the intersection of this line and an isoergotropic curve. Notably, the EMC region is significantly reduced for the anisotropic Pauli channel compared to the gADC case. In particular, unlike the gADC, no pair of pure states exhibits EMC under anisotropic Pauli noise. In contrast, when $m_z<0$, the second term in Eq.~(\ref{eq:pauli_mzg}) becomes relevant, modifying both the shape and extent of the EMC region. In this regime, the boundary of the EMC region is jointly determined by the isoergotropic curve and the constraint, \textcolor{black}{$m_x^2/m_z=\mathrm{const}$}. These results clearly demonstrate the sensitivity of EMC to the initial state parameters in the presence of anisotropic Pauli noise.

% So, When $\mathcal{E}_1(0)>\mathcal{E}_2(0)$ then $m_{x1}<m_{x2}$ must be satisfied for a crossing to exist. So $\mathcal{C}(\rho_1)<\mathcal{C}(\rho_2)$ must be satisfied.

% The asymptotic decay in the two relevant noise regimes is summarized in Table~\ref{tab:ergotropy-asymptotics}.

% \begin{table}[h]
% \centering
% \begin{tabular}{c c}
% \hline
% \textbf{Noise regime} & \textbf{Asymptotic ergotropy} \\
% \hline
% $\gamma_\perp < \gamma_z$ 
% & $\sim (m_z+|m_z|)\, e^{-4 \gamma_\perp t}$ \\[6pt]
% $\gamma_\perp > \gamma_z$ 
% & $\sim |m_x|\, e^{-2(\gamma_\perp+\gamma_z)t}$ \\
% \hline
% \end{tabular}
% \caption{Asymptotic behavior of the ergotropy for different anisotropic Pauli noise regimes.}
% \label{tab:ergotropy-asymptotics}
% \end{table}

% First, consider the regime $\gamma_\perp<\gamma_z$. In the long-time limit $t\to\infty$, 

% Next, consider the regime $\gamma_\perp>\gamma_z$. 
% \end{proof}

% Let us now focus on the case when \(m_z<0\). In this 

\begin{figure}
    \centering
    \includegraphics[width=\columnwidth]{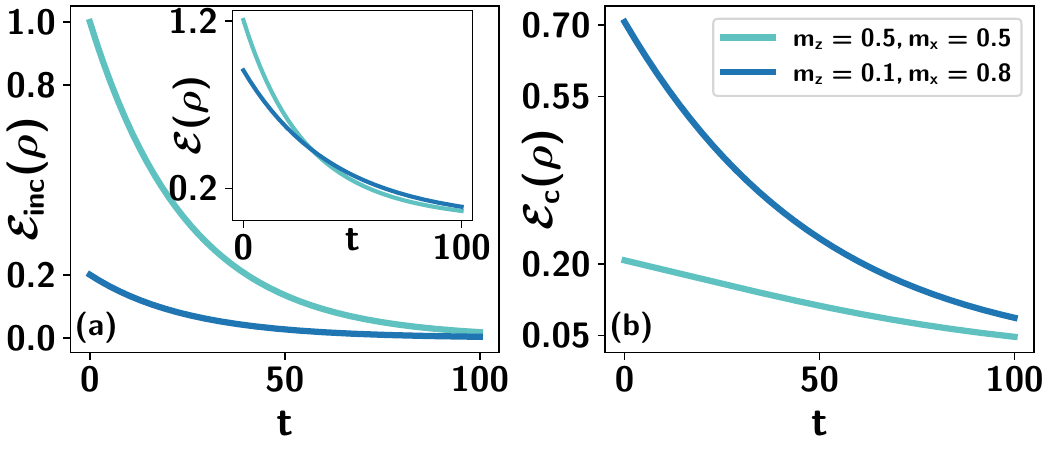}
    \caption{\textbf{Incoherent and coherent ergotropies against time for anisotropic Pauli channel.} (a) The incoherent ergotropy decreases exponentially, which does not show any crossings. 
    \textcolor{black}{ In the inset, we plot the total ergotropies corresponding the two initial state \(\rho_1\) and \(\rho_2\) which shows a EME.} (b) coherent ergotropy is plotted against \(t\). Unlike the gADC case, for anisotropic Pauli channel, coherent ergotropy decreases sharply. An exchange of behavior for the two different states is observed in coherent ergotropy which delays the saturation to the steady value of total ergotropy. Other parameters of the systems are \(h_z=1\), \(\gamma_\perp=0.01\) and \(\gamma_z=0.001\). }
    \label{fig:qubit_inco_pauli}
\end{figure}

\emph{Coherent and incoherent ergotropy in anisotropic Pauli channel.} Beyond the gADC, a similar qualitative behavior is observed for the anisotropic Pauli channel which can also be observed from Eq. (\ref{eq:pauli_ergotropy}). In this case as well, a behavioral exchange between the coherent and incoherent contributions to ergotropy occurs for pairs of states that exhibit EMC. Unlike the gADC, however, the coherent ergotropy does not display nonanalytic behavior in time, instead, it decreases sharply. Meanwhile, the coherent part of the ergotropy relaxes more slowly, while the incoherent part relaxes faster. The combined effect of these contrasting relaxation behaviors gives rise to ergotropic Mpemba crossings, as illustrated in Fig.~\ref{fig:qubit_inco_pauli}(a) and (b).

\section{Non-Markovian evolution of a qubit battery and the ergotropy}
\label{app:non_markovian}

In this section, we provide the details of the non-Markovian evolution of a single qubit where the entire system and environment are evolved by the interaction Hamiltonian~\cite{Kamin2020}, given in Eq.~(\ref{eq:non_markov_hint}). Now, we take an ansatz, given in Eq.~(\ref{eq:nm_ansatz}) and put it in the Schrodinger equation which is given as
\begin{align}
    i\dot{\nu}(t) &= \sum_k \eta_k(t) g_k e^{i(2h_z - \omega_k)t} \nonumber\\
    i \dot{\eta}_k(t) &= \nu(t) {g_k}^* e^{-i(2h_z - \omega_k)t},
\end{align}
where we have taken \(\hbar=1\). Now, the solution of the above coupled equation can be obtained easily  as
\begin{align}
    \dot{\nu}(t) &= -i \sum_k \eta_k(t) g_k e^{i(2h_z - \omega_k)t} \nonumber\\
    &= - \sum_k g_k {g_k}^* \int_0^t \nu(t') e^{-i(2h_z - \omega_k)t'} dt' \cdot e^{i(2h_z - \omega_k)t} \nonumber\\
    &= - \sum_k |g_k|^2 \int_0^t \nu(t') e^{i(2h_z - \omega_k)(t - t')} dt' \nonumber\\
    &= - \int_0^t K(t - t') \nu(t') dt',
\end{align}
where in the second line, we have used the fact that \(\eta_k(t) = -i \, {g_k}^* \int_0^t \nu(t') e^{-i(2h_z - \omega_k)t'} \, dt'\). Now, performing a Laplace transformation~\footnote{\textcolor{black}{The Laplace transform is taken using the formula \(f(s)=\int_{0}^{\infty} f(t) e^{-st} dt\).} } on both sides of the above equation, we obtain
\begin{eqnarray}
    &s\tilde{\nu}(s) - \nu(0) = -\tilde{K}(s) \tilde{\nu}(s) \nonumber\\
    &\Rightarrow \tilde{\nu}(s)[s + \tilde{K}(s)] = \nu(0) \nonumber\\
    & \tilde{\nu}(s) = \frac{\nu(0)}{s + \tilde{K}(s)} \nonumber\\&
    \qquad=\frac{\alpha}{s + \tilde{K}(s)}.
\end{eqnarray}
Now, in order to obtain \(\nu\), \textcolor{black}{we again perform an inverse Laplace transform,} which is given as
\begin{align}
    \abs{\nu} &= e^{-\frac{\lambda t}{2}} \left| \alpha \left( \cosh\left( \frac{\zeta t}{2} \right) + \frac{\lambda - i\Delta}{\zeta} \sinh\left( \frac{\zeta t}{2} \right) \right) \right|,
\end{align}
where \(\zeta^2 = (\lambda - i\Delta)^2 - 2\gamma \lambda\).
\textcolor{black}{
%It can be shown \(|\nu|^2<1\) for any time \(t\). 
Using the normalization condition of the evolved state in Eq.~(\ref{eq:nm_ansatz}), we obtain 
\begin{eqnarray}
&&|\alpha|^2|\nu|^2+\sum_k|\eta_k|^2+|\beta|^2=1,\nonumber\\
&&\Rightarrow |\alpha|^2|\nu|^2 = |\alpha|^2-\sum_k|\eta_k|^2,\nonumber\\
&&\Rightarrow|\alpha|^2|\nu|^2<|\alpha|^2
\Rightarrow
|\nu|^2<1,
\label{eq:nu_lessthan_one}
\end{eqnarray}
where, in the second line, we use \(|\alpha|^2+|\beta|^2=1\).
}
Now tracing out the environment, the evolved state of the system is given in Eq.~(\ref{eq:evolved_non_markov}) and the eigenvalues of the evolved state are given as
\begin{align}
    \frac{1}{2} \left( 1 - \sqrt{1 - 4\abs{\alpha\nu}^2 + 4\abs{\alpha\nu}^4 + 4\abs{\beta}^2 \abs{\alpha\nu}^2 } \right) \nonumber\\
    \frac{1}{2} \left( 1 + \sqrt{1 - 4\abs{\alpha\nu}^2 + 4\abs{\alpha\nu}^4 + 4\abs{\beta}^2 \abs{\alpha\nu}^2 } \right).
\end{align}
Using the eigenvalues of the evolved state, we can easily find out the ergotropy of the state,
% Energy at time $t$ is given by 
% \begin{equation}
% \rho(t) = h_z \left(-1 + 2|\alpha\nu|^2\right)
% \end{equation}
% And energy of the passive state is 
% \begin{equation}
%     E_{passive}= -h_z\sqrt{1 - 4|\alpha\nu|^2 + 4|\alpha\nu|^4 + 4|\beta|^2|\alpha\nu|^2}
% \end{equation}
% The ergotropy is given by 
\begin{align}
    \mathcal{E}(t)=& h_z \left(-1 + 2|\alpha\nu|^2\right) \nonumber\\&+ h_z \sqrt{1 - 4|\alpha\nu|^2  4|\alpha\nu|^4 + 4|\beta|^2|\alpha\nu|^2}.
\end{align}
Now as described in Sec.~\ref{sec:nonmarkov_mpemba}, we can easily generalize this formulation to a mixed initial state, where an arbitrary single qubit mixed state is given as 
\[
\rho_0 = \frac{1}{2}\left( I + \vec{m}\cdot\vec{\sigma} \right)
= \lambda_+ \ket{\psi_+}\bra{\psi_+}
+ \lambda_- \ket{\psi_-}\bra{\psi_-},
\]
with \(\lambda_{\pm} = \frac{1}{2}\left(1 \pm |\vec{m}|\right),\) and
\(
\ket{\psi_+} = \cos\!\left(\frac{\theta}{2}\right)\ket{0}
+ e^{i\phi}\sin\!\left(\frac{\theta}{2}\right)\ket{1}\equiv\alpha_+|0\rangle+\beta_+|1\rangle,
\) and 
\(
\ket{\psi_-} = \sin\!\left(\frac{\theta}{2}\right)\ket{0}
- e^{i\phi}\cos\!\left(\frac{\theta}{2}\right)\ket{1}\equiv\alpha_-|0\rangle+\beta_-|1\rangle
\).
Using the linarity of unitary and partial trace operation, we can add up the solutions for  $\ket{\psi_+}$ and $\ket{\psi_-}$ and the evolved state is given as
\[
\rho(t) = \lambda_+ \rho_+(t) + \lambda_- \rho_-(t),
\]
where \(\rho_+(0) = \ket{\psi_+}\bra{\psi_+}\) and \(\rho_-(0) = \ket{\psi_-}\bra{\psi_-}\) and the total evolved state is given as
\begin{widetext}
\begin{equation}
    \rho(t) =
\begin{pmatrix}
(\lambda_+ |\alpha_+|^2 + \lambda_- |\alpha_-|^2)|\nu|^2
&
(\lambda_+ \alpha_+\beta_+ + \lambda_- \alpha_- \beta_-)\nu
\\[6pt]
(\lambda_+ \alpha_+^* \beta_+^* + \lambda_- \alpha_-^* \beta_-^*)\nu^*
&
1 - (\lambda_+ |\alpha_+|^2 + \lambda_- |\alpha_-|^2)|\nu|^2
\end{pmatrix}.
\end{equation}
The eigenvalues of the evolved state are
\[
\lambda_1=\frac{1}{2}\left(
1 - \sqrt{
\left(2\left( \lambda_+ |\alpha_+|^2 + \lambda_- |\alpha_-|^2 \right)|\nu|^2 - 1\right)^2
+ 4 \left| (\lambda_+ \alpha_+\beta_+ + \lambda_- \alpha_- \beta_-)\nu \right|^2
}
\right),
\]

\[
\lambda_2=\frac{1}{2}\left(
1 + \sqrt{
\left(2\left( \lambda_+ |\alpha_+|^2 + \lambda_- |\alpha_-|^2 \right)|\nu|^2 - 1\right)^2
+ 4 \left| (\lambda_+ \alpha_+\beta_+ + \lambda_- \alpha_- \beta_-)\nu \right|^2
}
\right).
\]
using the eigenvalues of the state, one can easily calculate the ergotropy of the evolved state which is given as

\begin{equation}
    \mathcal{E}
=  h_z \left( 2\left( \lambda_+ |\alpha_+|^2 + \lambda_- |\alpha_-|^2 \right)|\nu|^2 - 1 \right)
+ h_z\sqrt{
\left(2\left( \lambda_+ |\alpha_+|^2 + \lambda_- |\alpha_-|^2 \right)|\nu|^2 - 1\right)^2
+ 4 \left| (\lambda_+ \alpha_+\beta_+ + \lambda_- \alpha_- \beta_-)\nu \right|^2
}.
\end{equation}
and \textcolor{black}{substituting the expressions of \(\lambda_{\pm}\), \(\alpha_{\pm}\) and \(\beta_{\pm}\), we obtain} the ergotropy in terms of magnetization of the state, which is given as
\begin{equation}
\mathcal{E}
=
h_z \left(
|\nu|^2 \left( 1 + |\vec m| \cos\theta \right) - 1
\right)
+ h_z \sqrt{
\left(
|\nu|^2 \left( 1 + |\vec m| \cos\theta \right) - 1
\right)^2
+ |\vec m|^2 \sin^2\theta \, |\nu|^2
}.
\label{eq:nonmarkov_ergo_m}
\end{equation}
\end{widetext}

%where

% \[
% \rho(t) =
% \begin{pmatrix}
% \lambda_+ |\nu_+|^2 + \lambda_- |\nu_-|^2
% &
% \lambda_+ \nu_+ \beta_+ + \lambda_- \nu_- \beta_-
% \\[6pt]
% \lambda_+ \nu_+^* \beta_+^* + \lambda_- \nu_-^* \beta_-^*
% &
% 1 - \left(\lambda_+ |\nu_+|^2 + \lambda_- |\nu_-|^2\right)
% \end{pmatrix}.
% \]

% Energy is
% \[
% E = h_z \left( 2\left( \lambda_+ |\alpha_+|^2 + \lambda_- |\alpha_-|^2 \right)|\nu|^2 - 1 \right).
% \]

% \[
% \frac{1}{2}\left(
% 1 - \sqrt{
% \left(1 - 2\lambda_+ |\nu_+|^2 - 2\lambda_- |\nu_-|^2\right)^2
% + 4 \left| \lambda_+ \nu_+ \beta_+ + \lambda_- \nu_- \beta_- \right|^2
% }
% \right),
% \]
% \[
% \frac{1}{2}\left(
% 1 + \sqrt{
% \left(1 - 2\lambda_+ |\nu_+|^2 - 2\lambda_- |\nu_-|^2\right)^2
% + 4 \left| \lambda_+ \nu_+ \beta_+ + \lambda_- \nu_- \beta_- \right|^2
% }
% \right).
% \]

\subsection{Proofs for Lemma 3 and 4} 
\label{app:proof_lemma34}
In this subsection, we prove the lemma 3 for non-Markovian ergotropic Mpemba effect which is given as
\textbf{Lemma 3.} \emph{For two isoergotropic states, the ergotropy of the state with a larger magnetization along the $z$ direction decays faster than that of the state with smaller $m_z$ for non-Markovian evolution.} 
\begin{proof}
In this proof, we again use the phase covariance of the channel and consider the ergotropy in the \(xz\)-plane. The ergotropy, in the \(xz\)-plane is given as
\begin{eqnarray}
&\mathcal{E}(t,m_x,m_z)
= \left(|\nu|^2(1+m_z)-1\right)
\nonumber \\
&+\sqrt{
\left(|\nu|^2(1+m_z)-1\right)^2 +
m_x^2 |\nu|^2
}.
\label{eq:nm_mx_mz}
\end{eqnarray}
Now using Eq.~(\ref{eq:iso_ergo_xz}), one can write that $m_x^2=\mathcal{E}_0^2-2\mathcal{E}_0 m_z$, where \(\mathcal{E}_0\) is the constant ergotropic surface. After substituting this into Eq.~(\ref{eq:nm_mx_mz}), we now prove that for a particular time $t$, the  ergotropy decays monotonically with the increase of $m_z$ on the isoergotropic surface. To do so, we compute the first order derivative of the ergotropy which is given as 
\begin{align} 
\frac{\partial \mathcal{E}(t,m_z)}{\partial m_z} = |\nu|^2 \left( 1+ \frac{A} {B} \right),
\end{align}
where \(A\equiv|\nu|^2(1+m_z)-1-\mathcal{E}_0 \) and \(B\equiv \sqrt{\left(|\nu|^2(1+m_z)-1\right)^2+
(\mathcal{E}_0^2-2\mathcal{E}_0 m_z)|\nu|^2}\). It is easy to find out that  \( |\nu|^2(1+m_z)-1-\mathcal{E}_0 = |\nu|^2-1+|\nu|^2 m_z-\mathcal{E}_0<0\), hence, \(A<0\), \textcolor{black}{where $|\nu|^2<1$ as shown in Eq. (\ref{eq:nu_lessthan_one})}. Now \(A^2-B^2=(\mathcal{E}_{0}^2+2\mathcal{E}_0)(1-|\nu|^2)>0\). So it is also clear that $\left|\frac{A}{B}\right|>1$. Hence, one can write
\[
 \frac{\partial \mathcal{E}(t,m_z)}{\partial m_z}<0,
\]
which proves that \(\mathcal{E}(t,m_z)\) monotonicity decreases with the increment of \(m_z\) on the isoergotropic surface.
\end{proof}

\textbf{Lemma 4.} \emph{For two non-isoergotropic states with identical $m_x$, the ergotropy of the state with a larger $m_z$ decays more slowly than that of the state with a smaller $m_z$.}
\begin{proof}
In order to prove the above statement, we fix the value of $m_x = M_x$ and computing the derivative with respect to $m_z$, we obtain
\begin{eqnarray}
\frac{\partial \mathcal{E}(t,m_z)}{\partial m_z}
&=& |\nu|^2 \left(
1 + \frac{|\nu|^2(1+m_z)-1}
{\sqrt{(|\nu|^2(1+m_z)-1)^2 + M_x^2 |\nu|^2}}
\right)\nonumber\\
    &\equiv& |\nu|^2 \left( 1 + \frac{C}{D} \right).
\end{eqnarray}
Now, it is easy to show that \(\left| \frac{C}{D} \right| < 1\) which indicates the following relation,
\[
\frac{\partial \mathcal{E}}{\partial m_z} > 0.
\]
This proves the lemma.
    
\end{proof}
% \emph{Odd number of crossings.} In order to prove the odd number of crossings of ergotropy curves during the evolution in the case of non-Markovian evolution. We evaluate the the behavior of ergotropy at large time limit ergotropy in eq.~(\ref{eq:nonmarkov_ergo_m}) can be written as
% % \[
% % \sqrt{
% % \left(
% % |\nu|^2 \left( 1 + |\vec m| \cos\theta \right) - 1
% % \right)^2
% % + |\vec m|^2 \sin^2\theta \, |\nu|^2
% % }
% % \]
% \begin{align}
% \mathcal{E}&\sim
% 1
% -
% |\nu|^2 \left( 1 + |\vec m| \cos\theta \right)
% +
% \frac{|\nu|^2}{2}
% |\vec m|^2 \sin^2\theta \nonumber\\& 
% \sim
% h_z \, \frac{|\nu|^2}{2} \, m_x^2,
% \end{align}
% which shows that of the initial state has large \(m_x\), it has lower ergotropy at the final time, hence, to obtain a crossing the coherence of the arbitrary state must be lowered than fixed state. 

\bibliography{ref.bib}

\begin{thebibliography}{97}%
\makeatletter
\providecommand \@ifxundefined [1]{%
 \@ifx{#1\undefined}
}%
\providecommand \@ifnum [1]{%
 \ifnum #1\expandafter \@firstoftwo
 \else \expandafter \@secondoftwo
 \fi
}%
\providecommand \@ifx [1]{%
 \ifx #1\expandafter \@firstoftwo
 \else \expandafter \@secondoftwo
 \fi
}%
\providecommand \natexlab [1]{#1}%
\providecommand \enquote  [1]{``#1''}%
\providecommand \bibnamefont  [1]{#1}%
\providecommand \bibfnamefont [1]{#1}%
\providecommand \citenamefont [1]{#1}%
\providecommand \href@noop [0]{\@secondoftwo}%
\providecommand \href [0]{\begingroup \@sanitize@url \@href}%
\providecommand \@href[1]{\@@startlink{#1}\@@href}%
\providecommand \@@href[1]{\endgroup#1\@@endlink}%
\providecommand \@sanitize@url [0]{\catcode `\\12\catcode `\$12\catcode
  `\&12\catcode `\#12\catcode `\^12\catcode `\_12\catcode `\%12\relax}%
\providecommand \@@startlink[1]{}%
\providecommand \@@endlink[0]{}%
\providecommand \url  [0]{\begingroup\@sanitize@url \@url }%
\providecommand \@url [1]{\endgroup\@href {#1}{\urlprefix }}%
\providecommand \urlprefix  [0]{URL }%
\providecommand \Eprint [0]{\href }%
\providecommand \doibase [0]{https://doi.org/}%
\providecommand \selectlanguage [0]{\@gobble}%
\providecommand \bibinfo  [0]{\@secondoftwo}%
\providecommand \bibfield  [0]{\@secondoftwo}%
\providecommand \translation [1]{[#1]}%
\providecommand \BibitemOpen [0]{}%
\providecommand \bibitemStop [0]{}%
\providecommand \bibitemNoStop [0]{.\EOS\space}%
\providecommand \EOS [0]{\spacefactor3000\relax}%
\providecommand \BibitemShut  [1]{\csname bibitem#1\endcsname}%
\let\auto@bib@innerbib\@empty
%</preamble>
\bibitem [{\citenamefont {Mpemba}\ and\ \citenamefont
  {Osborne}(1969)}]{Mpemba1969}%
  \BibitemOpen
  \bibfield  {author} {\bibinfo {author} {\bibfnamefont {E.~B.}\ \bibnamefont
  {Mpemba}}\ and\ \bibinfo {author} {\bibfnamefont {D.~G.}\ \bibnamefont
  {Osborne}},\ }\bibfield  {title} {\bibinfo {title} {Cool?},\ }\href
  {https://doi.org/10.1088/0031-9120/4/3/312} {\bibfield  {journal} {\bibinfo
  {journal} {Physics Education}\ }\textbf {\bibinfo {volume} {4}},\ \bibinfo
  {pages} {172} (\bibinfo {year} {1969})}\BibitemShut {NoStop}%
\bibitem [{\citenamefont {Lu}\ and\ \citenamefont {Raz}(2017)}]{Lu2017}%
  \BibitemOpen
  \bibfield  {author} {\bibinfo {author} {\bibfnamefont {Z.}~\bibnamefont
  {Lu}}\ and\ \bibinfo {author} {\bibfnamefont {O.}~\bibnamefont {Raz}},\
  }\bibfield  {title} {\bibinfo {title} {Nonequilibrium thermodynamics of the
  markovian mpemba effect and its inverse},\ }\href
  {https://doi.org/10.1073/pnas.1701264114} {\bibfield  {journal} {\bibinfo
  {journal} {Proceedings of the National Academy of Sciences}\ }\textbf
  {\bibinfo {volume} {114}},\ \bibinfo {pages} {5083–5088} (\bibinfo {year}
  {2017})}\BibitemShut {NoStop}%
\bibitem [{\citenamefont {Liu}\ \emph {et~al.}(2024{\natexlab{a}})\citenamefont
  {Liu}, \citenamefont {Zhang}, \citenamefont {Yin},\ and\ \citenamefont
  {Zhang}}]{liu_prl_2024}%
  \BibitemOpen
  \bibfield  {author} {\bibinfo {author} {\bibfnamefont {S.}~\bibnamefont
  {Liu}}, \bibinfo {author} {\bibfnamefont {H.-K.}\ \bibnamefont {Zhang}},
  \bibinfo {author} {\bibfnamefont {S.}~\bibnamefont {Yin}},\ and\ \bibinfo
  {author} {\bibfnamefont {S.-X.}\ \bibnamefont {Zhang}},\ }\bibfield  {title}
  {\bibinfo {title} {Symmetry restoration and quantum mpemba effect in
  symmetric random circuits},\ }\href
  {https://doi.org/10.1103/PhysRevLett.133.140405} {\bibfield  {journal}
  {\bibinfo  {journal} {Phys. Rev. Lett.}\ }\textbf {\bibinfo {volume} {133}},\
  \bibinfo {pages} {140405} (\bibinfo {year} {2024}{\natexlab{a}})}\BibitemShut
  {NoStop}%
\bibitem [{\citenamefont {Turkeshi}\ \emph {et~al.}(2025)\citenamefont
  {Turkeshi}, \citenamefont {Calabrese},\ and\ \citenamefont
  {De~Luca}}]{Turkeshi2025}%
  \BibitemOpen
  \bibfield  {author} {\bibinfo {author} {\bibfnamefont {X.}~\bibnamefont
  {Turkeshi}}, \bibinfo {author} {\bibfnamefont {P.}~\bibnamefont
  {Calabrese}},\ and\ \bibinfo {author} {\bibfnamefont {A.}~\bibnamefont
  {De~Luca}},\ }\bibfield  {title} {\bibinfo {title} {Quantum mpemba effect in
  random circuits},\ }\href {https://doi.org/10.1103/5d6p-8d1b} {\bibfield
  {journal} {\bibinfo  {journal} {Phys. Rev. Lett.}\ }\textbf {\bibinfo
  {volume} {135}},\ \bibinfo {pages} {040403} (\bibinfo {year}
  {2025})}\BibitemShut {NoStop}%
\bibitem [{\citenamefont {Yu}\ \emph {et~al.}(2025)\citenamefont {Yu},
  \citenamefont {Liu},\ and\ \citenamefont {Zhang}}]{Yu2025}%
  \BibitemOpen
  \bibfield  {author} {\bibinfo {author} {\bibfnamefont {H.}~\bibnamefont
  {Yu}}, \bibinfo {author} {\bibfnamefont {S.}~\bibnamefont {Liu}},\ and\
  \bibinfo {author} {\bibfnamefont {S.-X.}\ \bibnamefont {Zhang}},\ }\bibfield
  {title} {\bibinfo {title} {Quantum mpemba effects from symmetry
  perspectives},\ }\bibfield  {journal} {\bibinfo  {journal} {AAPPS Bulletin}\
  }\textbf {\bibinfo {volume} {35}},\ \href
  {https://doi.org/10.1007/s43673-025-00157-7} {10.1007/s43673-025-00157-7}
  (\bibinfo {year} {2025})\BibitemShut {NoStop}%
\bibitem [{\citenamefont {Ares}\ \emph {et~al.}(2023)\citenamefont {Ares},
  \citenamefont {Murciano},\ and\ \citenamefont {Calabrese}}]{Ares2023}%
  \BibitemOpen
  \bibfield  {author} {\bibinfo {author} {\bibfnamefont {F.}~\bibnamefont
  {Ares}}, \bibinfo {author} {\bibfnamefont {S.}~\bibnamefont {Murciano}},\
  and\ \bibinfo {author} {\bibfnamefont {P.}~\bibnamefont {Calabrese}},\
  }\bibfield  {title} {\bibinfo {title} {Entanglement asymmetry as a probe of
  symmetry breaking},\ }\bibfield  {journal} {\bibinfo  {journal} {Nature
  Communications}\ }\textbf {\bibinfo {volume} {14}},\ \href
  {https://doi.org/10.1038/s41467-023-37747-8} {10.1038/s41467-023-37747-8}
  (\bibinfo {year} {2023})\BibitemShut {NoStop}%
\bibitem [{\citenamefont {Yamashika}\ \emph {et~al.}(2024)\citenamefont
  {Yamashika}, \citenamefont {Ares},\ and\ \citenamefont
  {Calabrese}}]{Yamashika2024}%
  \BibitemOpen
  \bibfield  {author} {\bibinfo {author} {\bibfnamefont {S.}~\bibnamefont
  {Yamashika}}, \bibinfo {author} {\bibfnamefont {F.}~\bibnamefont {Ares}},\
  and\ \bibinfo {author} {\bibfnamefont {P.}~\bibnamefont {Calabrese}},\
  }\bibfield  {title} {\bibinfo {title} {Entanglement asymmetry and quantum
  mpemba effect in two-dimensional free-fermion systems},\ }\href
  {https://doi.org/10.1103/PhysRevB.110.085126} {\bibfield  {journal} {\bibinfo
   {journal} {Phys. Rev. B}\ }\textbf {\bibinfo {volume} {110}},\ \bibinfo
  {pages} {085126} (\bibinfo {year} {2024})}\BibitemShut {NoStop}%
\bibitem [{\citenamefont {Murciano}\ \emph {et~al.}(2024)\citenamefont
  {Murciano}, \citenamefont {Ares}, \citenamefont {Klich},\ and\ \citenamefont
  {Calabrese}}]{Murciano2024}%
  \BibitemOpen
  \bibfield  {author} {\bibinfo {author} {\bibfnamefont {S.}~\bibnamefont
  {Murciano}}, \bibinfo {author} {\bibfnamefont {F.}~\bibnamefont {Ares}},
  \bibinfo {author} {\bibfnamefont {I.}~\bibnamefont {Klich}},\ and\ \bibinfo
  {author} {\bibfnamefont {P.}~\bibnamefont {Calabrese}},\ }\bibfield  {title}
  {\bibinfo {title} {Entanglement asymmetry and quantum mpemba effect in the xy
  spin chain},\ }\href {https://doi.org/10.1088/1742-5468/ad17b4} {\bibfield
  {journal} {\bibinfo  {journal} {Journal of Statistical Mechanics: Theory and
  Experiment}\ }\textbf {\bibinfo {volume} {2024}},\ \bibinfo {pages} {013103}
  (\bibinfo {year} {2024})}\BibitemShut {NoStop}%
\bibitem [{\citenamefont {Ares}\ \emph {et~al.}(2025)\citenamefont {Ares},
  \citenamefont {Calabrese},\ and\ \citenamefont {Murciano}}]{Ares2025}%
  \BibitemOpen
  \bibfield  {author} {\bibinfo {author} {\bibfnamefont {F.}~\bibnamefont
  {Ares}}, \bibinfo {author} {\bibfnamefont {P.}~\bibnamefont {Calabrese}},\
  and\ \bibinfo {author} {\bibfnamefont {S.}~\bibnamefont {Murciano}},\
  }\bibfield  {title} {\bibinfo {title} {The quantum mpemba effects},\ }\href
  {https://doi.org/10.1038/s42254-025-00838-0} {\bibfield  {journal} {\bibinfo
  {journal} {Nature Reviews Physics}\ }\textbf {\bibinfo {volume} {7}},\
  \bibinfo {pages} {451–460} (\bibinfo {year} {2025})}\BibitemShut {NoStop}%
\bibitem [{\citenamefont {Carollo}\ \emph
  {et~al.}(2021{\natexlab{a}})\citenamefont {Carollo}, \citenamefont
  {Lasanta},\ and\ \citenamefont {Lesanovsky}}]{Carollo2021}%
  \BibitemOpen
  \bibfield  {author} {\bibinfo {author} {\bibfnamefont {F.}~\bibnamefont
  {Carollo}}, \bibinfo {author} {\bibfnamefont {A.}~\bibnamefont {Lasanta}},\
  and\ \bibinfo {author} {\bibfnamefont {I.}~\bibnamefont {Lesanovsky}},\
  }\bibfield  {title} {\bibinfo {title} {Exponentially accelerated approach to
  stationarity in markovian open quantum systems through the mpemba effect},\
  }\href {https://doi.org/10.1103/PhysRevLett.127.060401} {\bibfield  {journal}
  {\bibinfo  {journal} {Phys. Rev. Lett.}\ }\textbf {\bibinfo {volume} {127}},\
  \bibinfo {pages} {060401} (\bibinfo {year} {2021}{\natexlab{a}})}\BibitemShut
  {NoStop}%
\bibitem [{\citenamefont {Chatterjee}\ \emph
  {et~al.}(2024{\natexlab{a}})\citenamefont {Chatterjee}, \citenamefont
  {Takada},\ and\ \citenamefont {Hayakawa}}]{chatterjee_pra_2024}%
  \BibitemOpen
  \bibfield  {author} {\bibinfo {author} {\bibfnamefont {A.~K.}\ \bibnamefont
  {Chatterjee}}, \bibinfo {author} {\bibfnamefont {S.}~\bibnamefont {Takada}},\
  and\ \bibinfo {author} {\bibfnamefont {H.}~\bibnamefont {Hayakawa}},\
  }\bibfield  {title} {\bibinfo {title} {Multiple quantum mpemba effect:
  Exceptional points and oscillations},\ }\href
  {https://doi.org/10.1103/PhysRevA.110.022213} {\bibfield  {journal} {\bibinfo
   {journal} {Phys. Rev. A}\ }\textbf {\bibinfo {volume} {110}},\ \bibinfo
  {pages} {022213} (\bibinfo {year} {2024}{\natexlab{a}})}\BibitemShut
  {NoStop}%
\bibitem [{\citenamefont {Moroder}\ \emph {et~al.}(2024)\citenamefont
  {Moroder}, \citenamefont {Culhane}, \citenamefont {Zawadzki},\ and\
  \citenamefont {Goold}}]{moroder_prl_2024}%
  \BibitemOpen
  \bibfield  {author} {\bibinfo {author} {\bibfnamefont {M.}~\bibnamefont
  {Moroder}}, \bibinfo {author} {\bibfnamefont {O.}~\bibnamefont {Culhane}},
  \bibinfo {author} {\bibfnamefont {K.}~\bibnamefont {Zawadzki}},\ and\
  \bibinfo {author} {\bibfnamefont {J.}~\bibnamefont {Goold}},\ }\bibfield
  {title} {\bibinfo {title} {Thermodynamics of the quantum mpemba effect},\
  }\href {https://doi.org/10.1103/PhysRevLett.133.140404} {\bibfield  {journal}
  {\bibinfo  {journal} {Phys. Rev. Lett.}\ }\textbf {\bibinfo {volume} {133}},\
  \bibinfo {pages} {140404} (\bibinfo {year} {2024})}\BibitemShut {NoStop}%
\bibitem [{\citenamefont {Nava}\ and\ \citenamefont
  {Egger}(2024{\natexlab{a}})}]{nava_prl_2024}%
  \BibitemOpen
  \bibfield  {author} {\bibinfo {author} {\bibfnamefont {A.}~\bibnamefont
  {Nava}}\ and\ \bibinfo {author} {\bibfnamefont {R.}~\bibnamefont {Egger}},\
  }\bibfield  {title} {\bibinfo {title} {Mpemba effects in open nonequilibrium
  quantum systems},\ }\href {https://doi.org/10.1103/PhysRevLett.133.136302}
  {\bibfield  {journal} {\bibinfo  {journal} {Phys. Rev. Lett.}\ }\textbf
  {\bibinfo {volume} {133}},\ \bibinfo {pages} {136302} (\bibinfo {year}
  {2024}{\natexlab{a}})}\BibitemShut {NoStop}%
\bibitem [{\citenamefont {Chatterjee}\ \emph
  {et~al.}(2024{\natexlab{b}})\citenamefont {Chatterjee}, \citenamefont
  {Takada},\ and\ \citenamefont {Hayakawa}}]{Chatterjee2024}%
  \BibitemOpen
  \bibfield  {author} {\bibinfo {author} {\bibfnamefont {A.~K.}\ \bibnamefont
  {Chatterjee}}, \bibinfo {author} {\bibfnamefont {S.}~\bibnamefont {Takada}},\
  and\ \bibinfo {author} {\bibfnamefont {H.}~\bibnamefont {Hayakawa}},\
  }\bibfield  {title} {\bibinfo {title} {Multiple quantum mpemba effect:
  Exceptional points and oscillations},\ }\href
  {https://doi.org/10.1103/PhysRevA.110.022213} {\bibfield  {journal} {\bibinfo
   {journal} {Phys. Rev. A}\ }\textbf {\bibinfo {volume} {110}},\ \bibinfo
  {pages} {022213} (\bibinfo {year} {2024}{\natexlab{b}})}\BibitemShut
  {NoStop}%
\bibitem [{\citenamefont {Qian}\ \emph {et~al.}(2025)\citenamefont {Qian},
  \citenamefont {Wang},\ and\ \citenamefont {Wang}}]{Qian2025}%
  \BibitemOpen
  \bibfield  {author} {\bibinfo {author} {\bibfnamefont {D.}~\bibnamefont
  {Qian}}, \bibinfo {author} {\bibfnamefont {H.}~\bibnamefont {Wang}},\ and\
  \bibinfo {author} {\bibfnamefont {J.}~\bibnamefont {Wang}},\ }\bibfield
  {title} {\bibinfo {title} {Intrinsic quantum mpemba effect in markovian
  systems and quantum circuits},\ }\href {https://doi.org/10.1103/qj8n-k5j2}
  {\bibfield  {journal} {\bibinfo  {journal} {Phys. Rev. B}\ }\textbf {\bibinfo
  {volume} {111}},\ \bibinfo {pages} {L220304} (\bibinfo {year}
  {2025})}\BibitemShut {NoStop}%
\bibitem [{\citenamefont {Caldas}\ and\ \citenamefont
  {Pires}(2025)}]{caldas2025}%
  \BibitemOpen
  \bibfield  {author} {\bibinfo {author} {\bibfnamefont {E.~L.}\ \bibnamefont
  {Caldas}}\ and\ \bibinfo {author} {\bibfnamefont {D.~P.}\ \bibnamefont
  {Pires}},\ }\href {https://arxiv.org/abs/2512.07561} {\bibinfo {title}
  {Exponentially accelerated relaxation and quantum mpemba effect in open
  quantum systems}} (\bibinfo {year} {2025}),\ \Eprint
  {https://arxiv.org/abs/2512.07561} {arXiv:2512.07561 [quant-ph]} \BibitemShut
  {NoStop}%
\bibitem [{\citenamefont {Saliba}\ and\ \citenamefont
  {Drumond}(2025)}]{saliba2025}%
  \BibitemOpen
  \bibfield  {author} {\bibinfo {author} {\bibfnamefont {R.~F.}\ \bibnamefont
  {Saliba}}\ and\ \bibinfo {author} {\bibfnamefont {R.~C.}\ \bibnamefont
  {Drumond}},\ }\href {https://arxiv.org/abs/2512.13509} {\bibinfo {title}
  {Unraveling the quantum mpemba effect on markovian open quantum systems}}
  (\bibinfo {year} {2025}),\ \Eprint {https://arxiv.org/abs/2512.13509}
  {arXiv:2512.13509 [quant-ph]} \BibitemShut {NoStop}%
\bibitem [{\citenamefont {Ulčakar}\ \emph {et~al.}(2025)\citenamefont
  {Ulčakar}, \citenamefont {Sharipov}, \citenamefont {Lagnese},\ and\
  \citenamefont {Lenarčič}}]{ulčakar2025}%
  \BibitemOpen
  \bibfield  {author} {\bibinfo {author} {\bibfnamefont {I.}~\bibnamefont
  {Ulčakar}}, \bibinfo {author} {\bibfnamefont {R.}~\bibnamefont {Sharipov}},
  \bibinfo {author} {\bibfnamefont {G.}~\bibnamefont {Lagnese}},\ and\ \bibinfo
  {author} {\bibfnamefont {Z.}~\bibnamefont {Lenarčič}},\ }\href
  {https://arxiv.org/abs/2511.16739} {\bibinfo {title} {Conserved quantities
  enable the quantum mpemba effect in weakly open systems}} (\bibinfo {year}
  {2025}),\ \Eprint {https://arxiv.org/abs/2511.16739} {arXiv:2511.16739
  [quant-ph]} \BibitemShut {NoStop}%
\bibitem [{\citenamefont {Strachan}\ \emph
  {et~al.}(2025{\natexlab{a}})\citenamefont {Strachan}, \citenamefont
  {Purkayastha},\ and\ \citenamefont {Clark}}]{Strachan_prl_2025}%
  \BibitemOpen
  \bibfield  {author} {\bibinfo {author} {\bibfnamefont {D.~J.}\ \bibnamefont
  {Strachan}}, \bibinfo {author} {\bibfnamefont {A.}~\bibnamefont
  {Purkayastha}},\ and\ \bibinfo {author} {\bibfnamefont {S.~R.}\ \bibnamefont
  {Clark}},\ }\bibfield  {title} {\bibinfo {title} {Non-markovian quantum
  mpemba effect},\ }\href {https://doi.org/10.1103/PhysRevLett.134.220403}
  {\bibfield  {journal} {\bibinfo  {journal} {Phys. Rev. Lett.}\ }\textbf
  {\bibinfo {volume} {134}},\ \bibinfo {pages} {220403} (\bibinfo {year}
  {2025}{\natexlab{a}})}\BibitemShut {NoStop}%
\bibitem [{\citenamefont
  {Longhi}(2025{\natexlab{a}})}]{Longhi_2025_systemenv_corr}%
  \BibitemOpen
  \bibfield  {author} {\bibinfo {author} {\bibfnamefont {S.}~\bibnamefont
  {Longhi}},\ }\bibfield  {title} {\bibinfo {title} {Quantum mpemba effect from
  initial system–reservoir entanglement},\ }\bibfield  {journal} {\bibinfo
  {journal} {APL Quantum}\ }\textbf {\bibinfo {volume} {2}},\ \href
  {https://doi.org/10.1063/5.0266143} {10.1063/5.0266143} (\bibinfo {year}
  {2025}{\natexlab{a}})\BibitemShut {NoStop}%
\bibitem [{\citenamefont {Kochsiek}\ \emph {et~al.}(2022)\citenamefont
  {Kochsiek}, \citenamefont {Carollo},\ and\ \citenamefont
  {Lesanovsky}}]{Kochsiek2022}%
  \BibitemOpen
  \bibfield  {author} {\bibinfo {author} {\bibfnamefont {S.}~\bibnamefont
  {Kochsiek}}, \bibinfo {author} {\bibfnamefont {F.}~\bibnamefont {Carollo}},\
  and\ \bibinfo {author} {\bibfnamefont {I.}~\bibnamefont {Lesanovsky}},\
  }\bibfield  {title} {\bibinfo {title} {Accelerating the approach of
  dissipative quantum spin systems towards stationarity through global spin
  rotations},\ }\href {https://doi.org/10.1103/PhysRevA.106.012207} {\bibfield
  {journal} {\bibinfo  {journal} {Phys. Rev. A}\ }\textbf {\bibinfo {volume}
  {106}},\ \bibinfo {pages} {012207} (\bibinfo {year} {2022})}\BibitemShut
  {NoStop}%
\bibitem [{\citenamefont {Dong}\ \emph {et~al.}(2025)\citenamefont {Dong},
  \citenamefont {Mu}, \citenamefont {Qin},\ and\ \citenamefont
  {Cui}}]{Dong2025}%
  \BibitemOpen
  \bibfield  {author} {\bibinfo {author} {\bibfnamefont {J.~W.}\ \bibnamefont
  {Dong}}, \bibinfo {author} {\bibfnamefont {H.~F.}\ \bibnamefont {Mu}},
  \bibinfo {author} {\bibfnamefont {M.}~\bibnamefont {Qin}},\ and\ \bibinfo
  {author} {\bibfnamefont {H.~T.}\ \bibnamefont {Cui}},\ }\bibfield  {title}
  {\bibinfo {title} {Quantum mpemba effect of localization in the dissipative
  mosaic model},\ }\href {https://doi.org/10.1103/PhysRevA.111.022215}
  {\bibfield  {journal} {\bibinfo  {journal} {Phys. Rev. A}\ }\textbf {\bibinfo
  {volume} {111}},\ \bibinfo {pages} {022215} (\bibinfo {year}
  {2025})}\BibitemShut {NoStop}%
\bibitem [{\citenamefont {Wei}\ \emph {et~al.}(2025)\citenamefont {Wei},
  \citenamefont {Xu}, \citenamefont {Jiang}, \citenamefont {Hu},\ and\
  \citenamefont {Pan}}]{wei2025}%
  \BibitemOpen
  \bibfield  {author} {\bibinfo {author} {\bibfnamefont {Z.}~\bibnamefont
  {Wei}}, \bibinfo {author} {\bibfnamefont {M.}~\bibnamefont {Xu}}, \bibinfo
  {author} {\bibfnamefont {X.-P.}\ \bibnamefont {Jiang}}, \bibinfo {author}
  {\bibfnamefont {H.}~\bibnamefont {Hu}},\ and\ \bibinfo {author}
  {\bibfnamefont {L.}~\bibnamefont {Pan}},\ }\href
  {https://arxiv.org/abs/2508.18906} {\bibinfo {title} {Quantum mpemba effect
  in dissipative spin chains at criticality}} (\bibinfo {year} {2025}),\
  \Eprint {https://arxiv.org/abs/2508.18906} {arXiv:2508.18906 [quant-ph]}
  \BibitemShut {NoStop}%
\bibitem [{\citenamefont {Das}\ \emph {et~al.}(2025)\citenamefont {Das},
  \citenamefont {Chaki}, \citenamefont {Ghosh},\ and\ \citenamefont
  {Sen}}]{das2025}%
  \BibitemOpen
  \bibfield  {author} {\bibinfo {author} {\bibfnamefont {A.}~\bibnamefont
  {Das}}, \bibinfo {author} {\bibfnamefont {P.}~\bibnamefont {Chaki}}, \bibinfo
  {author} {\bibfnamefont {P.}~\bibnamefont {Ghosh}},\ and\ \bibinfo {author}
  {\bibfnamefont {U.}~\bibnamefont {Sen}},\ }\href
  {https://arxiv.org/abs/2512.24839} {\bibinfo {title} {Role reversal in
  quantum mpemba effect}} (\bibinfo {year} {2025}),\ \Eprint
  {https://arxiv.org/abs/2512.24839} {arXiv:2512.24839 [quant-ph]} \BibitemShut
  {NoStop}%
\bibitem [{\citenamefont {Chatterjee}\ \emph {et~al.}(2023)\citenamefont
  {Chatterjee}, \citenamefont {Takada},\ and\ \citenamefont
  {Hayakawa}}]{Chatterjee2023}%
  \BibitemOpen
  \bibfield  {author} {\bibinfo {author} {\bibfnamefont {A.~K.}\ \bibnamefont
  {Chatterjee}}, \bibinfo {author} {\bibfnamefont {S.}~\bibnamefont {Takada}},\
  and\ \bibinfo {author} {\bibfnamefont {H.}~\bibnamefont {Hayakawa}},\
  }\bibfield  {title} {\bibinfo {title} {Quantum mpemba effect in a quantum dot
  with reservoirs},\ }\href {https://doi.org/10.1103/PhysRevLett.131.080402}
  {\bibfield  {journal} {\bibinfo  {journal} {Phys. Rev. Lett.}\ }\textbf
  {\bibinfo {volume} {131}},\ \bibinfo {pages} {080402} (\bibinfo {year}
  {2023})}\BibitemShut {NoStop}%
\bibitem [{\citenamefont {Wang}\ and\ \citenamefont {Wang}(2024)}]{Wang2024}%
  \BibitemOpen
  \bibfield  {author} {\bibinfo {author} {\bibfnamefont {X.}~\bibnamefont
  {Wang}}\ and\ \bibinfo {author} {\bibfnamefont {J.}~\bibnamefont {Wang}},\
  }\bibfield  {title} {\bibinfo {title} {Mpemba effects in nonequilibrium open
  quantum systems},\ }\href {https://doi.org/10.1103/PhysRevResearch.6.033330}
  {\bibfield  {journal} {\bibinfo  {journal} {Phys. Rev. Res.}\ }\textbf
  {\bibinfo {volume} {6}},\ \bibinfo {pages} {033330} (\bibinfo {year}
  {2024})}\BibitemShut {NoStop}%
\bibitem [{\citenamefont {Nava}\ and\ \citenamefont
  {Egger}(2024{\natexlab{b}})}]{Nava2024}%
  \BibitemOpen
  \bibfield  {author} {\bibinfo {author} {\bibfnamefont {A.}~\bibnamefont
  {Nava}}\ and\ \bibinfo {author} {\bibfnamefont {R.}~\bibnamefont {Egger}},\
  }\bibfield  {title} {\bibinfo {title} {Mpemba effects in open nonequilibrium
  quantum systems},\ }\href {https://doi.org/10.1103/PhysRevLett.133.136302}
  {\bibfield  {journal} {\bibinfo  {journal} {Phys. Rev. Lett.}\ }\textbf
  {\bibinfo {volume} {133}},\ \bibinfo {pages} {136302} (\bibinfo {year}
  {2024}{\natexlab{b}})}\BibitemShut {NoStop}%
\bibitem [{\citenamefont {Longhi}(2024)}]{Longhi2024}%
  \BibitemOpen
  \bibfield  {author} {\bibinfo {author} {\bibfnamefont {S.}~\bibnamefont
  {Longhi}},\ }\bibfield  {title} {\bibinfo {title} {Bosonic mpemba effect with
  non-classical states of light},\ }\bibfield  {journal} {\bibinfo  {journal}
  {APL Quantum}\ }\textbf {\bibinfo {volume} {1}},\ \href
  {https://doi.org/10.1063/5.0234457} {10.1063/5.0234457} (\bibinfo {year}
  {2024})\BibitemShut {NoStop}%
\bibitem [{\citenamefont {Longhi}(2025{\natexlab{b}})}]{Longhi2025}%
  \BibitemOpen
  \bibfield  {author} {\bibinfo {author} {\bibfnamefont {S.}~\bibnamefont
  {Longhi}},\ }\bibfield  {title} {\bibinfo {title} {Mpemba effect and
  super-accelerated thermalization in the damped quantum harmonic oscillator},\
  }\href {https://doi.org/10.22331/q-2025-03-26-1677} {\bibfield  {journal}
  {\bibinfo  {journal} {Quantum}\ }\textbf {\bibinfo {volume} {9}},\ \bibinfo
  {pages} {1677} (\bibinfo {year} {2025}{\natexlab{b}})}\BibitemShut {NoStop}%
\bibitem [{\citenamefont {Strachan}\ \emph
  {et~al.}(2025{\natexlab{b}})\citenamefont {Strachan}, \citenamefont
  {Purkayastha},\ and\ \citenamefont {Clark}}]{Strachan2025}%
  \BibitemOpen
  \bibfield  {author} {\bibinfo {author} {\bibfnamefont {D.~J.}\ \bibnamefont
  {Strachan}}, \bibinfo {author} {\bibfnamefont {A.}~\bibnamefont
  {Purkayastha}},\ and\ \bibinfo {author} {\bibfnamefont {S.~R.}\ \bibnamefont
  {Clark}},\ }\bibfield  {title} {\bibinfo {title} {Non-markovian quantum
  mpemba effect},\ }\href {https://doi.org/10.1103/PhysRevLett.134.220403}
  {\bibfield  {journal} {\bibinfo  {journal} {Phys. Rev. Lett.}\ }\textbf
  {\bibinfo {volume} {134}},\ \bibinfo {pages} {220403} (\bibinfo {year}
  {2025}{\natexlab{b}})}\BibitemShut {NoStop}%
\bibitem [{\citenamefont {Li}\ \emph {et~al.}(2025{\natexlab{a}})\citenamefont
  {Li}, \citenamefont {Li},\ and\ \citenamefont {Li}}]{Li2025}%
  \BibitemOpen
  \bibfield  {author} {\bibinfo {author} {\bibfnamefont {Y.}~\bibnamefont
  {Li}}, \bibinfo {author} {\bibfnamefont {W.}~\bibnamefont {Li}},\ and\
  \bibinfo {author} {\bibfnamefont {X.}~\bibnamefont {Li}},\ }\bibfield
  {title} {\bibinfo {title} {Ergotropic mpemba effect in non-markovian quantum
  systems},\ }\href {https://doi.org/10.1103/5xrr-x2rm} {\bibfield  {journal}
  {\bibinfo  {journal} {Phys. Rev. A}\ }\textbf {\bibinfo {volume} {112}},\
  \bibinfo {pages} {032209} (\bibinfo {year} {2025}{\natexlab{a}})}\BibitemShut
  {NoStop}%
\bibitem [{\citenamefont {Zhang}\ \emph {et~al.}(2026)\citenamefont {Zhang},
  \citenamefont {Luo},\ and\ \citenamefont {Wu}}]{zhang2026}%
  \BibitemOpen
  \bibfield  {author} {\bibinfo {author} {\bibfnamefont {Z.-Z.}\ \bibnamefont
  {Zhang}}, \bibinfo {author} {\bibfnamefont {H.-G.}\ \bibnamefont {Luo}},\
  and\ \bibinfo {author} {\bibfnamefont {W.}~\bibnamefont {Wu}},\ }\href
  {https://arxiv.org/abs/2511.13173} {\bibinfo {title} {Quantum mpemba effect
  induced by non-markovian exceptional point}} (\bibinfo {year} {2026}),\
  \Eprint {https://arxiv.org/abs/2511.13173} {arXiv:2511.13173 [quant-ph]}
  \BibitemShut {NoStop}%
\bibitem [{\citenamefont {Bao}(2026)}]{bao2026}%
  \BibitemOpen
  \bibfield  {author} {\bibinfo {author} {\bibfnamefont {R.}~\bibnamefont
  {Bao}},\ }\bibfield  {title} {\bibinfo {title} {Initial-state typicality in
  quantum relaxation},\ }\href {https://doi.org/10.1103/wgr5-lb6b} {\bibfield
  {journal} {\bibinfo  {journal} {Phys. Rev. Lett.}\ ,\ } (\bibinfo {year}
  {2026})}\BibitemShut {NoStop}%
\bibitem [{\citenamefont {Li}\ \emph {et~al.}(2025{\natexlab{b}})\citenamefont
  {Li}, \citenamefont {Li},\ and\ \citenamefont {Yan}}]{li_arxiv_2025}%
  \BibitemOpen
  \bibfield  {author} {\bibinfo {author} {\bibfnamefont {X.}~\bibnamefont
  {Li}}, \bibinfo {author} {\bibfnamefont {Y.}~\bibnamefont {Li}},\ and\
  \bibinfo {author} {\bibfnamefont {Y.}~\bibnamefont {Yan}},\ }\href
  {https://arxiv.org/abs/2511.16996} {\bibinfo {title} {Canonical quantum
  mpemba effect in a dissipative qubit}} (\bibinfo {year}
  {2025}{\natexlab{b}}),\ \Eprint {https://arxiv.org/abs/2511.16996}
  {arXiv:2511.16996 [quant-ph]} \BibitemShut {NoStop}%
\bibitem [{\citenamefont {Bagui}\ \emph {et~al.}(2025)\citenamefont {Bagui},
  \citenamefont {Chatterjee},\ and\ \citenamefont {Agarwalla}}]{bagui2025}%
  \BibitemOpen
  \bibfield  {author} {\bibinfo {author} {\bibfnamefont {P.}~\bibnamefont
  {Bagui}}, \bibinfo {author} {\bibfnamefont {A.}~\bibnamefont {Chatterjee}},\
  and\ \bibinfo {author} {\bibfnamefont {B.~K.}\ \bibnamefont {Agarwalla}},\
  }\href {https://arxiv.org/abs/2512.02709} {\bibinfo {title} {Detection of
  mpemba effect through good observables in open quantum systems}} (\bibinfo
  {year} {2025}),\ \Eprint {https://arxiv.org/abs/2512.02709} {arXiv:2512.02709
  [cond-mat.stat-mech]} \BibitemShut {NoStop}%
\bibitem [{\citenamefont {Solanki}\ \emph {et~al.}(2025)\citenamefont
  {Solanki}, \citenamefont {Lesanovsky},\ and\ \citenamefont
  {Perfetto}}]{solanki2025}%
  \BibitemOpen
  \bibfield  {author} {\bibinfo {author} {\bibfnamefont {P.}~\bibnamefont
  {Solanki}}, \bibinfo {author} {\bibfnamefont {I.}~\bibnamefont
  {Lesanovsky}},\ and\ \bibinfo {author} {\bibfnamefont {G.}~\bibnamefont
  {Perfetto}},\ }\href {https://arxiv.org/abs/2512.10005} {\bibinfo {title}
  {Universal relaxation speedup in open quantum systems through transient
  conditional and unconditional resetting}} (\bibinfo {year} {2025}),\ \Eprint
  {https://arxiv.org/abs/2512.10005} {arXiv:2512.10005 [cond-mat.stat-mech]}
  \BibitemShut {NoStop}%
\bibitem [{\citenamefont {Lejeune}\ \emph {et~al.}(2026)\citenamefont
  {Lejeune}, \citenamefont {Papič}, \citenamefont {Goold}, \citenamefont
  {Binder}, \citenamefont {Damanet},\ and\ \citenamefont
  {Moroder}}]{lejeune2026}%
  \BibitemOpen
  \bibfield  {author} {\bibinfo {author} {\bibfnamefont {T.}~\bibnamefont
  {Lejeune}}, \bibinfo {author} {\bibfnamefont {M.}~\bibnamefont {Papič}},
  \bibinfo {author} {\bibfnamefont {J.}~\bibnamefont {Goold}}, \bibinfo
  {author} {\bibfnamefont {F.~C.}\ \bibnamefont {Binder}}, \bibinfo {author}
  {\bibfnamefont {F.}~\bibnamefont {Damanet}},\ and\ \bibinfo {author}
  {\bibfnamefont {M.}~\bibnamefont {Moroder}},\ }\href
  {https://arxiv.org/abs/2602.03765} {\bibinfo {title} {Accelerating qubit
  reset through the mpemba effect}} (\bibinfo {year} {2026}),\ \Eprint
  {https://arxiv.org/abs/2602.03765} {arXiv:2602.03765 [quant-ph]} \BibitemShut
  {NoStop}%
\bibitem [{\citenamefont {Joshi}\ \emph {et~al.}(2024)\citenamefont {Joshi},
  \citenamefont {Franke}, \citenamefont {Rath}, \citenamefont {Ares},
  \citenamefont {Murciano}, \citenamefont {Kranzl}, \citenamefont {Blatt},
  \citenamefont {Zoller}, \citenamefont {Vermersch}, \citenamefont {Calabrese},
  \citenamefont {Roos},\ and\ \citenamefont {Joshi}}]{joshi_prl_2024}%
  \BibitemOpen
  \bibfield  {author} {\bibinfo {author} {\bibfnamefont {L.~K.}\ \bibnamefont
  {Joshi}}, \bibinfo {author} {\bibfnamefont {J.}~\bibnamefont {Franke}},
  \bibinfo {author} {\bibfnamefont {A.}~\bibnamefont {Rath}}, \bibinfo {author}
  {\bibfnamefont {F.}~\bibnamefont {Ares}}, \bibinfo {author} {\bibfnamefont
  {S.}~\bibnamefont {Murciano}}, \bibinfo {author} {\bibfnamefont
  {F.}~\bibnamefont {Kranzl}}, \bibinfo {author} {\bibfnamefont
  {R.}~\bibnamefont {Blatt}}, \bibinfo {author} {\bibfnamefont
  {P.}~\bibnamefont {Zoller}}, \bibinfo {author} {\bibfnamefont
  {B.}~\bibnamefont {Vermersch}}, \bibinfo {author} {\bibfnamefont
  {P.}~\bibnamefont {Calabrese}}, \bibinfo {author} {\bibfnamefont {C.~F.}\
  \bibnamefont {Roos}},\ and\ \bibinfo {author} {\bibfnamefont {M.~K.}\
  \bibnamefont {Joshi}},\ }\bibfield  {title} {\bibinfo {title} {Observing the
  quantum mpemba effect in quantum simulations},\ }\href
  {https://doi.org/10.1103/PhysRevLett.133.010402} {\bibfield  {journal}
  {\bibinfo  {journal} {Phys. Rev. Lett.}\ }\textbf {\bibinfo {volume} {133}},\
  \bibinfo {pages} {010402} (\bibinfo {year} {2024})}\BibitemShut {NoStop}%
\bibitem [{\citenamefont {Chatterjee}\ \emph {et~al.}(2025)\citenamefont
  {Chatterjee}, \citenamefont {Khan}, \citenamefont {Jain},\ and\ \citenamefont
  {Mahesh}}]{chatterjee2025}%
  \BibitemOpen
  \bibfield  {author} {\bibinfo {author} {\bibfnamefont {A.}~\bibnamefont
  {Chatterjee}}, \bibinfo {author} {\bibfnamefont {S.}~\bibnamefont {Khan}},
  \bibinfo {author} {\bibfnamefont {S.}~\bibnamefont {Jain}},\ and\ \bibinfo
  {author} {\bibfnamefont {T.~S.}\ \bibnamefont {Mahesh}},\ }\href
  {https://arxiv.org/abs/2509.13451} {\bibinfo {title} {Direct experimental
  observation of quantum mpemba effect without bath engineering}} (\bibinfo
  {year} {2025}),\ \Eprint {https://arxiv.org/abs/2509.13451} {arXiv:2509.13451
  [quant-ph]} \BibitemShut {NoStop}%
\bibitem [{\citenamefont {Schnepper}\ \emph {et~al.}(2025)\citenamefont
  {Schnepper}, \citenamefont {de~Oliveira}, \citenamefont {Vieira},
  \citenamefont {Zawadzki},\ and\ \citenamefont {Serra}}]{schnepper2025}%
  \BibitemOpen
  \bibfield  {author} {\bibinfo {author} {\bibfnamefont {B.~P.}\ \bibnamefont
  {Schnepper}}, \bibinfo {author} {\bibfnamefont {J.~L.~D.}\ \bibnamefont
  {de~Oliveira}}, \bibinfo {author} {\bibfnamefont {C.~H.~S.}\ \bibnamefont
  {Vieira}}, \bibinfo {author} {\bibfnamefont {K.}~\bibnamefont {Zawadzki}},\
  and\ \bibinfo {author} {\bibfnamefont {R.~M.}\ \bibnamefont {Serra}},\ }\href
  {https://arxiv.org/abs/2511.14552} {\bibinfo {title} {Experimental
  observation and application of the genuine quantum mpemba effect}} (\bibinfo
  {year} {2025}),\ \Eprint {https://arxiv.org/abs/2511.14552} {arXiv:2511.14552
  [quant-ph]} \BibitemShut {NoStop}%
\bibitem [{\citenamefont {Alicki}\ and\ \citenamefont
  {Fannes}(2013)}]{Alicki2013}%
  \BibitemOpen
  \bibfield  {author} {\bibinfo {author} {\bibfnamefont {R.}~\bibnamefont
  {Alicki}}\ and\ \bibinfo {author} {\bibfnamefont {M.}~\bibnamefont
  {Fannes}},\ }\bibfield  {title} {\bibinfo {title} {Entanglement boost for
  extractable work from ensembles of quantum batteries},\ }\href
  {https://doi.org/10.1103/PhysRevE.87.042123} {\bibfield  {journal} {\bibinfo
  {journal} {Phys. Rev. E}\ }\textbf {\bibinfo {volume} {87}},\ \bibinfo
  {pages} {042123} (\bibinfo {year} {2013})}\BibitemShut {NoStop}%
\bibitem [{\citenamefont {Campaioli}\ \emph {et~al.}(2024)\citenamefont
  {Campaioli}, \citenamefont {Gherardini}, \citenamefont {Quach}, \citenamefont
  {Polini},\ and\ \citenamefont {Andolina}}]{quantum_battery_review}%
  \BibitemOpen
  \bibfield  {author} {\bibinfo {author} {\bibfnamefont {F.}~\bibnamefont
  {Campaioli}}, \bibinfo {author} {\bibfnamefont {S.}~\bibnamefont
  {Gherardini}}, \bibinfo {author} {\bibfnamefont {J.~Q.}\ \bibnamefont
  {Quach}}, \bibinfo {author} {\bibfnamefont {M.}~\bibnamefont {Polini}},\ and\
  \bibinfo {author} {\bibfnamefont {G.~M.}\ \bibnamefont {Andolina}},\
  }\bibfield  {title} {\bibinfo {title} {Colloquium: Quantum batteries},\
  }\href {https://doi.org/10.1103/RevModPhys.96.031001} {\bibfield  {journal}
  {\bibinfo  {journal} {Rev. Mod. Phys.}\ }\textbf {\bibinfo {volume} {96}},\
  \bibinfo {pages} {031001} (\bibinfo {year} {2024})}\BibitemShut {NoStop}%
\bibitem [{\citenamefont {Ferraro}\ \emph {et~al.}(2026)\citenamefont
  {Ferraro}, \citenamefont {Cavaliere}, \citenamefont {Genoni}, \citenamefont
  {Benenti},\ and\ \citenamefont {Sassetti}}]{Ferraro2026}%
  \BibitemOpen
  \bibfield  {author} {\bibinfo {author} {\bibfnamefont {D.}~\bibnamefont
  {Ferraro}}, \bibinfo {author} {\bibfnamefont {F.}~\bibnamefont {Cavaliere}},
  \bibinfo {author} {\bibfnamefont {M.~G.}\ \bibnamefont {Genoni}}, \bibinfo
  {author} {\bibfnamefont {G.}~\bibnamefont {Benenti}},\ and\ \bibinfo {author}
  {\bibfnamefont {M.}~\bibnamefont {Sassetti}},\ }\bibfield  {title} {\bibinfo
  {title} {Opportunities and challenges of quantum batteries},\ }\href
  {https://doi.org/10.1038/s42254-025-00906-5} {\bibfield  {journal} {\bibinfo
  {journal} {Nature Reviews Physics}\ }\textbf {\bibinfo {volume} {8}},\
  \bibinfo {pages} {115–127} (\bibinfo {year} {2026})}\BibitemShut {NoStop}%
\bibitem [{\citenamefont {Binder}\ \emph {et~al.}(2015)\citenamefont {Binder},
  \citenamefont {Vinjanampathy}, \citenamefont {Modi},\ and\ \citenamefont
  {Goold}}]{Binder2015}%
  \BibitemOpen
  \bibfield  {author} {\bibinfo {author} {\bibfnamefont {F.~C.}\ \bibnamefont
  {Binder}}, \bibinfo {author} {\bibfnamefont {S.}~\bibnamefont
  {Vinjanampathy}}, \bibinfo {author} {\bibfnamefont {K.}~\bibnamefont
  {Modi}},\ and\ \bibinfo {author} {\bibfnamefont {J.}~\bibnamefont {Goold}},\
  }\bibfield  {title} {\bibinfo {title} {Quantacell: powerful charging of
  quantum batteries},\ }\href {https://doi.org/10.1088/1367-2630/17/7/075015}
  {\bibfield  {journal} {\bibinfo  {journal} {New Journal of Physics}\ }\textbf
  {\bibinfo {volume} {17}},\ \bibinfo {pages} {075015} (\bibinfo {year}
  {2015})}\BibitemShut {NoStop}%
\bibitem [{\citenamefont {Campaioli}\ \emph {et~al.}(2017)\citenamefont
  {Campaioli}, \citenamefont {Pollock}, \citenamefont {Binder}, \citenamefont
  {C\'eleri}, \citenamefont {Goold}, \citenamefont {Vinjanampathy},\ and\
  \citenamefont {Modi}}]{Campaioli2017}%
  \BibitemOpen
  \bibfield  {author} {\bibinfo {author} {\bibfnamefont {F.}~\bibnamefont
  {Campaioli}}, \bibinfo {author} {\bibfnamefont {F.~A.}\ \bibnamefont
  {Pollock}}, \bibinfo {author} {\bibfnamefont {F.~C.}\ \bibnamefont {Binder}},
  \bibinfo {author} {\bibfnamefont {L.}~\bibnamefont {C\'eleri}}, \bibinfo
  {author} {\bibfnamefont {J.}~\bibnamefont {Goold}}, \bibinfo {author}
  {\bibfnamefont {S.}~\bibnamefont {Vinjanampathy}},\ and\ \bibinfo {author}
  {\bibfnamefont {K.}~\bibnamefont {Modi}},\ }\bibfield  {title} {\bibinfo
  {title} {Enhancing the charging power of quantum batteries},\ }\href
  {https://doi.org/10.1103/PhysRevLett.118.150601} {\bibfield  {journal}
  {\bibinfo  {journal} {Phys. Rev. Lett.}\ }\textbf {\bibinfo {volume} {118}},\
  \bibinfo {pages} {150601} (\bibinfo {year} {2017})}\BibitemShut {NoStop}%
\bibitem [{\citenamefont {Ferraro}\ \emph {et~al.}(2018)\citenamefont
  {Ferraro}, \citenamefont {Campisi}, \citenamefont {Andolina}, \citenamefont
  {Pellegrini},\ and\ \citenamefont {Polini}}]{Ferraro2018}%
  \BibitemOpen
  \bibfield  {author} {\bibinfo {author} {\bibfnamefont {D.}~\bibnamefont
  {Ferraro}}, \bibinfo {author} {\bibfnamefont {M.}~\bibnamefont {Campisi}},
  \bibinfo {author} {\bibfnamefont {G.~M.}\ \bibnamefont {Andolina}}, \bibinfo
  {author} {\bibfnamefont {V.}~\bibnamefont {Pellegrini}},\ and\ \bibinfo
  {author} {\bibfnamefont {M.}~\bibnamefont {Polini}},\ }\bibfield  {title}
  {\bibinfo {title} {High-power collective charging of a solid-state quantum
  battery},\ }\href {https://doi.org/10.1103/PhysRevLett.120.117702} {\bibfield
   {journal} {\bibinfo  {journal} {Phys. Rev. Lett.}\ }\textbf {\bibinfo
  {volume} {120}},\ \bibinfo {pages} {117702} (\bibinfo {year}
  {2018})}\BibitemShut {NoStop}%
\bibitem [{\citenamefont {Grazi}\ \emph {et~al.}(2025)\citenamefont {Grazi},
  \citenamefont {Johannesson}, \citenamefont {Ferraro},\ and\ \citenamefont
  {Ziani}}]{grazi2025}%
  \BibitemOpen
  \bibfield  {author} {\bibinfo {author} {\bibfnamefont {R.}~\bibnamefont
  {Grazi}}, \bibinfo {author} {\bibfnamefont {H.}~\bibnamefont {Johannesson}},
  \bibinfo {author} {\bibfnamefont {D.}~\bibnamefont {Ferraro}},\ and\ \bibinfo
  {author} {\bibfnamefont {N.~T.}\ \bibnamefont {Ziani}},\ }\href
  {https://arxiv.org/abs/2512.14521} {\bibinfo {title} {Finite-time protocols
  stabilize charging in noisy ising quantum batteries}} (\bibinfo {year}
  {2025}),\ \Eprint {https://arxiv.org/abs/2512.14521} {arXiv:2512.14521
  [quant-ph]} \BibitemShut {NoStop}%
\bibitem [{\citenamefont {Andolina}\ \emph {et~al.}(2019)\citenamefont
  {Andolina}, \citenamefont {Keck}, \citenamefont {Mari}, \citenamefont
  {Campisi}, \citenamefont {Giovannetti},\ and\ \citenamefont
  {Polini}}]{Andolina_prl_2019}%
  \BibitemOpen
  \bibfield  {author} {\bibinfo {author} {\bibfnamefont {G.~M.}\ \bibnamefont
  {Andolina}}, \bibinfo {author} {\bibfnamefont {M.}~\bibnamefont {Keck}},
  \bibinfo {author} {\bibfnamefont {A.}~\bibnamefont {Mari}}, \bibinfo {author}
  {\bibfnamefont {M.}~\bibnamefont {Campisi}}, \bibinfo {author} {\bibfnamefont
  {V.}~\bibnamefont {Giovannetti}},\ and\ \bibinfo {author} {\bibfnamefont
  {M.}~\bibnamefont {Polini}},\ }\bibfield  {title} {\bibinfo {title}
  {Extractable work, the role of correlations, and asymptotic freedom in
  quantum batteries},\ }\href {https://doi.org/10.1103/PhysRevLett.122.047702}
  {\bibfield  {journal} {\bibinfo  {journal} {Phys. Rev. Lett.}\ }\textbf
  {\bibinfo {volume} {122}},\ \bibinfo {pages} {047702} (\bibinfo {year}
  {2019})}\BibitemShut {NoStop}%
\bibitem [{\citenamefont {Francica}\ \emph {et~al.}(2020)\citenamefont
  {Francica}, \citenamefont {Binder}, \citenamefont {Guarnieri}, \citenamefont
  {Mitchison}, \citenamefont {Goold},\ and\ \citenamefont
  {Plastina}}]{Francica2020}%
  \BibitemOpen
  \bibfield  {author} {\bibinfo {author} {\bibfnamefont {G.}~\bibnamefont
  {Francica}}, \bibinfo {author} {\bibfnamefont {F.~C.}\ \bibnamefont
  {Binder}}, \bibinfo {author} {\bibfnamefont {G.}~\bibnamefont {Guarnieri}},
  \bibinfo {author} {\bibfnamefont {M.~T.}\ \bibnamefont {Mitchison}}, \bibinfo
  {author} {\bibfnamefont {J.}~\bibnamefont {Goold}},\ and\ \bibinfo {author}
  {\bibfnamefont {F.}~\bibnamefont {Plastina}},\ }\bibfield  {title} {\bibinfo
  {title} {Quantum coherence and ergotropy},\ }\href
  {https://doi.org/10.1103/PhysRevLett.125.180603} {\bibfield  {journal}
  {\bibinfo  {journal} {Phys. Rev. Lett.}\ }\textbf {\bibinfo {volume} {125}},\
  \bibinfo {pages} {180603} (\bibinfo {year} {2020})}\BibitemShut {NoStop}%
\bibitem [{\citenamefont {Shi}\ \emph {et~al.}(2022)\citenamefont {Shi},
  \citenamefont {Ding}, \citenamefont {Wan}, \citenamefont {Wang},\ and\
  \citenamefont {Yang}}]{Shi2022}%
  \BibitemOpen
  \bibfield  {author} {\bibinfo {author} {\bibfnamefont {H.-L.}\ \bibnamefont
  {Shi}}, \bibinfo {author} {\bibfnamefont {S.}~\bibnamefont {Ding}}, \bibinfo
  {author} {\bibfnamefont {Q.-K.}\ \bibnamefont {Wan}}, \bibinfo {author}
  {\bibfnamefont {X.-H.}\ \bibnamefont {Wang}},\ and\ \bibinfo {author}
  {\bibfnamefont {W.-L.}\ \bibnamefont {Yang}},\ }\bibfield  {title} {\bibinfo
  {title} {Entanglement, coherence, and extractable work in quantum
  batteries},\ }\href {https://doi.org/10.1103/PhysRevLett.129.130602}
  {\bibfield  {journal} {\bibinfo  {journal} {Phys. Rev. Lett.}\ }\textbf
  {\bibinfo {volume} {129}},\ \bibinfo {pages} {130602} (\bibinfo {year}
  {2022})}\BibitemShut {NoStop}%
\bibitem [{\citenamefont {Vigneshwar}\ and\ \citenamefont
  {Sankaranarayanan}(2025)}]{vigneshwar2025}%
  \BibitemOpen
  \bibfield  {author} {\bibinfo {author} {\bibfnamefont {B.}~\bibnamefont
  {Vigneshwar}}\ and\ \bibinfo {author} {\bibfnamefont {R.}~\bibnamefont
  {Sankaranarayanan}},\ }\href {https://arxiv.org/abs/2512.14497} {\bibinfo
  {title} {Nonlocal contributions to ergotropy: A thermodynamic perspective}}
  (\bibinfo {year} {2025}),\ \Eprint {https://arxiv.org/abs/2512.14497}
  {arXiv:2512.14497 [quant-ph]} \BibitemShut {NoStop}%
\bibitem [{\citenamefont {Juli\`a-Farr\'e}\ \emph {et~al.}(2020)\citenamefont
  {Juli\`a-Farr\'e}, \citenamefont {Salamon}, \citenamefont {Riera},
  \citenamefont {Bera},\ and\ \citenamefont {Lewenstein}}]{Farre2020}%
  \BibitemOpen
  \bibfield  {author} {\bibinfo {author} {\bibfnamefont {S.}~\bibnamefont
  {Juli\`a-Farr\'e}}, \bibinfo {author} {\bibfnamefont {T.}~\bibnamefont
  {Salamon}}, \bibinfo {author} {\bibfnamefont {A.}~\bibnamefont {Riera}},
  \bibinfo {author} {\bibfnamefont {M.~N.}\ \bibnamefont {Bera}},\ and\
  \bibinfo {author} {\bibfnamefont {M.}~\bibnamefont {Lewenstein}},\ }\bibfield
   {title} {\bibinfo {title} {Bounds on the capacity and power of quantum
  batteries},\ }\href {https://doi.org/10.1103/PhysRevResearch.2.023113}
  {\bibfield  {journal} {\bibinfo  {journal} {Phys. Rev. Res.}\ }\textbf
  {\bibinfo {volume} {2}},\ \bibinfo {pages} {023113} (\bibinfo {year}
  {2020})}\BibitemShut {NoStop}%
\bibitem [{\citenamefont {Yang}\ \emph {et~al.}(2023)\citenamefont {Yang},
  \citenamefont {Yang}, \citenamefont {Alimuddin}, \citenamefont {Salvia},
  \citenamefont {Fei}, \citenamefont {Zhao}, \citenamefont {Nimmrichter},\ and\
  \citenamefont {Luo}}]{Yang2023}%
  \BibitemOpen
  \bibfield  {author} {\bibinfo {author} {\bibfnamefont {X.}~\bibnamefont
  {Yang}}, \bibinfo {author} {\bibfnamefont {Y.-H.}\ \bibnamefont {Yang}},
  \bibinfo {author} {\bibfnamefont {M.}~\bibnamefont {Alimuddin}}, \bibinfo
  {author} {\bibfnamefont {R.}~\bibnamefont {Salvia}}, \bibinfo {author}
  {\bibfnamefont {S.-M.}\ \bibnamefont {Fei}}, \bibinfo {author} {\bibfnamefont
  {L.-M.}\ \bibnamefont {Zhao}}, \bibinfo {author} {\bibfnamefont
  {S.}~\bibnamefont {Nimmrichter}},\ and\ \bibinfo {author} {\bibfnamefont
  {M.-X.}\ \bibnamefont {Luo}},\ }\bibfield  {title} {\bibinfo {title} {Battery
  capacity of energy-storing quantum systems},\ }\href
  {https://doi.org/10.1103/PhysRevLett.131.030402} {\bibfield  {journal}
  {\bibinfo  {journal} {Phys. Rev. Lett.}\ }\textbf {\bibinfo {volume} {131}},\
  \bibinfo {pages} {030402} (\bibinfo {year} {2023})}\BibitemShut {NoStop}%
\bibitem [{\citenamefont {Mohan}\ and\ \citenamefont {Pati}(2021)}]{Mohan2021}%
  \BibitemOpen
  \bibfield  {author} {\bibinfo {author} {\bibfnamefont {B.}~\bibnamefont
  {Mohan}}\ and\ \bibinfo {author} {\bibfnamefont {A.~K.}\ \bibnamefont
  {Pati}},\ }\bibfield  {title} {\bibinfo {title} {Reverse quantum speed limit:
  How slowly a quantum battery can discharge},\ }\href
  {https://doi.org/10.1103/PhysRevA.104.042209} {\bibfield  {journal} {\bibinfo
   {journal} {Phys. Rev. A}\ }\textbf {\bibinfo {volume} {104}},\ \bibinfo
  {pages} {042209} (\bibinfo {year} {2021})}\BibitemShut {NoStop}%
\bibitem [{\citenamefont {Mohan}\ and\ \citenamefont
  {Pati}(2022)}]{Mohan_2022}%
  \BibitemOpen
  \bibfield  {author} {\bibinfo {author} {\bibfnamefont {B.}~\bibnamefont
  {Mohan}}\ and\ \bibinfo {author} {\bibfnamefont {A.~K.}\ \bibnamefont
  {Pati}},\ }\bibfield  {title} {\bibinfo {title} {Quantum speed limits for
  observables},\ }\bibfield  {journal} {\bibinfo  {journal} {Physical Review
  A}\ }\textbf {\bibinfo {volume} {106}},\ \href
  {https://doi.org/10.1103/physreva.106.042436} {10.1103/physreva.106.042436}
  (\bibinfo {year} {2022})\BibitemShut {NoStop}%
\bibitem [{\citenamefont {Shrimali}\ \emph {et~al.}(2024)\citenamefont
  {Shrimali}, \citenamefont {Panda},\ and\ \citenamefont
  {Pati}}]{Shrimali2024}%
  \BibitemOpen
  \bibfield  {author} {\bibinfo {author} {\bibfnamefont {D.}~\bibnamefont
  {Shrimali}}, \bibinfo {author} {\bibfnamefont {B.}~\bibnamefont {Panda}},\
  and\ \bibinfo {author} {\bibfnamefont {A.~K.}\ \bibnamefont {Pati}},\
  }\bibfield  {title} {\bibinfo {title} {Stronger speed limit for observables:
  Tighter bound for the capacity of entanglement, the modular hamiltonian, and
  the charging of a quantum battery},\ }\href
  {https://doi.org/10.1103/PhysRevA.110.022425} {\bibfield  {journal} {\bibinfo
   {journal} {Phys. Rev. A}\ }\textbf {\bibinfo {volume} {110}},\ \bibinfo
  {pages} {022425} (\bibinfo {year} {2024})}\BibitemShut {NoStop}%
\bibitem [{\citenamefont {Ghosh}\ \emph {et~al.}(2021)\citenamefont {Ghosh},
  \citenamefont {Chanda}, \citenamefont {Mal},\ and\ \citenamefont
  {Sen(De)}}]{Ghosh2021}%
  \BibitemOpen
  \bibfield  {author} {\bibinfo {author} {\bibfnamefont {S.}~\bibnamefont
  {Ghosh}}, \bibinfo {author} {\bibfnamefont {T.}~\bibnamefont {Chanda}},
  \bibinfo {author} {\bibfnamefont {S.}~\bibnamefont {Mal}},\ and\ \bibinfo
  {author} {\bibfnamefont {A.}~\bibnamefont {Sen(De)}},\ }\bibfield  {title}
  {\bibinfo {title} {Fast charging of a quantum battery assisted by noise},\
  }\href {https://doi.org/10.1103/PhysRevA.104.032207} {\bibfield  {journal}
  {\bibinfo  {journal} {Phys. Rev. A}\ }\textbf {\bibinfo {volume} {104}},\
  \bibinfo {pages} {032207} (\bibinfo {year} {2021})}\BibitemShut {NoStop}%
\bibitem [{\citenamefont {Zakavati}\ \emph {et~al.}(2021)\citenamefont
  {Zakavati}, \citenamefont {Tabesh},\ and\ \citenamefont
  {Salimi}}]{Zakavati2021}%
  \BibitemOpen
  \bibfield  {author} {\bibinfo {author} {\bibfnamefont {S.}~\bibnamefont
  {Zakavati}}, \bibinfo {author} {\bibfnamefont {F.~T.}\ \bibnamefont
  {Tabesh}},\ and\ \bibinfo {author} {\bibfnamefont {S.}~\bibnamefont
  {Salimi}},\ }\bibfield  {title} {\bibinfo {title} {Bounds on charging power
  of open quantum batteries},\ }\href
  {https://doi.org/10.1103/PhysRevE.104.054117} {\bibfield  {journal} {\bibinfo
   {journal} {Phys. Rev. E}\ }\textbf {\bibinfo {volume} {104}},\ \bibinfo
  {pages} {054117} (\bibinfo {year} {2021})}\BibitemShut {NoStop}%
\bibitem [{\citenamefont {Arjmandi}\ \emph {et~al.}(2022)\citenamefont
  {Arjmandi}, \citenamefont {Mohammadi},\ and\ \citenamefont
  {Santos}}]{Arjmandi2022}%
  \BibitemOpen
  \bibfield  {author} {\bibinfo {author} {\bibfnamefont {M.~B.}\ \bibnamefont
  {Arjmandi}}, \bibinfo {author} {\bibfnamefont {H.}~\bibnamefont
  {Mohammadi}},\ and\ \bibinfo {author} {\bibfnamefont {A.~C.}\ \bibnamefont
  {Santos}},\ }\bibfield  {title} {\bibinfo {title} {Enhancing self-discharging
  process with disordered quantum batteries},\ }\href
  {https://doi.org/10.1103/PhysRevE.105.054115} {\bibfield  {journal} {\bibinfo
   {journal} {Phys. Rev. E}\ }\textbf {\bibinfo {volume} {105}},\ \bibinfo
  {pages} {054115} (\bibinfo {year} {2022})}\BibitemShut {NoStop}%
\bibitem [{\citenamefont {Sen}\ and\ \citenamefont {Sen}(2023)}]{sen2023}%
  \BibitemOpen
  \bibfield  {author} {\bibinfo {author} {\bibfnamefont {K.}~\bibnamefont
  {Sen}}\ and\ \bibinfo {author} {\bibfnamefont {U.}~\bibnamefont {Sen}},\
  }\href {https://arxiv.org/abs/2302.07166} {\bibinfo {title} {Noisy quantum
  batteries}} (\bibinfo {year} {2023}),\ \Eprint
  {https://arxiv.org/abs/2302.07166} {arXiv:2302.07166 [quant-ph]} \BibitemShut
  {NoStop}%
\bibitem [{\citenamefont {Liu}\ \emph {et~al.}(2024{\natexlab{b}})\citenamefont
  {Liu}, \citenamefont {Wang}, \citenamefont {Fan}, \citenamefont {Wu},\ and\
  \citenamefont {Liu}}]{Liu2024}%
  \BibitemOpen
  \bibfield  {author} {\bibinfo {author} {\bibfnamefont {S.-Q.}\ \bibnamefont
  {Liu}}, \bibinfo {author} {\bibfnamefont {L.}~\bibnamefont {Wang}}, \bibinfo
  {author} {\bibfnamefont {H.}~\bibnamefont {Fan}}, \bibinfo {author}
  {\bibfnamefont {F.-L.}\ \bibnamefont {Wu}},\ and\ \bibinfo {author}
  {\bibfnamefont {S.-Y.}\ \bibnamefont {Liu}},\ }\bibfield  {title} {\bibinfo
  {title} {Better performance of quantum batteries in different environments
  compared to closed batteries},\ }\href
  {https://doi.org/10.1103/PhysRevA.109.042411} {\bibfield  {journal} {\bibinfo
   {journal} {Phys. Rev. A}\ }\textbf {\bibinfo {volume} {109}},\ \bibinfo
  {pages} {042411} (\bibinfo {year} {2024}{\natexlab{b}})}\BibitemShut
  {NoStop}%
\bibitem [{\citenamefont {Ahuja}\ \emph {et~al.}(2025)\citenamefont {Ahuja},
  \citenamefont {Konar},\ and\ \citenamefont {De}}]{ahuja2025}%
  \BibitemOpen
  \bibfield  {author} {\bibinfo {author} {\bibfnamefont {S.}~\bibnamefont
  {Ahuja}}, \bibinfo {author} {\bibfnamefont {T.~K.}\ \bibnamefont {Konar}},\
  and\ \bibinfo {author} {\bibfnamefont {A.~S.}\ \bibnamefont {De}},\ }\href
  {https://arxiv.org/abs/2509.25109} {\bibinfo {title} {Enhancing
  work-extraction in quantum batteries via correlated reservoirs}} (\bibinfo
  {year} {2025}),\ \Eprint {https://arxiv.org/abs/2509.25109} {arXiv:2509.25109
  [quant-ph]} \BibitemShut {NoStop}%
\bibitem [{\citenamefont {Tirone}\ \emph {et~al.}(2023)\citenamefont {Tirone},
  \citenamefont {Salvia}, \citenamefont {Chessa},\ and\ \citenamefont
  {Giovannetti}}]{Tirone2023}%
  \BibitemOpen
  \bibfield  {author} {\bibinfo {author} {\bibfnamefont {S.}~\bibnamefont
  {Tirone}}, \bibinfo {author} {\bibfnamefont {R.}~\bibnamefont {Salvia}},
  \bibinfo {author} {\bibfnamefont {S.}~\bibnamefont {Chessa}},\ and\ \bibinfo
  {author} {\bibfnamefont {V.}~\bibnamefont {Giovannetti}},\ }\bibfield
  {title} {\bibinfo {title} {Work extraction processes from noisy quantum
  batteries: The role of nonlocal resources},\ }\href
  {https://doi.org/10.1103/PhysRevLett.131.060402} {\bibfield  {journal}
  {\bibinfo  {journal} {Phys. Rev. Lett.}\ }\textbf {\bibinfo {volume} {131}},\
  \bibinfo {pages} {060402} (\bibinfo {year} {2023})}\BibitemShut {NoStop}%
\bibitem [{\citenamefont {Tirone}\ \emph {et~al.}(2024)\citenamefont {Tirone},
  \citenamefont {Salvia}, \citenamefont {Chessa},\ and\ \citenamefont
  {Giovannetti}}]{Tirone2024}%
  \BibitemOpen
  \bibfield  {author} {\bibinfo {author} {\bibfnamefont {S.}~\bibnamefont
  {Tirone}}, \bibinfo {author} {\bibfnamefont {R.}~\bibnamefont {Salvia}},
  \bibinfo {author} {\bibfnamefont {S.}~\bibnamefont {Chessa}},\ and\ \bibinfo
  {author} {\bibfnamefont {V.}~\bibnamefont {Giovannetti}},\ }\bibfield
  {title} {\bibinfo {title} {{Quantum work capacitances: Ultimate limits for
  energy extraction on noisy quantum batteries}},\ }\href
  {https://doi.org/10.21468/SciPostPhys.17.2.041} {\bibfield  {journal}
  {\bibinfo  {journal} {SciPost Phys.}\ }\textbf {\bibinfo {volume} {17}},\
  \bibinfo {pages} {041} (\bibinfo {year} {2024})}\BibitemShut {NoStop}%
\bibitem [{\citenamefont {Tirone}\ \emph {et~al.}(2025)\citenamefont {Tirone},
  \citenamefont {Salvia}, \citenamefont {Chessa},\ and\ \citenamefont
  {Giovannetti}}]{Tirone2025}%
  \BibitemOpen
  \bibfield  {author} {\bibinfo {author} {\bibfnamefont {S.}~\bibnamefont
  {Tirone}}, \bibinfo {author} {\bibfnamefont {R.}~\bibnamefont {Salvia}},
  \bibinfo {author} {\bibfnamefont {S.}~\bibnamefont {Chessa}},\ and\ \bibinfo
  {author} {\bibfnamefont {V.}~\bibnamefont {Giovannetti}},\ }\bibfield
  {title} {\bibinfo {title} {Quantum work extraction efficiency for noisy
  quantum batteries: The role of coherence},\ }\href
  {https://doi.org/10.1103/PhysRevA.111.012204} {\bibfield  {journal} {\bibinfo
   {journal} {Phys. Rev. A}\ }\textbf {\bibinfo {volume} {111}},\ \bibinfo
  {pages} {012204} (\bibinfo {year} {2025})}\BibitemShut {NoStop}%
\bibitem [{\citenamefont {Sarkar}\ \emph {et~al.}(2025)\citenamefont {Sarkar},
  \citenamefont {Chaki}, \citenamefont {Ghosh},\ and\ \citenamefont
  {Sen}}]{sarkar2025}%
  \BibitemOpen
  \bibfield  {author} {\bibinfo {author} {\bibfnamefont {A.}~\bibnamefont
  {Sarkar}}, \bibinfo {author} {\bibfnamefont {P.}~\bibnamefont {Chaki}},
  \bibinfo {author} {\bibfnamefont {P.}~\bibnamefont {Ghosh}},\ and\ \bibinfo
  {author} {\bibfnamefont {U.}~\bibnamefont {Sen}},\ }\href
  {https://arxiv.org/abs/2505.16851} {\bibinfo {title} {Fluctuation in energy
  extraction from quantum batteries: How open should the system be to control
  it?}} (\bibinfo {year} {2025}),\ \Eprint {https://arxiv.org/abs/2505.16851}
  {arXiv:2505.16851 [quant-ph]} \BibitemShut {NoStop}%
\bibitem [{\citenamefont {Kamin}\ \emph {et~al.}(2020)\citenamefont {Kamin},
  \citenamefont {Tabesh}, \citenamefont {Salimi}, \citenamefont {Kheirandish},\
  and\ \citenamefont {Santos}}]{Kamin2020}%
  \BibitemOpen
  \bibfield  {author} {\bibinfo {author} {\bibfnamefont {F.~H.}\ \bibnamefont
  {Kamin}}, \bibinfo {author} {\bibfnamefont {F.~T.}\ \bibnamefont {Tabesh}},
  \bibinfo {author} {\bibfnamefont {S.}~\bibnamefont {Salimi}}, \bibinfo
  {author} {\bibfnamefont {F.}~\bibnamefont {Kheirandish}},\ and\ \bibinfo
  {author} {\bibfnamefont {A.~C.}\ \bibnamefont {Santos}},\ }\bibfield  {title}
  {\bibinfo {title} {Non-markovian effects on charging and self-discharging
  process of quantum batteries},\ }\href
  {https://doi.org/10.1088/1367-2630/ab9ee2} {\bibfield  {journal} {\bibinfo
  {journal} {New Journal of Physics}\ }\textbf {\bibinfo {volume} {22}},\
  \bibinfo {pages} {083007} (\bibinfo {year} {2020})}\BibitemShut {NoStop}%
\bibitem [{\citenamefont {Santos}(2021)}]{Santos2021}%
  \BibitemOpen
  \bibfield  {author} {\bibinfo {author} {\bibfnamefont {A.~C.}\ \bibnamefont
  {Santos}},\ }\bibfield  {title} {\bibinfo {title} {Quantum advantage of
  two-level batteries in the self-discharging process},\ }\href
  {https://doi.org/10.1103/PhysRevE.103.042118} {\bibfield  {journal} {\bibinfo
   {journal} {Phys. Rev. E}\ }\textbf {\bibinfo {volume} {103}},\ \bibinfo
  {pages} {042118} (\bibinfo {year} {2021})}\BibitemShut {NoStop}%
\bibitem [{\citenamefont {Xu}\ \emph {et~al.}(2024)\citenamefont {Xu},
  \citenamefont {Li}, \citenamefont {Zhu},\ and\ \citenamefont {Liu}}]{Xu2024}%
  \BibitemOpen
  \bibfield  {author} {\bibinfo {author} {\bibfnamefont {K.}~\bibnamefont
  {Xu}}, \bibinfo {author} {\bibfnamefont {H.-G.}\ \bibnamefont {Li}}, \bibinfo
  {author} {\bibfnamefont {H.-J.}\ \bibnamefont {Zhu}},\ and\ \bibinfo {author}
  {\bibfnamefont {W.-M.}\ \bibnamefont {Liu}},\ }\bibfield  {title} {\bibinfo
  {title} {Inhibiting the self-discharging process of quantum batteries in
  non-markovian noises},\ }\href {https://doi.org/10.1103/PhysRevE.109.054132}
  {\bibfield  {journal} {\bibinfo  {journal} {Phys. Rev. E}\ }\textbf {\bibinfo
  {volume} {109}},\ \bibinfo {pages} {054132} (\bibinfo {year}
  {2024})}\BibitemShut {NoStop}%
\bibitem [{\citenamefont {Morrone}\ \emph {et~al.}(2023)\citenamefont
  {Morrone}, \citenamefont {Rossi},\ and\ \citenamefont
  {Genoni}}]{Morrone2023}%
  \BibitemOpen
  \bibfield  {author} {\bibinfo {author} {\bibfnamefont {D.}~\bibnamefont
  {Morrone}}, \bibinfo {author} {\bibfnamefont {M.~A.}\ \bibnamefont {Rossi}},\
  and\ \bibinfo {author} {\bibfnamefont {M.~G.}\ \bibnamefont {Genoni}},\
  }\bibfield  {title} {\bibinfo {title} {Daemonic ergotropy in continuously
  monitored open quantum batteries},\ }\href
  {https://doi.org/10.1103/PhysRevApplied.20.044073} {\bibfield  {journal}
  {\bibinfo  {journal} {Phys. Rev. Appl.}\ }\textbf {\bibinfo {volume} {20}},\
  \bibinfo {pages} {044073} (\bibinfo {year} {2023})}\BibitemShut {NoStop}%
\bibitem [{\citenamefont {Hadipour}\ and\ \citenamefont
  {Haseli}(2025)}]{Maryam2025}%
  \BibitemOpen
  \bibfield  {author} {\bibinfo {author} {\bibfnamefont {M.}~\bibnamefont
  {Hadipour}}\ and\ \bibinfo {author} {\bibfnamefont {S.}~\bibnamefont
  {Haseli}},\ }\href {https://arxiv.org/abs/2502.05508} {\bibinfo {title}
  {Nonequilibrium quantum batteries: Amplified work extraction through thermal
  bath modulation}} (\bibinfo {year} {2025}),\ \Eprint
  {https://arxiv.org/abs/2502.05508} {arXiv:2502.05508 [quant-ph]} \BibitemShut
  {NoStop}%
\bibitem [{\citenamefont {Lu}\ \emph {et~al.}(2025)\citenamefont {Lu},
  \citenamefont {Tian}, \citenamefont {L\"u},\ and\ \citenamefont
  {Shang}}]{topological_quantumbattery}%
  \BibitemOpen
  \bibfield  {author} {\bibinfo {author} {\bibfnamefont {Z.-G.}\ \bibnamefont
  {Lu}}, \bibinfo {author} {\bibfnamefont {G.}~\bibnamefont {Tian}}, \bibinfo
  {author} {\bibfnamefont {X.-Y.}\ \bibnamefont {L\"u}},\ and\ \bibinfo
  {author} {\bibfnamefont {C.}~\bibnamefont {Shang}},\ }\href
  {https://doi.org/10.1103/PhysRevLett.134.180401} {\bibinfo {title}
  {Topological quantum batteries}} (\bibinfo {year} {2025})\BibitemShut
  {NoStop}%
\bibitem [{\citenamefont {Vigneshwar}\ and\ \citenamefont
  {Sankaranarayanan}(2026)}]{Vigneshwar2026}%
  \BibitemOpen
  \bibfield  {author} {\bibinfo {author} {\bibfnamefont {B.}~\bibnamefont
  {Vigneshwar}}\ and\ \bibinfo {author} {\bibfnamefont {R.}~\bibnamefont
  {Sankaranarayanan}},\ }\bibfield  {title} {\bibinfo {title} {Noise resilience
  of spin quantum battery in the presence of dm interactions},\ }\href
  {https://doi.org/10.1088/1751-8121/ae30b8} {\bibfield  {journal} {\bibinfo
  {journal} {Journal of Physics A: Mathematical and Theoretical}\ }\textbf
  {\bibinfo {volume} {59}},\ \bibinfo {pages} {015302} (\bibinfo {year}
  {2026})}\BibitemShut {NoStop}%
\bibitem [{\citenamefont {Mondal}\ and\ \citenamefont
  {Sen}(2025)}]{mondal2025}%
  \BibitemOpen
  \bibfield  {author} {\bibinfo {author} {\bibfnamefont {S.}~\bibnamefont
  {Mondal}}\ and\ \bibinfo {author} {\bibfnamefont {U.}~\bibnamefont {Sen}},\
  }\href {https://arxiv.org/abs/2507.15811} {\bibinfo {title} {Mpemba effect in
  self-contained quantum refrigerators: accelerated cooling}} (\bibinfo {year}
  {2025}),\ \Eprint {https://arxiv.org/abs/2507.15811} {arXiv:2507.15811
  [quant-ph]} \BibitemShut {NoStop}%
\bibitem [{\citenamefont {Chattopadhyay}\ \emph {et~al.}(2026)\citenamefont
  {Chattopadhyay}, \citenamefont {Santos},\ and\ \citenamefont
  {Misra}}]{chattopadhyay2026}%
  \BibitemOpen
  \bibfield  {author} {\bibinfo {author} {\bibfnamefont {P.}~\bibnamefont
  {Chattopadhyay}}, \bibinfo {author} {\bibfnamefont {J.~F.~G.}\ \bibnamefont
  {Santos}},\ and\ \bibinfo {author} {\bibfnamefont {A.}~\bibnamefont
  {Misra}},\ }\href {https://arxiv.org/abs/2601.05046} {\bibinfo {title}
  {Anomaly to resource: The mpemba effect in quantum thermometry}} (\bibinfo
  {year} {2026}),\ \Eprint {https://arxiv.org/abs/2601.05046} {arXiv:2601.05046
  [quant-ph]} \BibitemShut {NoStop}%
\bibitem [{\citenamefont {Van~Vu}\ and\ \citenamefont
  {Hayakawa}(2025)}]{van_prl_2025}%
  \BibitemOpen
  \bibfield  {author} {\bibinfo {author} {\bibfnamefont {T.}~\bibnamefont
  {Van~Vu}}\ and\ \bibinfo {author} {\bibfnamefont {H.}~\bibnamefont
  {Hayakawa}},\ }\bibfield  {title} {\bibinfo {title} {Thermomajorization
  mpemba effect},\ }\href {https://doi.org/10.1103/PhysRevLett.134.107101}
  {\bibfield  {journal} {\bibinfo  {journal} {Phys. Rev. Lett.}\ }\textbf
  {\bibinfo {volume} {134}},\ \bibinfo {pages} {107101} (\bibinfo {year}
  {2025})}\BibitemShut {NoStop}%
\bibitem [{\citenamefont {Summer}\ \emph {et~al.}(2026)\citenamefont {Summer},
  \citenamefont {Moroder}, \citenamefont {Bettmann}, \citenamefont {Turkeshi},
  \citenamefont {Marvian},\ and\ \citenamefont {Goold}}]{summer_prx_2026}%
  \BibitemOpen
  \bibfield  {author} {\bibinfo {author} {\bibfnamefont {A.}~\bibnamefont
  {Summer}}, \bibinfo {author} {\bibfnamefont {M.}~\bibnamefont {Moroder}},
  \bibinfo {author} {\bibfnamefont {L.~P.}\ \bibnamefont {Bettmann}}, \bibinfo
  {author} {\bibfnamefont {X.}~\bibnamefont {Turkeshi}}, \bibinfo {author}
  {\bibfnamefont {I.}~\bibnamefont {Marvian}},\ and\ \bibinfo {author}
  {\bibfnamefont {J.}~\bibnamefont {Goold}},\ }\bibfield  {title} {\bibinfo
  {title} {A resource-theoretical unification of mpemba effects: classical and
  quantum},\ }\href {https://doi.org/10.1103/rbt4-psfd} {\bibfield  {journal}
  {\bibinfo  {journal} {Phys. Rev. X}\ ,\ } (\bibinfo {year}
  {2026})}\BibitemShut {NoStop}%
\bibitem [{\citenamefont {Feyisa}\ \emph {et~al.}(2026)\citenamefont {Feyisa},
  \citenamefont {Ku}, \citenamefont {You},\ and\ \citenamefont
  {Jen}}]{feyisa2026}%
  \BibitemOpen
  \bibfield  {author} {\bibinfo {author} {\bibfnamefont {C.~G.}\ \bibnamefont
  {Feyisa}}, \bibinfo {author} {\bibfnamefont {H.-Y.}\ \bibnamefont {Ku}},
  \bibinfo {author} {\bibfnamefont {J.~S.}\ \bibnamefont {You}},\ and\ \bibinfo
  {author} {\bibfnamefont {H.~H.}\ \bibnamefont {Jen}},\ }\href
  {https://arxiv.org/abs/2604.07926} {\bibinfo {title} {Informational mpemba
  effect for fast state purification in non-hermitian system}} (\bibinfo {year}
  {2026}),\ \Eprint {https://arxiv.org/abs/2604.07926} {arXiv:2604.07926
  [quant-ph]} \BibitemShut {NoStop}%
\bibitem [{\citenamefont {Moroder}\ \emph {et~al.}(2026)\citenamefont
  {Moroder}, \citenamefont {Binder},\ and\ \citenamefont
  {Goold}}]{moroder2026}%
  \BibitemOpen
  \bibfield  {author} {\bibinfo {author} {\bibfnamefont {M.}~\bibnamefont
  {Moroder}}, \bibinfo {author} {\bibfnamefont {F.~C.}\ \bibnamefont
  {Binder}},\ and\ \bibinfo {author} {\bibfnamefont {J.}~\bibnamefont
  {Goold}},\ }\href {https://arxiv.org/abs/2603.24183} {\bibinfo {title}
  {Digitally optimized initializations for fast thermodynamic computing}}
  (\bibinfo {year} {2026}),\ \Eprint {https://arxiv.org/abs/2603.24183}
  {arXiv:2603.24183 [cond-mat.stat-mech]} \BibitemShut {NoStop}%
\bibitem [{\citenamefont {Medina}\ \emph {et~al.}(2025)\citenamefont {Medina},
  \citenamefont {Culhane}, \citenamefont {Binder}, \citenamefont {Landi},\ and\
  \citenamefont {Goold}}]{Medina2024}%
  \BibitemOpen
  \bibfield  {author} {\bibinfo {author} {\bibfnamefont {I.}~\bibnamefont
  {Medina}}, \bibinfo {author} {\bibfnamefont {O.}~\bibnamefont {Culhane}},
  \bibinfo {author} {\bibfnamefont {F.~C.}\ \bibnamefont {Binder}}, \bibinfo
  {author} {\bibfnamefont {G.~T.}\ \bibnamefont {Landi}},\ and\ \bibinfo
  {author} {\bibfnamefont {J.}~\bibnamefont {Goold}},\ }\bibfield  {title}
  {\bibinfo {title} {Anomalous discharging of quantum batteries: The ergotropic
  mpemba effect},\ }\href {https://doi.org/10.1103/PhysRevLett.134.220402}
  {\bibfield  {journal} {\bibinfo  {journal} {Phys. Rev. Lett.}\ }\textbf
  {\bibinfo {volume} {134}},\ \bibinfo {pages} {220402} (\bibinfo {year}
  {2025})}\BibitemShut {NoStop}%
\bibitem [{\citenamefont {Joshi}\ and\ \citenamefont
  {Mahesh}(2022)}]{nmr_battery}%
  \BibitemOpen
  \bibfield  {author} {\bibinfo {author} {\bibfnamefont {J.}~\bibnamefont
  {Joshi}}\ and\ \bibinfo {author} {\bibfnamefont {T.~S.}\ \bibnamefont
  {Mahesh}},\ }\bibfield  {title} {\bibinfo {title} {Experimental investigation
  of a quantum battery using star-topology nmr spin systems},\ }\href
  {https://doi.org/10.1103/PhysRevA.106.042601} {\bibfield  {journal} {\bibinfo
   {journal} {Phys. Rev. A}\ }\textbf {\bibinfo {volume} {106}},\ \bibinfo
  {pages} {042601} (\bibinfo {year} {2022})}\BibitemShut {NoStop}%
\bibitem [{\citenamefont {Hu}\ \emph {et~al.}(2022)\citenamefont {Hu},
  \citenamefont {Qiu}, \citenamefont {Souza}, \citenamefont {Yuan},
  \citenamefont {Zhou}, \citenamefont {Zhang}, \citenamefont {Chu},
  \citenamefont {Pan}, \citenamefont {Hu}, \citenamefont {Li}, \citenamefont
  {Xu}, \citenamefont {Zhong}, \citenamefont {Liu}, \citenamefont {Yan},
  \citenamefont {Tan}, \citenamefont {Bachelard}, \citenamefont {Villas-Boas},
  \citenamefont {Santos},\ and\ \citenamefont {Yu}}]{superconductQBexp}%
  \BibitemOpen
  \bibfield  {author} {\bibinfo {author} {\bibfnamefont {C.-K.}\ \bibnamefont
  {Hu}}, \bibinfo {author} {\bibfnamefont {J.}~\bibnamefont {Qiu}}, \bibinfo
  {author} {\bibfnamefont {P.~J.~P.}\ \bibnamefont {Souza}}, \bibinfo {author}
  {\bibfnamefont {J.}~\bibnamefont {Yuan}}, \bibinfo {author} {\bibfnamefont
  {Y.}~\bibnamefont {Zhou}}, \bibinfo {author} {\bibfnamefont {L.}~\bibnamefont
  {Zhang}}, \bibinfo {author} {\bibfnamefont {J.}~\bibnamefont {Chu}}, \bibinfo
  {author} {\bibfnamefont {X.}~\bibnamefont {Pan}}, \bibinfo {author}
  {\bibfnamefont {L.}~\bibnamefont {Hu}}, \bibinfo {author} {\bibfnamefont
  {J.}~\bibnamefont {Li}}, \bibinfo {author} {\bibfnamefont {Y.}~\bibnamefont
  {Xu}}, \bibinfo {author} {\bibfnamefont {Y.}~\bibnamefont {Zhong}}, \bibinfo
  {author} {\bibfnamefont {S.}~\bibnamefont {Liu}}, \bibinfo {author}
  {\bibfnamefont {F.}~\bibnamefont {Yan}}, \bibinfo {author} {\bibfnamefont
  {D.}~\bibnamefont {Tan}}, \bibinfo {author} {\bibfnamefont {R.}~\bibnamefont
  {Bachelard}}, \bibinfo {author} {\bibfnamefont {C.~J.}\ \bibnamefont
  {Villas-Boas}}, \bibinfo {author} {\bibfnamefont {A.~C.}\ \bibnamefont
  {Santos}},\ and\ \bibinfo {author} {\bibfnamefont {D.}~\bibnamefont {Yu}},\
  }\bibfield  {title} {\bibinfo {title} {Optimal charging of a superconducting
  quantum battery},\ }\href {https://doi.org/10.1088/2058-9565/ac8444}
  {\bibfield  {journal} {\bibinfo  {journal} {Quantum Science and Technology}\
  }\textbf {\bibinfo {volume} {7}},\ \bibinfo {pages} {045018} (\bibinfo {year}
  {2022})}\BibitemShut {NoStop}%
\bibitem [{\citenamefont {Carollo}\ \emph
  {et~al.}(2021{\natexlab{b}})\citenamefont {Carollo}, \citenamefont
  {Lasanta},\ and\ \citenamefont {Lesanovsky}}]{carollo_prl_2021}%
  \BibitemOpen
  \bibfield  {author} {\bibinfo {author} {\bibfnamefont {F.}~\bibnamefont
  {Carollo}}, \bibinfo {author} {\bibfnamefont {A.}~\bibnamefont {Lasanta}},\
  and\ \bibinfo {author} {\bibfnamefont {I.}~\bibnamefont {Lesanovsky}},\
  }\bibfield  {title} {\bibinfo {title} {Exponentially accelerated approach to
  stationarity in markovian open quantum systems through the mpemba effect},\
  }\href {https://doi.org/10.1103/PhysRevLett.127.060401} {\bibfield  {journal}
  {\bibinfo  {journal} {Phys. Rev. Lett.}\ }\textbf {\bibinfo {volume} {127}},\
  \bibinfo {pages} {060401} (\bibinfo {year} {2021}{\natexlab{b}})}\BibitemShut
  {NoStop}%
\bibitem [{\citenamefont {Allahverdyan}\ \emph {et~al.}(2004)\citenamefont
  {Allahverdyan}, \citenamefont {Balian},\ and\ \citenamefont
  {Nieuwenhuizen}}]{Allahverdyan2004}%
  \BibitemOpen
  \bibfield  {author} {\bibinfo {author} {\bibfnamefont {A.~E.}\ \bibnamefont
  {Allahverdyan}}, \bibinfo {author} {\bibfnamefont {R.}~\bibnamefont
  {Balian}},\ and\ \bibinfo {author} {\bibfnamefont {T.~M.}\ \bibnamefont
  {Nieuwenhuizen}},\ }\bibfield  {title} {\bibinfo {title} {Maximal work
  extraction from finite quantum systems},\ }\href
  {https://doi.org/10.1209/epl/i2004-10101-2} {\bibfield  {journal} {\bibinfo
  {journal} {Europhysics Letters (EPL)}\ }\textbf {\bibinfo {volume} {67}},\
  \bibinfo {pages} {565–571} (\bibinfo {year} {2004})}\BibitemShut {NoStop}%
\bibitem [{\citenamefont {Malavazi}\ \emph
  {et~al.}(2025{\natexlab{a}})\citenamefont {Malavazi}, \citenamefont {Ahmadi},
  \citenamefont {Horodecki},\ and\ \citenamefont {Dieguez}}]{malavazi2025}%
  \BibitemOpen
  \bibfield  {author} {\bibinfo {author} {\bibfnamefont {A.~H.~A.}\
  \bibnamefont {Malavazi}}, \bibinfo {author} {\bibfnamefont {B.}~\bibnamefont
  {Ahmadi}}, \bibinfo {author} {\bibfnamefont {P.}~\bibnamefont {Horodecki}},\
  and\ \bibinfo {author} {\bibfnamefont {P.~R.}\ \bibnamefont {Dieguez}},\
  }\href {https://arxiv.org/abs/2510.25549} {\bibinfo {title}
  {Charge-preserving operations in quantum batteries}} (\bibinfo {year}
  {2025}{\natexlab{a}}),\ \Eprint {https://arxiv.org/abs/2510.25549}
  {arXiv:2510.25549 [quant-ph]} \BibitemShut {NoStop}%
\bibitem [{\citenamefont {Nielsen}\ and\ \citenamefont
  {Chuang}(2010)}]{nielsen2010}%
  \BibitemOpen
  \bibfield  {author} {\bibinfo {author} {\bibfnamefont {M.~A.}\ \bibnamefont
  {Nielsen}}\ and\ \bibinfo {author} {\bibfnamefont {I.~L.}\ \bibnamefont
  {Chuang}},\ }\href@noop {} {\emph {\bibinfo {title} {Quantum computation and
  quantum information}}}\ (\bibinfo  {publisher} {Cambridge university press},\
  \bibinfo {year} {2010})\BibitemShut {NoStop}%
\bibitem [{\citenamefont {Breuer}\ and\ \citenamefont
  {Petruccione}(2002)}]{breuer2002}%
  \BibitemOpen
  \bibfield  {author} {\bibinfo {author} {\bibfnamefont {H.~P.}\ \bibnamefont
  {Breuer}}\ and\ \bibinfo {author} {\bibfnamefont {F.}~\bibnamefont
  {Petruccione}},\ }\href@noop {} {\emph {\bibinfo {title} {The Theory of Open
  Quantum Systems}}}\ (\bibinfo  {publisher} {Oxford University Press,
  Oxford},\ \bibinfo {year} {2002})\BibitemShut {NoStop}%
\bibitem [{\citenamefont {Rivas}\ and\ \citenamefont
  {Huelga}(2012)}]{rivas2012}%
  \BibitemOpen
  \bibfield  {author} {\bibinfo {author} {\bibfnamefont {A.}~\bibnamefont
  {Rivas}}\ and\ \bibinfo {author} {\bibfnamefont {S.~F.}\ \bibnamefont
  {Huelga}},\ }\href@noop {} {\emph {\bibinfo {title} {Open Quantum Systems: An
  Introduction}}}\ (\bibinfo  {publisher} {SpringerBriefs in Physics, Springer,
  Spain},\ \bibinfo {year} {2012})\BibitemShut {NoStop}%
\bibitem [{\citenamefont {Malavazi}\ \emph
  {et~al.}(2025{\natexlab{b}})\citenamefont {Malavazi}, \citenamefont {Sagar},
  \citenamefont {Ahmadi},\ and\ \citenamefont {Dieguez}}]{Malavazi_prxq_2025}%
  \BibitemOpen
  \bibfield  {author} {\bibinfo {author} {\bibfnamefont {A.~H.}\ \bibnamefont
  {Malavazi}}, \bibinfo {author} {\bibfnamefont {R.}~\bibnamefont {Sagar}},
  \bibinfo {author} {\bibfnamefont {B.}~\bibnamefont {Ahmadi}},\ and\ \bibinfo
  {author} {\bibfnamefont {P.~R.}\ \bibnamefont {Dieguez}},\ }\bibfield
  {title} {\bibinfo {title} {Two-time weak-measurement protocol for ergotropy
  protection in open quantum batteries},\ }\href
  {https://doi.org/10.1103/bv4w-jr6q} {\bibfield  {journal} {\bibinfo
  {journal} {PRX Energy}\ }\textbf {\bibinfo {volume} {4}},\ \bibinfo {pages}
  {023011} (\bibinfo {year} {2025}{\natexlab{b}})}\BibitemShut {NoStop}%
\bibitem [{\citenamefont {Furtado}\ and\ \citenamefont
  {Santos}(2025)}]{Furtado2025}%
  \BibitemOpen
  \bibfield  {author} {\bibinfo {author} {\bibfnamefont {J.}~\bibnamefont
  {Furtado}}\ and\ \bibinfo {author} {\bibfnamefont {A.~C.}\ \bibnamefont
  {Santos}},\ }\bibfield  {title} {\bibinfo {title} {Enhanced quantum mpemba
  effect with squeezed thermal reservoirs},\ }\href
  {https://doi.org/https://doi.org/10.1016/j.aop.2025.170135} {\bibfield
  {journal} {\bibinfo  {journal} {Annals of Physics}\ }\textbf {\bibinfo
  {volume} {480}},\ \bibinfo {pages} {170135} (\bibinfo {year}
  {2025})}\BibitemShut {NoStop}%
\bibitem [{\citenamefont {Lorenzo}\ \emph {et~al.}(2013)\citenamefont
  {Lorenzo}, \citenamefont {Plastina},\ and\ \citenamefont
  {Paternostro}}]{Lorenzo2013}%
  \BibitemOpen
  \bibfield  {author} {\bibinfo {author} {\bibfnamefont {S.}~\bibnamefont
  {Lorenzo}}, \bibinfo {author} {\bibfnamefont {F.}~\bibnamefont {Plastina}},\
  and\ \bibinfo {author} {\bibfnamefont {M.}~\bibnamefont {Paternostro}},\
  }\bibfield  {title} {\bibinfo {title} {Geometrical characterization of
  non-markovianity},\ }\href {https://doi.org/10.1103/PhysRevA.88.020102}
  {\bibfield  {journal} {\bibinfo  {journal} {Phys. Rev. A}\ }\textbf {\bibinfo
  {volume} {88}},\ \bibinfo {pages} {020102} (\bibinfo {year}
  {2013})}\BibitemShut {NoStop}%
\bibitem [{\citenamefont {Linden}\ \emph {et~al.}(2010)\citenamefont {Linden},
  \citenamefont {Popescu},\ and\ \citenamefont {Skrzypczyk}}]{Linden2010}%
  \BibitemOpen
  \bibfield  {author} {\bibinfo {author} {\bibfnamefont {N.}~\bibnamefont
  {Linden}}, \bibinfo {author} {\bibfnamefont {S.}~\bibnamefont {Popescu}},\
  and\ \bibinfo {author} {\bibfnamefont {P.}~\bibnamefont {Skrzypczyk}},\
  }\bibfield  {title} {\bibinfo {title} {How small can thermal machines be? the
  smallest possible refrigerator},\ }\href
  {https://doi.org/10.1103/PhysRevLett.105.130401} {\bibfield  {journal}
  {\bibinfo  {journal} {Phys. Rev. Lett.}\ }\textbf {\bibinfo {volume} {105}},\
  \bibinfo {pages} {130401} (\bibinfo {year} {2010})}\BibitemShut {NoStop}%
\bibitem [{\citenamefont {Campisi}\ \emph {et~al.}(2015)\citenamefont
  {Campisi}, \citenamefont {Pekola},\ and\ \citenamefont
  {Fazio}}]{Campisi2015}%
  \BibitemOpen
  \bibfield  {author} {\bibinfo {author} {\bibfnamefont {M.}~\bibnamefont
  {Campisi}}, \bibinfo {author} {\bibfnamefont {J.}~\bibnamefont {Pekola}},\
  and\ \bibinfo {author} {\bibfnamefont {R.}~\bibnamefont {Fazio}},\ }\bibfield
   {title} {\bibinfo {title} {Nonequilibrium fluctuations in quantum heat
  engines: theory, example, and possible solid state experiments},\ }\href
  {https://doi.org/10.1088/1367-2630/17/3/035012} {\bibfield  {journal}
  {\bibinfo  {journal} {New Journal of Physics}\ }\textbf {\bibinfo {volume}
  {17}},\ \bibinfo {pages} {035012} (\bibinfo {year} {2015})}\BibitemShut
  {NoStop}%
\bibitem [{\citenamefont {Allahverdyan}\ \emph {et~al.}(2010)\citenamefont
  {Allahverdyan}, \citenamefont {Hovhannisyan},\ and\ \citenamefont
  {Mahler}}]{Allahverdyan2010}%
  \BibitemOpen
  \bibfield  {author} {\bibinfo {author} {\bibfnamefont {A.~E.}\ \bibnamefont
  {Allahverdyan}}, \bibinfo {author} {\bibfnamefont {K.}~\bibnamefont
  {Hovhannisyan}},\ and\ \bibinfo {author} {\bibfnamefont {G.}~\bibnamefont
  {Mahler}},\ }\bibfield  {title} {\bibinfo {title} {Optimal refrigerator},\
  }\href {https://doi.org/10.1103/PhysRevE.81.051129} {\bibfield  {journal}
  {\bibinfo  {journal} {Phys. Rev. E}\ }\textbf {\bibinfo {volume} {81}},\
  \bibinfo {pages} {051129} (\bibinfo {year} {2010})}\BibitemShut {NoStop}%
\bibitem [{\citenamefont {Adhikary}(2026)}]{adhikary2026}%
  \BibitemOpen
  \bibfield  {author} {\bibinfo {author} {\bibfnamefont {K.}~\bibnamefont
  {Adhikary}},\ }\href {https://arxiv.org/abs/2604.11313} {\bibinfo {title}
  {Engineered non-gaussian coherence as a thermodynamic resource for quantum
  batteries}} (\bibinfo {year} {2026}),\ \Eprint
  {https://arxiv.org/abs/2604.11313} {arXiv:2604.11313 [quant-ph]} \BibitemShut
  {NoStop}%
\bibitem [{\citenamefont {Srivastav}\ \emph {et~al.}(2025)\citenamefont
  {Srivastav}, \citenamefont {Pandey}, \citenamefont {Mohan},\ and\
  \citenamefont {Pati}}]{Srivastav_pra_2025}%
  \BibitemOpen
  \bibfield  {author} {\bibinfo {author} {\bibfnamefont {A.}~\bibnamefont
  {Srivastav}}, \bibinfo {author} {\bibfnamefont {V.}~\bibnamefont {Pandey}},
  \bibinfo {author} {\bibfnamefont {B.}~\bibnamefont {Mohan}},\ and\ \bibinfo
  {author} {\bibfnamefont {A.~K.}\ \bibnamefont {Pati}},\ }\bibfield  {title}
  {\bibinfo {title} {Family of exact and inexact quantum speed limits for
  completely positive and trace-preserving dynamics},\ }\href
  {https://doi.org/10.1103/npkr-c4vf} {\bibfield  {journal} {\bibinfo
  {journal} {Phys. Rev. A}\ }\textbf {\bibinfo {volume} {112}},\ \bibinfo
  {pages} {052204} (\bibinfo {year} {2025})}\BibitemShut {NoStop}%
\bibitem [{\citenamefont {Gyamfi}(2020)}]{Gyamfi2020}%
  \BibitemOpen
  \bibfield  {author} {\bibinfo {author} {\bibfnamefont {J.~A.}\ \bibnamefont
  {Gyamfi}},\ }\bibfield  {title} {\bibinfo {title} {Fundamentals of quantum
  mechanics in liouville space},\ }\href
  {https://doi.org/10.1088/1361-6404/ab9fdd} {\bibfield  {journal} {\bibinfo
  {journal} {European Journal of Physics}\ }\textbf {\bibinfo {volume} {41}},\
  \bibinfo {pages} {063002} (\bibinfo {year} {2020})}\BibitemShut {NoStop}%
\end{thebibliography}%

\end{document}